\documentclass{GenThesis}

\newcommand{\ket}[1]{\ensuremath{|#1\rangle}\xspace}
\newcommand{\bra}[1]{\ensuremath{\langle #1|}\xspace}

\newcommand{\phd}{{\phantom{\dagger}}}

\renewcommand\Im{\operatorname{Im}}

\title{Statics and dynamics of weakly coupled antiferromagnetic spin-$1/2$ ladders in a magnetic field}

\author{Pierre Bouillot}
\faculty{Sciences}
\professora{T. Giamarchi}
\professorb{C. Kollath}
\sectionshort{Physique}
\cityofbirth{Vernier (GE)}
\country{Suisse}
\thesisnumber{4356}
\sectionp{Section de Physique}
\department{D\'{e}partement de la mati\`ere condens\'ee}
\university{Universit\'{e} de Gen\`{e}ve}
\yeargrad{2011}
\degree{Philosophi\ae Doctor (PhD)}
\degreedate{30/09/2011}
\memberA{}{}{}
\memberB{}{}{}

\approvaldate{5 Septembre 2011}
\dean{Jean-Marc TRISCONE}

\usepackage[Sonny]{fncychap}        
\usepackage{epigraph}

\begin{document}


\maketitle       
\fancyend


\frontmatter

\begin{dedication} 

\`a ma famille

\end{dedication}


\fancyend

\begin{acknowledgements}      

This PhD thesis would not have been possible without the help and the support of the people around me. Although it is practically impossible to mention them all, I would like in particular thank:

My two supervisors T. Giamarchi and C. Kollath for their constant support and availability for professional or personal discussions. Thanks to their very pedagogical explanations the quantum world became easier to access. I also thank them for their hard work proofreading my articles and thesis, and for their indulgence for my poor English. Their hard work and their strong motivation have set an example that I hope to reach someday.

A. Läuchli for his help and for welcoming me at IRMMA. He gave me the opportunity to meet the people from IRRMA and EPFL with who I had fruitful discussions.

C. Berthod for his technical support without which I would have been blocked hours on several programming issues. 

J.-C. Caux, R. Citro, A. Furusaki, S. C. Furuya, E. Orignac, M. Oshikawa for theoretical discussions about different parts of this thesis. 

C. Berthier, M. Horvati\'c, M. Klanj\v{s}ek, H. Mayaffre, C. Rüegg, D. Schmidiger, B. Thielemann, S. Ward and A. Zheludev for showing me the experimental point of view on quantum magnetism. They helped me to connect with success the experimental physics with its theoretical description. In addition, thanks to the visit of their impressive experimental facilities, they allowed me to understand how the experimental techniques are implemented.

My office mates E. Agoritsas, P. Barmettler, S. Bustingorry, P. Chudzi\'nski, T. Ewart, A. Iucci, A. Kantian, A. Klauser, A. Kosenkov, V. Lecompte, G. Leon, A. Lobos,  D. Poletti A. Tokuno and M. Zvonarev for their kind support and friendship along these years at the university.

My teaching colleagues J.-P. Eckmann, P. A. Jacquet, Y. Lisunova, P. Paruch, N. Reyren, B. Ziegler and all the students who made the Mechanics classes so interesting and rewarding.

D. Bichsel for his numerous precious advices all during my professional life.

The Swiss National Science Fundation under MaNEP and Division II, and the financial and academic support of the Université de Genève, in particular the Condensed Matter Physics and the Theoretical Physics Departments and their staff. 

And last but not the least, all my family, my close friends and Philippe for their presence, their love and their daily support. I dedicate this PhD thesis to you without who I could not have found the motivation to work so hard for accomplishing it. 

\end{acknowledgements}


\fancyend

\setcounter{page}{1}


\chapter*{Résumé en français}        
Dans cette thèse, nous étudions les propriétés statiques et dynamiques des échelles de spin-$1/2$ soumises à un champ magnétique. Faiblement couplées, ces échelles permettent d'étudier à la fois la physique du liquide de Luttinger (LL) apparaissant dans de nombreux systèmes unidimensionnels (1D) et la condensation de Bose-Einstein (BEC) qui est un effet typiquement tridimensionnel (3D). Notre travail a été en grande partie motivé par le composé $\mathrm{(C_5H_{12}N)_2CuBr_4}$ (BPCB) récemment synthétisé. Ce matériau est considéré comme ayant une simple structure d'échelles couplées où des interactions plus complexes (frustration, anisotropie ou Dzyloshinskii-Moriya) restent faibles. De plus, ses couplages d'échange sont suffisamment faibles pour rendre l'ensemble de son diagramme de phase expérimentalement accessible en appliquant un champ magnétique.

Pour étudier ces systèmes, nous utilisons une combinaison de méthodes analytiques (théorie du liquide de Luttinger et technique de bosonisation) et numériques (groupe de renormalisation de la matrice densité (DMRG)). La première est une théorie des champs permettant de décrire la physique de basse énergie de nombreux systèmes 1D sans bande interdite telle que la phase aimantée des échelles de spin-$1/2$. La seconde est une méthode variationnelle particulièrement bien adaptée aux systèmes 1D et permettant de calculer leurs propriétés à température nulle. Elle permet également d'extraire les paramètres du LL à partir du calcul des corrélations statiques et de l'aimantation pour obtenir une description quantitative de la physique de basse énergie. La méthode DMRG a été récemment étendue au calcul des propriétés à température finie ainsi que leur évolution temporelle. Cette dernière extension est notamment utilisée, dans ce travail, pour calculer les fonctions de corrélation dynamiques. Le faible couplage inter-échelles est, quant à lui, pris en compte en utilisant une approximation de champ moyen.

Dans un premier temps, nous explorons le diagramme de phase des échelles non-couplées en étudiant leurs propriétés thermodynamiques. Nous calculons leur magnétisation et leur chaleur spécifique. Ces deux quantités révèlent des caractéristiques de basse température fidèles à la description du LL. Au contraire, leur comportement à haute température est principalement dicté par les excitations de triplets à haute énergie. On détermine également à partir des extréma de ces deux quantités la limite de validité approximative de la description du LL. La comparaison de nos calculs avec les mesures effectuées sur BPCB dans un domaine complet de température et champ magnétique est excellente. Ceci confirme la simple structure d'échelle de ce composé. De plus, le domaine de validité de la description du LL pour BPCB est un ordre de grandeur plus grand que la température de transition de l'ordre 3D à basse température. Ce qui laisse ainsi un large domaine de température pour tester le liquide de Luttinger. 

Pour ce faire, nous calculons les prédictions du LL du temps de relaxation mesuré par résonance magnétique nucléaire (NMR). De plus, la prise en compte du couplage inter-échelles par une approximation de champ moyen permet d'accéder, à l'aide de la description du LL des échelles isolées, à la température critique et l'ordre 3D transverse antiferromagnétique associé. Ces derniers caractérisent la transition BEC. La mesure de ces trois quantités à l'aide d'expériences NMR et de diffraction de neutrons sont toutes en très bon accord avec les prédictions du  LL. Elles permettent donc de tester trois différentes fonctions de corrélations calculées à partir des mêmes paramètres du LL. Ceci fournit le premier test {\it quantitatif} de la théorie du liquide de Luttinger.

Dans un second temps, nous avons étudié les fonctions de corrélation dynamiques à température nulle des échelles non couplées. Les excitations fournissent d'importantes informations sur le système et permettent ainsi de le caractériser en détail. En particulier, nous présentons leur intéressante évolution avec le champ magnétique appliqué et pour différents couplages. Le continuum
apparaissant à basse énergie est qualitativement décrit à l'aide d'une approximation par une chaîne de spin-$1/2$ anisotrope équivalente à l'approximation de fort couplage de l'échelle. Cette approximation n'est en revanche pas valable pour la description des excitations de moyenne et haute énergie. En effet, ces dernières nécessitent la prise en compte des triplets de haute énergie négligés par cette approximation. Fait intéressant, les excitations de moyenne énergie peuvent être décrites par un modèle t-J et présentent donc des caractéristiques typiques des systèmes itinérants.  On vérifie de plus que l'évaluation numérique de ces corrélations valable à moyenne et haute énergie convergent correctement sur la description donnée par le LL restreint aux excitations de basse énergies. 

Les mesures de diffusion de neutron inélastique (INS) étant directement reliées aux corrélations dynamiques, on fournit une prédiction complète des spectres mesurés sur BPCB. Il est gratifiant de noter que la résolution en énergie et quantité de mouvement de nos calculs est actuellement meilleure que celle des expériences. Les mesures se limitant pour l'instant aux excitations de basse énergie, il est difficile d'y distinguer la différence avec la prédiction fournie par une échelle de spin-$1/2$ et celle donnée par l'approximation de fort couplage. Des mesures du spectre de moyenne et haute énergie comportant des excitations caractéristiques du modèle sous-jacent permettraient de raffiner l'étude expérimentale sur BPCB.

Plus généralement, d'un point de vue conceptuel ou en lien avec BPCB plusieurs points nécessitent une étude plus étendue.

La prise en compte du couplage inter-échelles et de la température pour les prédictions théoriques devrait être étendue aux quantités dynamiques. En effet, il serait intéressant d'étudier leur impact sur les excitations du système. Des phénomènes tels que le déplacement ou l'élargissement des excitations avec la température ont été observés dans d'autres systèmes magnétiques. Une étude détaillée près des champs critiques serait particulièrement recommandée. Dans ces régimes, le système subit une transition dimensionnelle entre un état 1D (où les excitations sont essentiellement fermioniques) et un état 3D (où les excitations ont une description bosonique). De plus, il serait important de prendre en compte la structure tridimensionnelle réelle du composé. Ceci nous permettrait également de comprendre plus en détail les déviations de l'ordre 3D mesuré expérimentalement sur BPCB avec notre prédiction utilisant une approximation de champ moyen.

Récemment, de faibles anisotropies ont été détectées sur BPCB à l'aide de mesures de résonance électron-spin. Les effets de ces anisotropies sur la physique des échelles étant actuellement peu connue, une étude approfondie de ces phénomènes serait pertinente. Elle permettrait notamment de comprendre certaines déviations entre les mesures sur BPCB et nos prédictions théoriques utilisant des échelles isotropes.

Récemment des singularités apparaissant dans les corrélations dynamiques et sortant de la description du LL ont été mises en évidence. Or, pour qu'une étude de ces effets sur les excitations des échelles soit possible, il faudrait améliorer la précision des corrélations calculées numériquement. L'optimisation des algorithmes existant ou le développement de nouvelles méthodes plus performantes pour atteindre ce but fait donc partie intégrante des extensions futures.

Etant donné que le composé BPCB est maintenant bien modélisé par de simples échelles de spin-$1/2$, il serait intéressant de lui ajouter des impuretés pour y étudier des phénomènes plus complexes. En effet, la présence de désordre permettrait l'étude du verre de Bose alors que l'ajout de porteurs de charges pourraient faire apparaître un état supraconducteur. D'un autre côté, notre démarche étant assez générale, elle pourrait être étendue à l'étude d'autres composés d'échelle tels que DIMPY récemment synthétisé ou des structures plus complexes.



\fancyend


\chapter*{Abstract}        

We investigate weakly coupled spin-$1/2$ ladders in a magnetic field. The work is motivated by recent experiments on the compound $\mathrm{(C_5H_{12}N)_2CuBr_4}$ (BPCB). We use a combination of numerical and analytical methods, in particular the density matrix renormalization group (DMRG) technique, to explore the phase diagram and the excitation spectra of such a system. We give detailed results on the temperature dependence of the magnetization and the specific heat, and the magnetic field dependence of the nuclear magnetic resonance (NMR) relaxation rate of single ladders. For coupled ladders, treating the weak interladder coupling within a mean field approach, we compute the transition temperature of triplet condensation and its corresponding antiferromagnetic order parameter. Existing experimental measurements are discussed and compared to our theoretical results. Furthermore we compute,
using time dependent DMRG, the dynamical correlations of a single spin
ladder. Our results allow to directly describe the inelastic neutron scattering cross
section up to high energies. We focus on the evolution of the spectra with the
magnetic field and compare their behavior for different couplings. The
characteristic features of the spectra are interpreted using different
analytical approaches such as the mapping onto a spin chain, a Luttinger
liquid (LL) or onto a t-J model. For values of parameters for which such
measurements exist, we compare our results to inelastic neutron scattering experiments on the
compound BPCB and find excellent agreement. We make additional predictions for the high energy part of the spectrum that are potentially testable in future experiments.



\fancyend

\setcounter{secnumdepth}{3} 
\setcounter{tocdepth}{3}    
\tableofcontents            
\fancyend









\mainmatter


\selectlanguage{english}


\chapter{Introduction}


In many condensed matter systems the quantum fluctuations and the interactions between particles play a crucial role.
Various important effects such as the high temperature superconductivity,  the fractional quantum Hall effect or Mott insulators arise in the so-called strongly correlated quantum systems. In these systems the interaction is comparable to other energy scales and thus needs to be treated on equal footing. In a Mott insulator, due to the Pauli principle, the interplay between interactions and kinetic energy can induce a
strong antiferromagnetic spin superexchange. Such exchange
leads to a remarkable dynamics for the spin
degrees of freedom. On a simple square lattice, the antiferromagnetic exchange can stabilize an antiferromagnetic order. By variations in dimensionality and connectivity of the lattice a variety of complex phenomena can arise~\cite{Auerbach_book_magnetism}, for instance, spin liquid~\cite{lecheminant_revue_1d}, Bose-Einstein condensation~\cite{giamarchi_ladder_coupled,nikuni00_tlcucl3_bec,ruegg03_tlcucl3}
(BEC), Luttinger liquid~\cite{giamarchi_book_1d,gogolin_1dbook} (LL) or Haldane gap~\cite{haldane_gap}. Recently, among those effects two fascinating situations in which the interaction strongly favors the formation of dimers have been explored in detail. 

The first situation concerns a
high dimensional system in which the
antiferromagnetic coupling can lead to a spin liquid state made of singlets along the dimers. In such a spin liquid the
application of a magnetic field leads to the creation of
triplons which are spin-$1$ excitations. The triplons which
behave essentially like itinerant bosons can condense leading
to a quantum phase transition that is in the universality class
of BEC. Such transitions have been explored experimentally and theoretically in a
large variety of materials, belonging to different structures
and dimensionalities~\cite{giamarchi_BEC_dimers_review}.
On the other hand, low dimensional systems behave quite
differently. Quantum fluctuations
are extreme, and no ordered state is usually possible. In many quasi one-dimensional systems the
ground state properties are described by
LL physics that predicts a {\it quasi} long range order. The elementary excitations are spin-$1/2$ excitations (spinons). They behave essentially as interacting spinless fermions. This typical behavior can be observed in spin ladder systems in the presence of a magnetic field. Although such systems have been studied theoretically intensively for many years in both zero~\cite{dagotto_ladder_review,gopalan_2ch,Weihong_spin_ladder,Troyer_thermo_ladder,barnes_ladder,reigrotzki_ladder_field,Zheng_bound_state_ladder,sushkov_ladder_boundstates,knetter_ladder} and finite magnetic field~\cite{giamarchi_ladder_coupled,mila_ladder_strongcoupling,orignac_BEC_NMR,chitra_spinchains_field,furusaki_correlations_ladder,Wang_thermo_ladder,wessel01_spinliquid_bec,normand_bond_spinladder,Maeda_spinchain_magnetization,Usami_LL_parameter,hikihara_LL_ladder_magneticfield,Tachiki_spin_ladder}, a {\it quantitive} description of the LL low energy physics remained to be performed specially for a direct comparison with experiments. 

Quite recently the remarkable ladder
compound~\cite{Patyal_BPCB}
$\mathrm{(C_5H_{12}N)_2CuBr_4}$,
usually called BPCB (also known as (Hpip)$_2$CuBr$_4$), has been investigated.
The compound BPCB
has been identified to be a very good realization of weakly
coupled spin ladders. The fact that the interladder coupling is
much smaller than the intraladder coupling leads to a clear
separation of energy scales. Due to this separation the compound offers
the exciting possibility to study both the phase with Luttinger
liquid properties typical for low dimensional systems \emph{and} the
BEC condensed phase typical for high dimensions. Additionally, the
magnetic field required for the realization of different phases
lies for this compound in the experimentally reachable range. Actually various experimental techniques such as nuclear magnetic resonance (NMR)~\cite{Klanjsek_NMR_3Dladder}, neutron diffraction\footnote{ND consists in elastic neutron scattering by opposition INS implies an energy transfer. This technique can be used to measure the long range magnetic orders and is shortly discussed in Sec.~\ref{sec:neutron_diffraction}.} (ND)~\cite{Thielemann_ND_3Dladder}, specific heat and magnetocalorific effect~\cite{Ruegg_thermo_ladder} are used to probe the static properties of different phases of this compound. In order to interpret correctly these experiments a quantitative theoretical description of weakly coupled spin ladders is thus strongly required.

The excitations of this compound have recently been
observed by inelastic neutron
scattering~\cite{Thielemann_INS_ladder,Savici_BPCB_INS} experiments (INS). These are directly related to the dynamical correlations of spin ladders. Although these dynamical correlations have been investigated intensively in absence of a magnetic field during the last decades~\cite{barnes_ladder,reigrotzki_ladder_field,Zheng_bound_state_ladder,sushkov_ladder_boundstates,knetter_ladder,shelton_spin_ladders}, a detailed analysis and a quantitative description of their magnetic field dependence specially for the high energy excitations is clearly missing. The direct investigation of such excitations is of high interest,
since they not only characterize well the spin
system, but the properties of the triplon/spinon excitations are
also closely related to the properties of some itinerant
bosonic/fermionic systems. Indeed using such mappings ~\cite{giamarchi_book_1d}
of spin systems to itinerant fermionic or bosonic systems, the quantum spin
systems can be used as quantum simulators to address some of
the issues of itinerant quantum systems.
One of their advantage compared to regular itinerant systems is the
fact that the Hamiltonian of a spin system is in general well characterized, since the spin exchange constants can be
directly measured. The exchange between the spins would
correspond to short range interactions, leading to very good
realization of some of the models of itinerant particles, for which the short range of the
interaction is usually only an approximation. In that respect
quantum spin systems play a role similar to the one of cold
atomic gases ~\cite{bloch_cold_atoms_optical_lattices_review},
in connection with the question of itinerant interacting systems.

In this thesis, we present an analysis of the properties of weakly coupled spin-$1/2$ ladders in a magnetic field. We consider both the low energy physics and the excitations providing a {\it quantitative} description necessary for an unbiased comparison with experiments. The main achieved results discussed in this thesis as well as in Ref.~\cite{Ruegg_thermo_ladder,Klanjsek_NMR_3Dladder,Thielemann_ND_3Dladder,Bouillot_ladder_statics_dynamics} are:
\begin{itemize}
\item Combining a LL analytical technique and numerical density matrix renormalization group (DMRG) methods, we provide a quantitative description of the static and dynamic properties of spin-$1/2$ ladders in a full range of temperature and energy.

\item We provide a detailed analysis of the dynamical correlations in a full range of magnetic field and couplings.

\item Taking into account a weak interladder coupling by a mean field approximation, we characterize the BEC of triplons appearing at low temperature in weakly coupled spin-$1/2$ ladders.

\item Comparing the experimental measurements on the compound BPCB to our theoretical computations, we confirm the weakly coupled spin-$1/2$ ladder structure of this material which provides the first quantitative test of the LL theory and shows a phase transition to a BEC of triplons.
\end{itemize}

\section{Plan of the thesis}

Although strongly based on Ref.~\cite{Bouillot_ladder_statics_dynamics}, this thesis contains various important technical aspects as well as introductions on the methods and broader discussions which were omitted in Ref.~\cite{Bouillot_ladder_statics_dynamics}. In addition it includes several comparisons with the experiments on BPCB from Ref.~\cite{Ruegg_thermo_ladder,Klanjsek_NMR_3Dladder,Thielemann_ND_3Dladder}. The plan of the thesis is as follows.

\begin{itemize}
\item In chapter~\ref{sec:coupledladder}, we define the model of weakly coupled spin ladders. Its basic excitations and phase diagram are introduced as well as the spin chain mapping which proves to be very helpful for the physical interpretations. The spin ladder compound BPCB is also characterized with a detailed discussion of its chemical structure as well as the resulting interactions. The chapter~\ref{sec:coupledladder} is a general introduction on spin-$1/2$ ladder systems and its experimental realizations. It can be easily skipped by an informed reader. 

\item The chapter~\ref{sec:methods} provides a description of various theoretical techniques suited for dealing with low dimensional systems. We introduce the DMRG methods as well as the LL theory focussing on their application on spin-$1/2$ ladder. In particular, we introduce the recent real-time variant of DMRG to obtain the
dynamics~\cite{Vidal_time_DMRG,white_time_DMRG,daley_time_DMRG,Schollwoeck_tDMRG}
in real time and the dynamical correlation functions. A similar
technique is also presented to obtain finite temperature
results~\cite{Verstraete_finiteT_DMRG,Zwolak_finiteT_DMRG,White_finT}. The effect of a weak interladder coupling is discussed using a mean field approximation. The chapter~\ref{sec:methods} is  technically oriented and presents how the achieved results shown in~\ref{sec:staticproperties} and~\ref{sec:dynamicalcorrelation} are computed.

\item In chapter~\ref{sec:staticproperties}, we give a detailed
characterization of the phase diagram of weakly coupled spin-$1/2$ ladders focusing on their static
properties (magnetization, specific heat, BEC critical temperature, order
parameter) and the NMR relaxation rate. This characterization is followed by a comparison with the measurements on BPCB.

\item The chapter~\ref{sec:dynamicalcorrelation} presents the computed
dynamical correlations of a single spin ladder at different magnetic fields and couplings. The numerical calculations are compared to previous
results (linked cluster expansion, spin chain mapping, weak
coupling approach) and analytical descriptions (LL, t-J model). The theoretical spectra
are compared to the low energy INS measurements on the compound BPCB and provide predictions for the high energy
part of the INS cross section. The effects of a low interladder coupling on the dynamics are briefly discussed as well as the ND technique for measuring long range magnetic order. 

\item In chapter~\ref{sec:conclusions}, we summarize our results and discuss further perspectives.

\item The appendix~\ref{sec:tjmodelmapping} presents the strong coupling expansion which provides a simplified picture for the understanding of many results presented in this thesis.

\item The appendix~\ref{sec:staticcorrfinite} summarizes the static correlations in spin-$1/2$ chains and ladders computed in Ref.~\cite{hikihara_LL_ladder_magneticfield} for finite size systems. These enter as a key point in the quantitative description of the low energy physics of spin-$1/2$ ladders.
\end{itemize}



\chapter{Spin-$1/2$ ladders}\label{sec:coupledladder} 

The physics of quantum spin systems depends strongly on their microscopic characteristics. In fact, a large variety of phenomena can appear depending on the local spin $S$, the type of interaction and the geometry of the system. The external environment is also very important. For instance, applying a magnetic field or pressure on the system can lead to even richer physics. Various combinations of internal constraints as well as external conditions have been investigated theoretically for many years~\cite{Auerbach_book_magnetism}. These studies have lead to several fundamental discoveries such as the specific properties of 1D spin chains which are gapped or gapless in case the local spins $S$ is integer or half-integer~\cite{haldane_gap} respectively or the strong connection between the high temperature superconductors and the 2D quantum magnetism on a square lattice~\cite{zhang_rice_singlet}. 

Motivated by these two important results, the spin-$1/2$ ladders lying between these two 1D and 2D limits, have been studied theoretically intensively in both zero~\cite{dagotto_ladder_review,gopalan_2ch,Weihong_spin_ladder,Troyer_thermo_ladder,barnes_ladder,reigrotzki_ladder_field,Zheng_bound_state_ladder,sushkov_ladder_boundstates,knetter_ladder} and finite magnetic field~\cite{giamarchi_ladder_coupled,mila_ladder_strongcoupling,orignac_BEC_NMR,chitra_spinchains_field,furusaki_correlations_ladder,Wang_thermo_ladder,wessel01_spinliquid_bec,normand_bond_spinladder,Maeda_spinchain_magnetization,Usami_LL_parameter,hikihara_LL_ladder_magneticfield,Tachiki_spin_ladder}. In addition the identification of the  $\mathrm{(C}_5\mathrm{H}_{12}\mathrm{N)}_2\mathrm{CuBr}_4$  (BPCB) compound~\cite{Patyal_BPCB} as a very good realization of weakly
coupled spin-$1/2$ ladders has even increased the interest in these systems. Indeed due to its particularly low energy couplings its complex phase diagram (shown in Fig.~\ref{fig:phasediagram}.b) is
fully accessible experimentally by tuning a magnetic field and has been explored with various techniques~\cite{Patyal_BPCB,watson_bpcb,Klanjsek_NMR_3Dladder,Thielemann_ND_3Dladder,Thielemann_INS_ladder,Savici_BPCB_INS,Ruegg_thermo_ladder,Anfuso_BPCB_magnetostriction,lorenz_thermalexp_magnetostriction,cizmar_esr_bpcb,Bouillot_ladder_statics_dynamics}.

In this chapter we first describe the weakly coupled spin-$1/2$ ladder model on which we will focus in this work. Next we remind briefly its main physical features and introduce the spin chain mapping which provides a simple interpretation for certain features. Finally we present the compound BPCB with an analysis of its chemical structure and the resulting interactions.

\section{Weakly coupled spin-$1/2$ ladders}\label{sec:model}

\begin{figure}[!h]
\begin{center}
\includegraphics[width=0.8\linewidth]{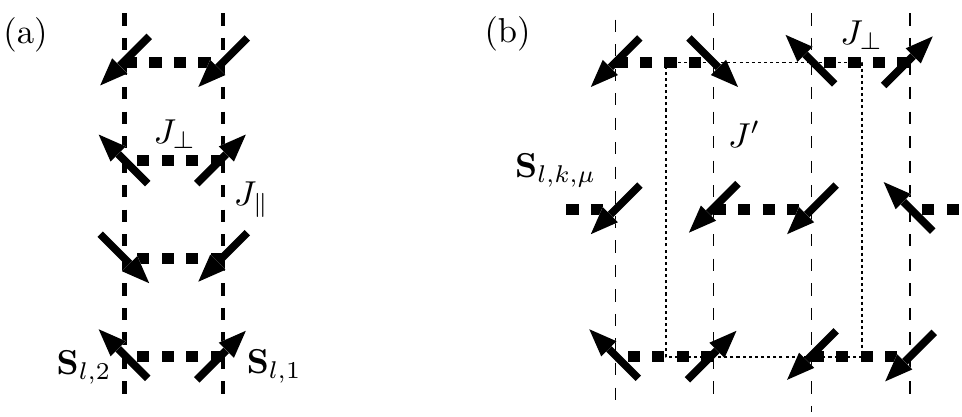}
\end{center}
\caption{(a) Schematic representation of a single ladder: ${\bf S}_{l,k}$ are the spin operators acting on the site  $l$ of the leg $k=1,2$. (b) Schematic representation of weakly coupled ladders (with the supposed interaction geometry of BPCB shown in Fig.~\ref{fig:structure}): ${\bf S}_{l,k,\mu}$ is the spin operator acting on the site  $l$ of the leg $k=1,2$ of the ladder $\mu$. In (b), the ladders are oriented perpendicular to the sheet. $J_ \perp$, $J_ \parallel$ and $J'$ are the coupling along the rungs and the legs, and the interladder coupling represented by thick, medium and thin dashed lines, respectively. The dotted rectangle represents an unit cell.
\label{fig:singleladder}}
\end{figure}

Various spin-$1/2$ ladder systems have been investigated including different coupling geometries with frustration or long range interaction as well as site dependent or anisotropic interactions~\cite{Weihong_spin_ladder,Japaridze_ladder,mila_ladder_strongcoupling,Usami_LL_parameter}. In this work we consider a simple ladder structure with isotropic Heisenberg couplings between nearest neighbors and no frustration as pictured in Fig.~\ref{fig:singleladder}.a. In addition, a weak interladder coupling is discussed with the assumed unfrustrated 3D coupling structure shown in Fig.~\ref{fig:singleladder}.b. The general Hamiltonian for these weakly coupled spin-$1/2$ ladders is 
\begin{equation}\label{equ:coupledladdershamiltonian}
H_{\textrm{3D}}=\sum_\mu H_{\mu}+J'\sum\mathbf{S}_{l,k,\mu}\cdot\mathbf{S}_{l',k',\mu'}
\end{equation}
where $H_\mu$ is the Hamiltonian of the single ladder $\mu$ and $J^\prime$ is the strength of the interladder coupling. The operator $\mathbf{S}_{l,k,\mu}=(S_{l,k,\mu}^x,S_{l,k,\mu}^y,S_{l,k,\mu}^z)$ acts at the site $l$ ($l=1,2,\ldots,L$) of the leg $k$ ($k=1,2$) of the ladder $\mu$. Often we will omit ladder indices from the subscripts of the operators (in particular, replace $\mathbf{S}_{l,k,\mu}$ with $\mathbf{S}_{l,k}$) to lighten notation. $S^\alpha_{l,k}$ ($\alpha=x,y,z$) are conventional spin-$1/2$ operators with (we mostly use $\hbar=k_B=1$)
\begin{equation}\label{equ:spincomm}
[S^\alpha_{l,k},S^\beta_{l,k}]=i\epsilon_{\alpha\beta\gamma} S^\gamma_{l,k},\quad\text{and}\quad S^\pm_{l,k}=S^x_{l,k}\pm iS^y_{l,k}
\end{equation}
$\alpha,\beta,\gamma = x,y,z$ and $\epsilon_{\alpha\beta\gamma}$ is the totally antisymmetric tensor.

The spin-$1/2$ Hamiltonian $H_\mu$ of the spin-$1/2$ two-leg ladder illustrated in Fig.~\ref{fig:singleladder}.a is
\begin{equation} \label{equ:spinladderhamiltonian}
H_\mu= J_\perp H_\perp + J_\parallel H_\parallel
\end{equation}
where $J_{\perp}$ ($J_{\parallel}$) is the coupling constant along the rungs (legs) and
\begin{equation}
H_\perp=\sum_{l}\mathbf{S}_{l,1}\cdot\mathbf{S}_{l,2}-h^z J_\perp^{-1} M^z,\quad
H_\parallel=\sum_{l,k} \mathbf{S}_{l,k}\cdot\mathbf{S}_{l+1,k} \label{equ:Hperp_parallel}
\end{equation}
The magnetic field, $h^z,$ is applied in the $z$ direction, and $M^z$ is the $z$-component of the total spin operator $\mathbf{M}=\sum_{l}(\mathbf{S}_{l,1} +\mathbf{S}_{l,2})$. Since $H_\mu$ has the symmetry $h^z\rightarrow-h^z$, $M^z\rightarrow-M^z$, we only consider $h^z\ge0$. The relation between $h^z$ and the physical magnetic field in experimental units is given in Eq.~\eqref{equ:experimentalhz}.

In this work we focus on the case of spin-$1/2$ antiferromagnetic
ladders weakly coupled to one another.
This means that the interladder coupling $J^\prime>0$ is much smaller than the intraladder couplings  $J_\parallel$ and $J_\perp$, i.e.
\begin{equation}
0<J^\prime \ll J_\parallel \text{ and } J_\perp. \label{equ:Jprimecond}
\end{equation}
Therefore, the interladder coupling $J^\prime$ will be treated perturbatively by a mean field approximation (see Sec.~\ref{sec:mean-field}) neglecting the microscopic details of the interladder interactions pictured in Fig.~\ref{fig:singleladder}.b for the supposed coupling structure of BPCB.

\subsection{Spin ladder to spin chain mapping} \label{sec:spinchainmap}

\begin{figure}[!h]
\begin{center}
\includegraphics[width=0.7\linewidth]{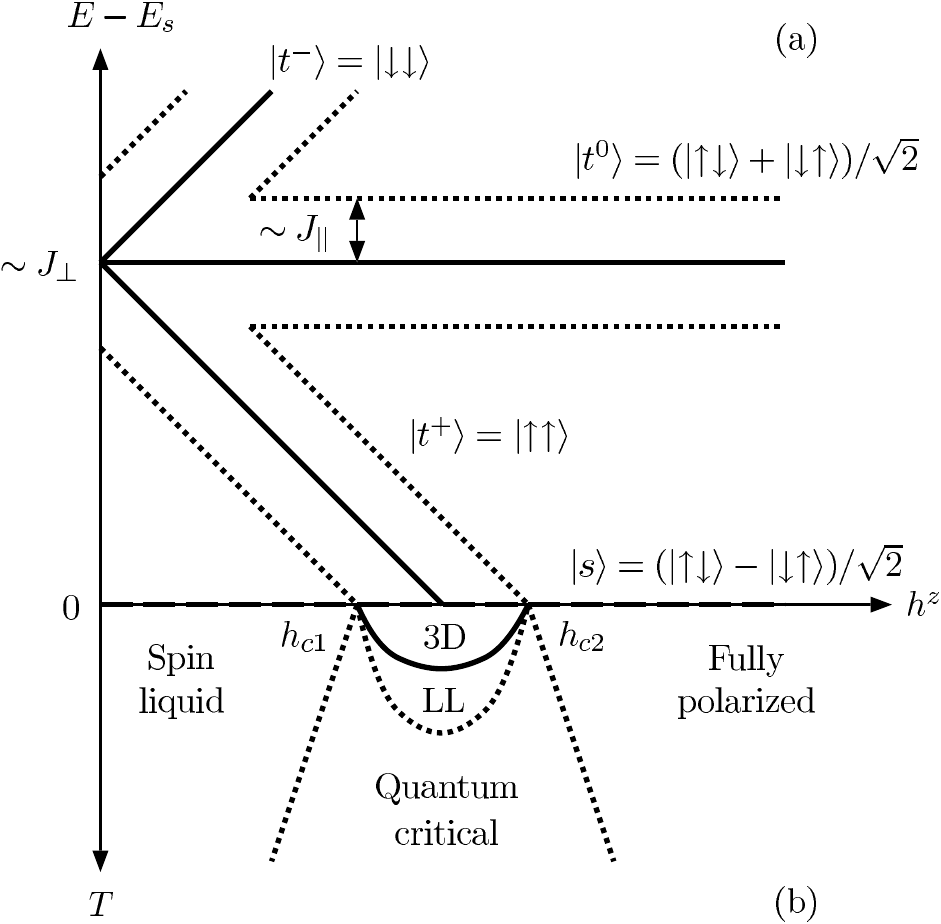}
\end{center}
\caption{(a) Energy of the triplets $|t^+\rangle$, $|t^0\rangle$, $|t^-\rangle$ (solid lines) and singlet $|s\rangle$ (dashed line)
versus the applied magnetic field in the absence of an interrung coupling ($J_\parallel=0$). The dotted lines represent the limits of the triplets
excitation band when $J_\parallel\neq0$. (b) Phase diagram of weakly coupled spin ladders: crossovers (dotted lines)
and phase transition (solid line) that only exists in the presence of an interladder coupling are sketched. [Taken from Ref.~\cite{Bouillot_ladder_statics_dynamics}]\label{fig:phasediagram}}
\end{figure}

The physical properties of a single ladder \eqref{equ:spinladderhamiltonian} are defined by the value of the dimensionless coupling
\begin{equation}\label{equ:couplingratio}
\gamma=\frac{J_\parallel}{J_\perp}.
\end{equation}
In the limit $J_\parallel=0$ (therefore $\gamma=0$) the rungs of the ladder are decoupled. We denote this \textit{decoupled bond limit} (DBL) hereafter. The four eigenstates of each decoupled rung are: the
singlet state
\begin{equation}
\ket{s}=\frac{\vert{\uparrow \downarrow}\rangle-\vert{\downarrow \uparrow}\rangle}{\sqrt{2}}
\label{equ:smult}
\end{equation}
with the energy $E_s=-3J_\perp/4,$ spin $S=0,$ and $z$-projection of the spin $S^z=0$, and three triplet states
\begin{equation}
\ket{t^+}=\vert{\uparrow\uparrow}\rangle, \quad
\ket{t^0}=\frac{\vert{\uparrow \downarrow}\rangle+\vert{\downarrow
\uparrow}\rangle}{\sqrt{2}}, \quad
\ket{t^-}=\vert{\downarrow\downarrow}\rangle
\label{equ:tmult}
\end{equation}
with $S=1,$ $S^z=1,0,-1$, and energies $E_{t^+}=J_\perp/4-h^z$,
$E_{t^0}=J_\perp/4$, $E_{t^-}=J_\perp/4+h^z$, respectively. The ground state is $\ket s$ below the critical value of the magnetic field, $h_c^{\rm DBL}= J_\perp$, and $\ket{t_+}$ above. The dependence of the energies on the magnetic field is shown in Fig.~\ref{fig:phasediagram}.a.

A small but finite $\gamma>0$ delocalizes triplets and creates bands of excitations with a bandwidth $\sim J_\parallel$ for each triplet branch. This leads to three distinct phases in the ladder system~\eqref{equ:spinladderhamiltonian} depending on the magnetic field:
\begin{itemize}
\item[(i)] {\it Spin liquid phase}\footnote{This phase is also called {\it quantum disordered}~\cite{giamarchi_BEC_dimers_review}. It appears in other antiferromagnetic systems such as frustrated antiferromagnets.}, which is characterized by a spin-singlet
ground state (see Sec.~\ref{sec:criticalfields}) and a gapped excitation spectrum (see Sec.~\ref{sec:spinliquidexcitations}). This phase appears for magnetic fields ranging from $0$ to $h_{c_1}.$

\item[(ii)] {\it Gapless phase,} which is characterized by a gapless excitation spectrum. It occurs between the critical fields $h_{c_1}$ and $h_{c_2}$. The ground state magnetization per rung, $m^z=\langle M^z\rangle/L$, increases from $0$ to $1$ for $h^z$ running from $h_{c_1}$ to $h_{c_2}$. The low energy physics can be described by the LL theory (see Sec.~\ref{sec:luttinger_liquid}).

\item[(iii)] {\it Fully polarized phase,} which is characterized by the fully polarized ground state and a gapped excitation spectrum. This phase appears above $h_{c2}$.
\end{itemize}

The transition between (i) and (ii) can also occur in several other gapped systems such as Haldane $S=1$ chains or frustrated chains~\cite{schu_spins,affleck_field,sachdev_qaf_magfield,chitra_spinchains_field}. In the gapless phase, the distance between the ground state and the bands $\ket{t^0}$ and $\ket{t^-},$ which is of the order of $ J_\perp$, is much larger than the width of the band $\ket{t^+} \sim J_\parallel$, since $\gamma \ll 1$. 

For small $\gamma$ the ladder problem can be reduced to a simpler spin chain problem. The essence of the {\it spin chain mapping}~\cite{tachiki_laddermapping,chaboussant_mapping,mila_ladder_strongcoupling,giamarchi_ladder_coupled} is to project out $\ket{t^0}$ and $\ket{t^-}$ bands from the Hilbert space of the model~\eqref{equ:spinladderhamiltonian}. The remaining states $\ket{s}$ and $\ket{t^+}$ are identified with the spin states
\begin{equation}
|\tilde\downarrow\rangle =|s\rangle,  \quad |\tilde\uparrow\rangle = |t^+\rangle. \label{equ:Hsreduced}
\end{equation}
The local spin operators ${\bf S}_{l,k}$ can therefore be identified in the reduced Hilbert space spanned by the states \eqref{equ:Hsreduced} with the new effective spin-$1/2$ operators $\tilde {\bf S}_{l}$:
\begin{equation}\label{equ:spinchainmaping}
\begin{array}{lll}
S^\pm_{l,k} = \frac{(-1)^k}{\sqrt{2}}\tilde S^\pm_l, \quad S^z_{l,k} = \frac{1}{4}\left(1+ 2 \tilde S^z_l\right).
\end{array}
\end{equation}
The Hamiltonian~\eqref{equ:spinladderhamiltonian} reduces to the Hamiltonian of the spin-$1/2$ XXZ Heisenberg chain
\begin{equation}\label{equ:strongcouplinghamiltonian}
H_{\text{XXZ}}=J_\parallel
\sum_{l}\left(\tilde S_l^x\tilde S_{l+1}^x+\tilde S_l^y\tilde S_{l+1}^y+\Delta\tilde S_l^z\tilde S_{l+1}^z\right)-\tilde h^z\tilde M^z +L\left(-\frac{J_\perp}{4}+\frac{J_\parallel}{8}-\frac{h^z}{2}\right).
\end{equation}
Here the pseudo spin magnetization is $\tilde M^z=\sum_{l}\tilde S^z_l$, the magnetic field $\tilde
h^z=h^z-J_\perp-J_\parallel/2$ and the anisotropy parameter
\begin{equation}
\Delta=\frac{1}{2}.
\end{equation}
Note that the spin chain mapping constitutes a part of a more general strong coupling expansion of the model~\eqref{equ:spinladderhamiltonian}, as discussed in the appendix~\ref{sec:tjmodelmapping}.

For the compound BPCB the parameter $\gamma$ is rather small
\begin{equation}\label{equ:couplingratio2}
\gamma\approx\frac{1}{3.55}\approx 0.282.
\end{equation}
and the spin chain mapping~\eqref{equ:strongcouplinghamiltonian} gives the values of many observables reasonably well. Some important effects in particular at high energy are, however, not captured by this approximation. For instance, due to their connection with high energy triplet excitations, several correlations cannot be described by this approximation. Other examples will be given in later chapters.

\subsection{Role of weak interladder coupling} \label{sec:weakinterladderc}

Let us now turn back to the more general Hamiltonian~\eqref{equ:coupledladdershamiltonian} and discuss the role of a weak interladder coupling $J^\prime$. The spin liquid and fully polarized phase are almost unaffected by the presence of $J^\prime$ whenever the gap in the excitation spectrum is larger than $J^\prime$ (see, e.g., Ref.~\cite{orignac_BEC_NMR} for more details). However, a new 3D antiferromagnetic order in the plane perpendicular to $h^z$ emerges in the gapless phase for $T~\lesssim J^\prime.$ The corresponding phase, called {\it 3D-ordered}, shows up at low enough temperatures $T_c$ in numerous experimental systems with reduced dimensionality and a
gapless spectrum~\cite{giamarchi_BEC_dimers_review}. This phase transition is in  the universality class of Bose-Einstein condensation and is discussed in more detail in Secs.~\ref{sec:mean-field} and~\ref{sec:coupledladderproperties}. For the temperature $T~\gtrsim J^\prime$ the ladders decouple from each other and the system undergoes a deconfinement transition into a Luttinger liquid regime (which will be described in Sec.~\ref{sec:luttinger_liquid}).  For $T~\gtrsim J_\parallel$ the rungs decouple from each other and the system becomes a (quantum critical) paramagnet. The transition from the 3D ordered to the LL~\cite{wessel01_spinliquid_bec} and the crossover from the LL to the quantum critical regime~\cite{Wang_thermo_ladder} induce specific features in several thermodynamic quantities such as the specific heat or the magnetization. These characteristics are pointed out in Secs.~\ref{sec:finiteTmagnetization} and~\ref{sec:specificheat} and used to locate the LL to quantum critical crossover. All the above mentioned phases are illustrated in Fig.~\ref{fig:phasediagram}.b.

\section{Experimental realizations of spin-$1/2$ ladders}\label{sec:bpcb}

The spin-$1/2$ ladder structure shown in Fig.~\ref{fig:singleladder}.a has been pointed out in several materials containing $\text{Cu}^{2+}$ ions with an unpaired external electronic orbital. Initially motivated by their strong connection with high temperature superconductors the inorganic compounds such as $\mathrm{SrCu_2O_3}$ \cite{azuma_srcuo} or $\mathrm{La_4Sr_{10}Cu_{24}O_{41}}$ \cite{Notbohm_cuprate_ladder} have been intensively investigated during the 90's. Although these materials show typical features of spin-$1/2$ ladders, their strong antiferromagnetic superexchange couplings ($\approx 1300\ \text{K}$) occurring through $\text{Cu}-\text{O}-\text{Cu}$ bonds induces a large spin gap, $h_{c1}$.  This big spin gap is of the order of few hundreds of Kelvin and thus prevents any investigation of the magnetic field effects.

During the last decade new organic compounds have shown similar spin ladder structures. Mediated by long organic chains, the antiferromagnetic superexchange couplings are usually much smaller than these observed in inorganic materials. Thereby, applying a magnetic field allows one in principle to explore the whole phase diagram shown in Fig.~\ref{fig:phasediagram}.b which was totally inaccessible experimentally for the inorganic materials. First investigated, the  compound  $\mathrm{Cu_2(C_5H_{12}N_2)_2Cl_4}$ \cite{chaboussant_ladder_strongcoupling,Calemczu_thermo_ladder} has finally shown significant deviations from the simple spin ladder structure (Fig.~\ref{fig:singleladder}.a). The presence of frustration or more complicated coupling paths~\cite{hammar_transition_cuhpcl,Stone_ladder} as well as Dzyaloshinskii-Moriya interactions~\cite{clemency_nmr_cuhpcl,caponni_thermo_cuhpcl} is actually debated.

\begin{figure}[!h]
\begin{center}
\includegraphics[width=0.7\linewidth]{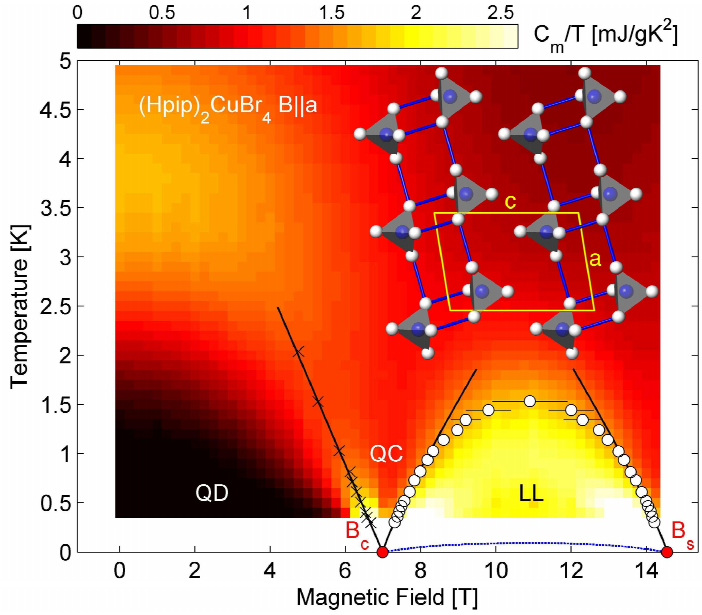}
\end{center}
\caption{False color representation of the measured ratio of the specific heat over the temperature, $c/T$, performed on BPCB versus the applied magnetic field, $h^z$, and the temperature, $T$ (see also Fig.~\ref{fig:chalspe_comp} for $c(T)$ plots for several magnetic fields). The experimentally determined phase diagram shows the various phases sketched in Fig.~\ref{fig:phasediagram}.b: spin liquid (called here quantum disorder (QD)), quantum critical (QC), and 
LL regimes. The two critical fields, $h_{c1}$ and $h_{c2}$, are denoted here by $B_{c}$ and $B_{s}$. Local maxima from the reduction of the triplet gap by the Zeeman effect are indicated by crosses. Circles denote the LL crossover based on measurements of the magnetocaloric effect (see Fig.~\ref{fig:comp_crossover}) using the $(\partial m^z/\partial T)|_{h^z}=0$ criterium discussed in Sec.~\ref{sec:finiteTmagnetization}. The dashed blue line indicates the onset of long-ranged order below approximatively $100\ \text{mK}$ (3D ordered phase in Fig.~\ref{fig:phasediagram}.b) shown in more details in Figs.~\ref{fig:INS_3D} and~\ref{fig:criticaltemperature}. In the inset, the lattice structure of BPCB in projection along the $\bf b$ axis, is depicted with $\text{Cu}$ atoms in blue and $\text{Br}$ in white. [Taken from Ref.~\cite{Ruegg_thermo_ladder}]
\label{fig:chalspe_BPCB}}
\end{figure}

\begin{figure}[!h]
\begin{center}
\includegraphics[width=0.8\linewidth]{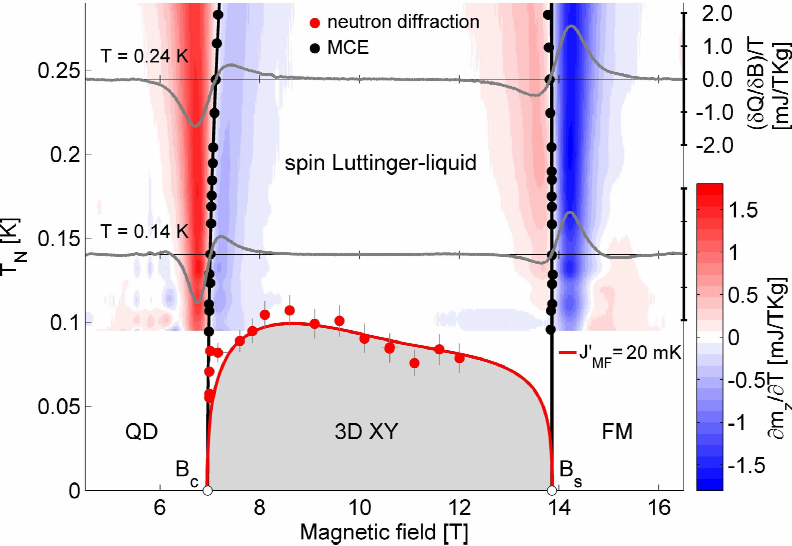}
\end{center}
\caption{Low temperature phase diagram of BPCB. The crossover temperature to the LL regime (black circles) is derived from the magnetocaloric effect represented by false color (see also Fig.~\ref{fig:comp_crossover}) using the $(\partial m^z/\partial T)|_{h^z}=0$ criterium discussed in Sec.~\ref{sec:finiteTmagnetization}. The phase transition to the 3D ordered phase, $T_c$, has been detected from the appearance of the magnetic order in ND measurements (red circles) (see also Fig.~\ref{fig:ND_measurements}). See Fig.~\ref{fig:criticaltemperature} for a comparison with the NMR determination of $T_c$. The red line is the theoretical prediction of $T_c$ discussed in Secs.~\ref{sec:mean-field} and~\ref{sec:coupledladderproperties}. [Taken from Ref.~\cite{Thielemann_ND_3Dladder}]\label{fig:INS_3D}}
\end{figure}

More recently the compound $\mathrm{(C_5H_{12}N)_2CuBr_4}$ first presented in Ref.~\cite{Patyal_BPCB} and commonly called BPCB or $\mathrm{(Hpip)_2CuBr_4}$ has been intensively investigated using different experimental methods such as nuclear magnetic
resonance~\cite{Klanjsek_NMR_3Dladder} (NMR), neutron
diffraction~\cite{Thielemann_ND_3Dladder} (ND), inelastic
neutron scattering~\cite{Thielemann_INS_ladder,Savici_BPCB_INS}
(INS), calorimetry~\cite{Ruegg_thermo_ladder}, magnetometry~\cite{watson_bpcb},
magnetostriction~\cite{Anfuso_BPCB_magnetostriction,lorenz_thermalexp_magnetostriction}, and electron spin resonance spectroscopy~\cite{cizmar_esr_bpcb} (ESR).  Except for small coupling anisotropies~\cite{cizmar_esr_bpcb}, which are briefly discussed in Secs.~\ref{sec:thermo_BPCB} and~\ref{sec:bpcb}, no significant deviations from the simple ladder structure (Fig.~\ref{fig:singleladder}.a) have been detected. In addition, a small interladder coupling $J'\ll J_\perp,J_\parallel$ has been pointed out in Refs.~\cite{Klanjsek_NMR_3Dladder,Thielemann_ND_3Dladder}.  Although the exact 3D interladder coupling structure~\cite{Thielemann_ND_3Dladder} is discussed in  Sec.~\ref{sec:thermo_BPCB}, this interladder coupling allows us to explore the transition from the 1D to the 3D regime which is experimentally accessible in BPCB. This compound is thus an extraordinary experimental tool for exploring the weakly coupled spin-$1/2$ ladder phase diagram depicted in Fig.~\ref{fig:phasediagram}. All the predicted phases have been observed in this compound and Figs.~\ref{fig:chalspe_BPCB} and~\ref{fig:INS_3D} show the experimental determination of its phase diagram from specific heat and neutron diffraction measurements, respectively. A detailed analysis of these experiments is presented in Sec.~\ref{sec:thermo_BPCB}. In this thesis, we focus mainly on this compound which provides a strong motivation and an experimental test for our theoretical investigation.

\begin{figure}[!h]
\begin{center}
\includegraphics[width=0.6\linewidth]{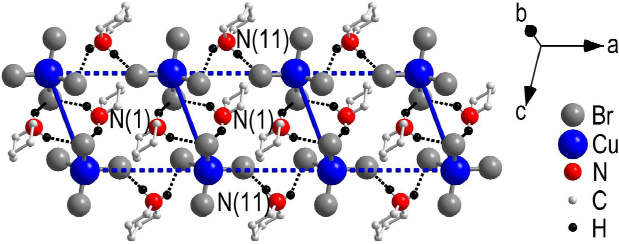}
\end{center}
\caption{Chemical structure of BPCB. The $10$ protons attached to the $\text{C}$ atoms are not shown. Solid thick blue
lines and dashed thick blue lines stand for the interaction path $J_\perp$ and $J_\parallel$, respectively. [Taken from Ref.~\cite{Klanjsek_NMR_3Dladder}] \label{fig:chemical_BPCB}}
\end{figure}

The compound $\mathrm{(C_5H_{12}N)_2CuBr_4}$ has the chemical structure represented in Fig. \ref{fig:chemical_BPCB}. This structure~\cite{Patyal_BPCB} is monoclinic, space group $P2_1/c$~\cite{Crystal_lattice_structures}, with the unit cell dimensions $a=8.487\ \text{\AA}$, $b=17.225\ \text{\AA}$ and $c=12.380\ \text{\AA}$ (${\bf a}$,
${\bf b}$ and ${\bf c}$ are the unit cell vectors of BPCB) and the angle $\beta=99.29^\circ$ between the $\bf a$ and $\bf c$ axes.

\begin{figure}[!h]
\begin{center}
\includegraphics[width=0.6\linewidth]{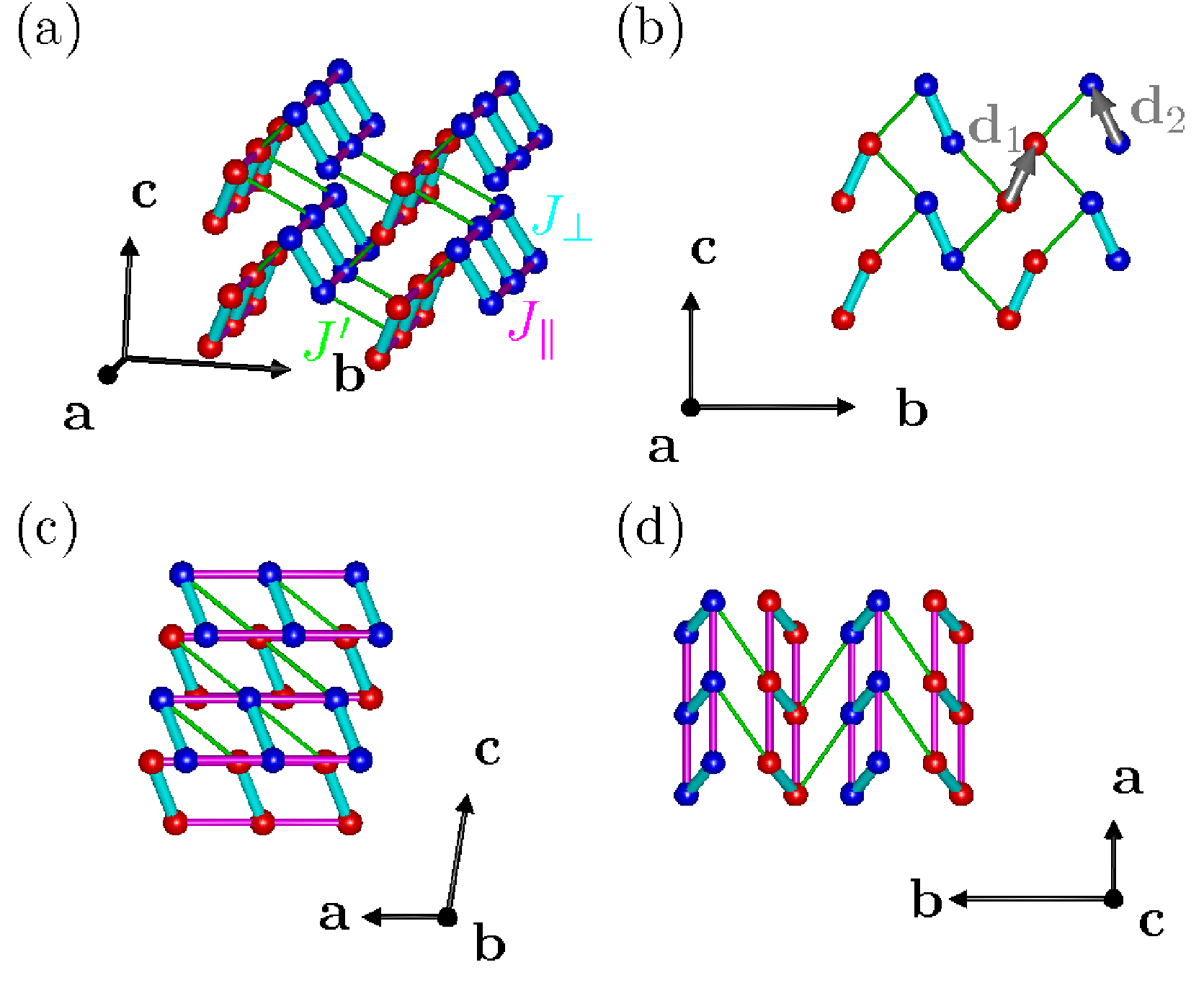}
\end{center}
\caption{Coupling structure of BPCB where the unpaired electron spins of the $\mathrm{Cu^{2+}}$ atoms in the first (second) type of ladders are pictured by red (blue) spheres. The $J_\perp$, $J_\parallel$ and $J'$ coupling paths are represented in turquoise, pink, and green, respectively. ${\bf a}$, ${\bf b}$ and ${\bf c}$ are the three unit cell vectors of the structure. Gray arrows are the rung vectors of the two types of ladders ${\bf d}_{1,2}$. (a) 3D structure. (b-d) Projection of the 3D structure onto the plane perpendicular to the ${\bf a}$, ${\bf b}$ and ${\bf c}$ axes. [Taken from Ref.~\cite{Bouillot_ladder_statics_dynamics}]
\label{fig:structure}}
\end{figure}

The magnetic properties of the compound are related to the
unpaired highest energy electronic orbital of the $\mathrm{Cu^{2+}}$
ions. Thus the corresponding spin structure (Figs.~\ref{fig:structure},~\ref{fig:chemical_BPCB} and~\ref{fig:singleladder}) matches with the
$\mathrm{Cu^{2+}}$ location~\cite{Patyal_BPCB,Klanjsek_NMR_3Dladder,Thielemann_INS_ladder}. The unpaired spins-$1/2$ interact together by antiferromagnetic superexchange coupling through the long organic chains and form two types
of equivalent weakly coupled ladders (Fig.~\ref{fig:structure}) along the ${\bf a}$ axis. The direction of the rung vectors of these ladders are
\begin{equation}\label{equ:rung_orientation}
\mathbf{d}_{1,2}=(0.3904,\pm0.1598,0.4842)
\end{equation}
in the primitive vector coordinates (Fig.~\ref{fig:structure}.b). Thus due to their different orientation the two types of ladders become slightly distinct when a magnetic field is applied breaking the structure symmetry.

The intraladder couplings from Eq.~\eqref{equ:spinladderhamiltonian} were determined to be $J_\perp\approx12.6-13.3~{\rm K}$, $J_\parallel\approx3.3-3.8~{\rm K}$ with different experimental techniques and at different experimental conditions~\cite{Klanjsek_NMR_3Dladder,Thielemann_ND_3Dladder,Thielemann_INS_ladder,Savici_BPCB_INS,Ruegg_thermo_ladder,watson_bpcb,Anfuso_BPCB_magnetostriction,lorenz_thermalexp_magnetostriction,Patyal_BPCB}. In this work, we use the values\footnote{These  parameters were the first outputs from the NMR measurements, which were later refined to the values from Ref.~\cite{Klanjsek_NMR_3Dladder}. Note that small changes in these values do not affect the main results of the calculations.}
\begin{equation}\label{equ:couplings}
J_{\perp}\approx12.6~\mathrm{K},\quad J_\parallel\approx3.55~\mathrm{K}.
\end{equation}
Recently, a slight anisotropy of the order of $5\%$ of $J_\perp$ has been discovered by ESR~\cite{cizmar_esr_bpcb} measurements. This anisotropy could explain the small discrepancies between the couplings found in different experiments. 

The magnetic field in Tesla is related to $h^z$ replacing
\begin{equation}\label{equ:experimentalhz}
h^z\rightarrow g\mu_B h^z
\end{equation}
in Eq.~\eqref{equ:spinladderhamiltonian} with $\mu_B$ being the Bohr magneton and $g$ being the Land\'e factor
of the unpaired copper electron spins. The latter depends on the orientation of the sample with respect to the magnetic field\footnote{More precisely, the Land\'e factor is different for each ladder forming the compound and varies with their orientation with respect to the magnetic field.}. For the orientation chosen in the NMR measurements~\cite{Klanjsek_NMR_3Dladder}, the Land\'e factor amounts to $g\approx2.126$. It can vary up to $\sim 10\%$ for other experimental setups~\cite{Patyal_BPCB}.

As one can see from the projection of the spin structure onto the plane perpendicular to the ${\bf a}$ axes (Fig. \ref{fig:structure}.b), each rung is expected to have $n_c=4$ interladder neighboring spins. The interladder coupling $J'$ has been experimentally determined to be~\cite{Klanjsek_NMR_3Dladder,Thielemann_ND_3Dladder}
\begin{equation}\label{equ:jprime}
J'\approx20-100~\mbox{mK}.
\end{equation}
As we will discuss in Sec.~\ref{sec:thermo_BPCB}, the exact 3D coupling structure shown in Fig.~\ref{fig:structure} and the precise value of $J'$ are actually debated~\cite{Klanjsek_NMR_3Dladder,Bouillot_ladder_statics_dynamics}.

\chapter{Methods}\label{sec:methods} 

Due to the exponential growth of the dimension of the Hilbert space with the number of sites, the theoretical investigation of quantum many body systems requires highly sophisticated techniques. In order to deal with weakly coupled spin-$1/2$ ladders, we focus here on methods suited for the treatment of quasi one-dimensional systems and their mean field extension such as the density
matrix renormalization group (DMRG) and the Luttinger liquid theory (LL) .

First, we give an overview of the so called density
matrix renormalization group (DMRG) or matrix product state
(MPS) method. This numerical method introduced by S. R. White in the beginning of the 90's is a very powerful technique to treat quantum many body physics in particular for one-dimensional systems. The DMRG allows one to investigate both static and dynamic properties at zero and finite temperature.

Second, we introduce an analytical low energy description for the gapless regime, the Luttinger liquid theory (LL). This quantum field theory is the cornerstone of the analytical description of many one-dimensional systems. In many situations, the bosonization in combination with a numerical determination of its parameters gives a quantitative description of the low energy physics. 

As we will see in the following these two methods are complementary and the choice depends mainly on the energy or temperature regime we want to focus on. Indeed, the DMRG provides a description of the high and intermediate energy properties as well as a description of the ground state. However, due to several numerical limitations, this method fails to describe the physics at very low energy. In contrast, the LL theory focuses on this regime and provides a quantitative description once its parameters are determined for the underlying model (using DMRG for example). 

Finally, we use a mean field approach to treat the weak interladder coupling in the case of weakly coupled spin-$1/2$ ladders both analytically and numerically. This approximation neglects the low quantum fluctuations related to the weak interactions, but fully takes into account the fluctuations along the ladders.

\section{DMRG}
The DMRG is a numerical method used to determine static and dynamic quantities at zero and finite temperature of quasi one-dimensional systems. This method was originally introduced by S.R.~White
~\cite{white_dmrg_letter,white_dmrg} to study static properties of one dimensional
systems. Since
usually the dimension of the total Hilbert space of a many-body
quantum system is too large to be treated exactly, the main idea of the DMRG algorithm is to describe the
important physics using a reduced effective space. This corresponds to a variational approach in a space of MPS wave functions. The DMRG has been proven very successful in many situations and has been generalized to compute dynamic properties of quantum systems using different approaches in frequency space ~\cite{Schollwock_DMRG,Hallberg_rev, Jeckelmann_rev}.
Recently the interest in this method even increased
after a successful generalization to time-dependent phenomena
and finite temperature situations
~\cite{Vidal_time_DMRG,daley_time_DMRG,white_time_DMRG,Verstraete_finiteT_DMRG,Zwolak_finiteT_DMRG,White_finT}. The real-time calculations further give an alternative route to determine dynamic correlation functions of the system~\cite{white_time_DMRG} which we use in the following.

In this section, we present an overview of the method providing first a short review of the basic ideas and focusing on its application to one-dimensional systems at zero temperature\footnote{We use in this description a ``classic'' approach of the DMRG (proposed by S.R.~White~\cite{white_dmrg}). Overviews over the more flexible MPS formulation can be found in ~\cite{Schollwock_DMRG,ScholwockDMRG_MPS,Verstraete_MPS}.}. Furthermore, we discuss the implementation of the time evolution and  the computation of momentum-frequency correlations. Finally, we extend the method to the simulation of finite temperatures. It allows us to compute thermodynamic quantities as the magnetization or the specific heat.

More details on the method, its extensions and its successful
applications can be found in Refs.~\cite{Schollwock_DMRG,Hallberg_rev,Jeckelmann_rev,ScholwockDMRG_MPS,Verstraete_MPS}. 

\subsection{Basic idea of DMRG}\label{sec:basicsDMRG}

\begin{figure}[h]
\begin{center}
\includegraphics[width=7.5cm]{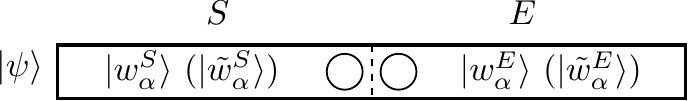}
\end{center}
\caption{DMRG decomposition of the Hilbert space in two blocks (the system $S$ and the environment $E$). The two circles symbolize the two sites lying at the boundary of each blocks for which the description is exact\label{fig:DMRG_basics}.}
\end{figure}

The basic idea of the DMRG technique consists in splitting the Hilbert space in two different blocks called {\it system} ($S$) and {\it environment} ($E$) (see Fig.~\ref{fig:DMRG_basics}). Each block contains a certain number of sites $L^S$ and $L^E$, respectively. Doing so, a state
\begin{equation}\label{equ:gen_decomp}
|\psi\rangle=\sum_{\alpha\beta} c_{\alpha\beta}\ |\tilde w_\alpha^S\tilde w_\beta^E\rangle
\end{equation}
of the global Hilbert space ($S\otimes E$ called {\it superblock}) can be decomposed in the basis formed by the tensor product of the states $|\tilde w_\alpha^S\rangle$ and $|\tilde w_\beta^E\rangle$ of the two blocks $S$ and $E$, respectively. For an exact description of the superblock, it is clear that the size of the basis of each block ($M^S$ and $M^E$) would grow exponentially with $L^S$ and $L^E$. Nevertheless, in order to make the computations accessible, it is possible to reduce the basis of each block keeping only $M\leq M^S,M^E$ states to approximate each of them. 

A powerful optimization of the truncated bases is provided by the singular value decomposition (SVD) in Ref.~\cite{QuantComp}. In linear algebra, the SVD of a matrix $C\in\mathbb{M}_{n,m}(\mathbb{C})$ having the elements $c_{\alpha\beta}$ \eqref{equ:gen_decomp} is a factorization of the form:
\begin{equation}\label{equ:SVDlin}
C=UWV
\end{equation}
where $W\in\mathbb{M}_{n,m}(\mathbb{R}_+)$ is upper diagonal, and $U\in\mathbb{M}_n(\mathbb{C})$ and $V\in\mathbb{M}_m(\mathbb{C})$ are unitary. We denote $w_{\alpha\beta}$ the elements of $W$ with $w_{\alpha\alpha}=w_\alpha\geq0$ and $w_{\alpha\beta}=0$ when $\alpha\neq \beta$. The elements of $U$ and $V$ are denoted $u_{\alpha\beta}$ and $v_{\alpha\beta}$, respectively. Using the SVD~\eqref{equ:SVDlin} in order to form a new basis
\begin{equation}
|w_\alpha^S\rangle=\sum_\beta u_{\beta\alpha}|\tilde w_\beta^S\rangle\quad,\quad
|w_\alpha^E\rangle=\sum_\beta v_{\alpha\beta}|\tilde w_\beta^E\rangle
\end{equation}
of each block, the description of the state $|\psi\rangle$~\eqref{equ:gen_decomp} simplifies to the so-called Schmidt decomposition~\cite{QuantComp}:
\begin{equation}\label{equ:SVD}
|\psi\rangle=\sum_\alpha w_\alpha|w_\alpha^Sw_\alpha^E\rangle
\end{equation}
where $w_\alpha$ are positive numbers with the normalization property $\sum_\alpha w_\alpha^2=1$ (so $\langle\psi|\psi\rangle=1$). This decomposition is easily computed through the reduced density matrices $\rho^S$ and $\rho^E$:
\begin{align}\label{equ:reduced_density}
\rho^S&=\mathrm{Tr}_E|\psi\rangle\langle\psi|=\sum_{\alpha}w_\alpha^2
|w_\alpha^S\rangle\langle w_\alpha^S|\notag\\
\rho^E&=\mathrm{Tr}_S|\psi\rangle\langle\psi|=\sum_{\alpha}w_\alpha^2
|w_\alpha^E\rangle\langle w_\alpha^E|.
\end{align}
The second equality is directly deduced from Eq.~\eqref{equ:SVD} and shows that the eigenbases $|w_\alpha^S\rangle$ and $|w_\alpha^E\rangle$ of $\rho^S$ and $\rho^E$ correspond to the basis $\ket{w^S_\alpha w^E_\alpha}$ of the Schmidt decomposition~\eqref{equ:SVD} respectively. Thus according to Eq.~\eqref{equ:reduced_density}, the Schmidt decomposition can be obtained by computing and diagonalizing the reduced density matrices $\rho^S$ and $\rho^E$.

Thereby we optimize the basis of each block for the description of $|\psi\rangle$ truncating the Schmidt decomposition~\eqref{equ:SVD} keeping only the states $\ket{w_\alpha^S}$ and $\ket{w_\alpha^E}$ corresponding to the $M$ largest $w_\alpha$:
\begin{equation}\label{equ:truncationDMRG}
\ket{\psi}\approx\sum_{\alpha=1}^M w_\alpha|w_\alpha^Sw_\alpha^E\rangle\equiv\ket{\tilde\psi}\quad\text{with}\quad\omega_1\geq\omega_2\geq\omega_3\geq\cdots\geq0.
\end{equation}
The quality of this basis for the description of $\ket{\psi}$ can be quantified by the so-called truncation error:
\begin{equation} 
\epsilon_M=\sum_{\alpha>M} w_\alpha^2=||\ket{\psi}-\ket{\tilde\psi}||^2
\end{equation}
which measures the difference between the exact state $\ket{\psi}$ and its approximation $\ket{\tilde\psi}$ using the optimized truncated basis. For a given $M$, this error is minimized by the truncation procedure~\eqref{equ:truncationDMRG} and clearly depends on the distribution of\footnote{The distribution of the $w_\alpha$ depends on the entanglement of $\ket{\psi}$ between the two blocks. Although a part of this entanglement is lost during the basis truncation~\eqref{equ:truncationDMRG} the latter procedure minimizes this loss for the Schmidt decomposition~\eqref{equ:SVD}.} $w_\alpha$. In the systems treated in this work the DMRG is essentially exact ($\epsilon_M\rightarrow0$) for a reasonable choice of $M$.

\subsection{Finite DMRG}\label{sec:finite_DMRG}

In order to deal with finite dimensional systems, we present the standard {\it finite} DMRG algorithm that allows one to compute the ground state $\ket{\psi_0}$ with open boundary conditions. 

\begin{figure}[!h]
\begin{center}
\includegraphics[width=7.5cm]{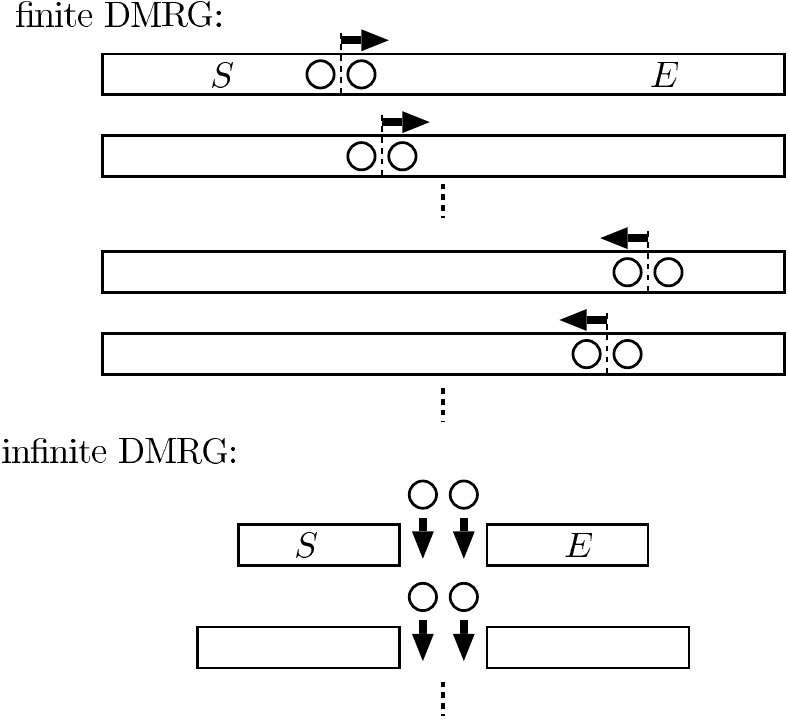}
\end{center}
\caption{Sketch of the finite and infinite DMRG procedures. The two circles symbolize the two sites lying at the boundary of each block for which the description is exact.\label{fig:DMRG_fin_inf}}
\end{figure}

This method consists in optimizing iteratively the bases $\ket{w^S_\alpha}$ and $\ket{w^E_\alpha}$ of dimension $M$ for each decomposition of the superblock following the truncation procedure described in Sec.~\ref{sec:basicsDMRG}. At each step of the method the edge of each block is shifted by one lattice site as shown in Fig.~\ref{fig:DMRG_fin_inf}. Next, $|\psi_0\rangle$ is computed in the actual two block decomposition using a Lanczos method~\cite{Matrix_Computations} and the corresponding optimized bases $\ket{w^S_\alpha}$ and $\ket{w^E_\alpha}$ are determined following Sec.~\ref{sec:basicsDMRG}. The approximate ground state $|\tilde\psi_0\rangle$ in this truncated basis~\eqref{equ:truncationDMRG} is then kept as the input state in the Lanczos method for the next step of the finite DMRG\footnote{$\ket{\tilde\psi_0}$ is considered as good approximation of $\ket{\psi_0}$ in the next two block decomposition of the system.}. To reach an optimal description of $\ket{\psi_0}$, before converging, this procedure  has to be repeated a few times~\cite{Schollwock_DMRG} passing through all the sites of the system. The observables are computed during the last DMRG sweep when they involve one of the two exactly known sites at the edge of the two block decomposition (see Fig.~\ref{fig:DMRG_fin_inf}). Once generated the operators are redefined at each step of the DMRG according to the new optimal basis before being combined and their expectation value computed.

A good starting state for the finite DMRG algorithm is provided by the infinite DMRG procedure. This consists in growing the system by introducing additionnal sites, two by two, in its center (as shown in Fig.~\ref{fig:DMRG_fin_inf}). This recursive procedure starts with a small enough system for which the ground state $\ket{\psi_0}$ can be computed exactly. Then the system is decomposed in two symmetrical blocks and their optimal $M$ states basis $\ket{w^S_\alpha}$ and $\ket{w^E_\alpha}$ with respect to $\ket{\psi_0}$ is determined as described in Sec.~\ref{sec:basicsDMRG}. These bases are used to approximate each block when adding the two next new sites at their edge. This choice of basis assumes that the two additional sites in the center do not change too much the ground state. This assumption is true while the thermodynamic limit is reached, but can be very bad at the beginning of the procedure, when the system is small. Nevertheless the infinite DMRG provides a usually good starting state for the finite DMRG.

\subsection{Time dependent DMRG}\label{sec:timeDMRG}

The
t-DMRG~\cite{Vidal_time_DMRG,daley_time_DMRG,white_time_DMRG,Schollwoeck_tDMRG}
(time dependent DMRG) method is based on the principle of the original DMRG (see Sec.~\ref{sec:basicsDMRG}). In order to deal with the time simulation, an effective reduced Hilbert space is chosen at each time step to describe the physics one is interested in. The implementation of this idea can be performed using different time-evolution algorithms. Here we use the second
order Trotter-Suzuki expansion for the time-evolution operator of a time-step $\delta t$~\cite{daley_time_DMRG,white_time_DMRG}:
\begin{equation}\label{equ:trotter}
e^{-iH\delta t}=\prod_{l\text{ odd}}e^{-iH_l\delta t/2}\prod_{l\text{ even}}e^{-iH_l\delta t}\prod_{l\text{ odd}}e^{-iH_l\delta t/2}+O(\delta t^3)
\end{equation}
where $H_l$ is the local Hamiltonian on the bond linking the
sites $l$ and $l+1$. This decomposition is valid when the total Hamiltonian $H$ decomposes in a sum of such terms $H_l$.

In order to apply the time-evolution \eqref{equ:trotter} on a given state we use the {\it sweep} procedure presented in Sec.~\ref{sec:finite_DMRG} for the finite DMRG. During the sweep, each term of~\eqref{equ:trotter} is applied on the system while it involves the two exactly known sites of the two block decomposition (see Fig.~\ref{fig:DMRG_fin_inf}). After each one of these steps, the optimized basis is updated as described in Sec.~\ref{sec:basicsDMRG}. Thus the t-DMRG adapts its effective description at each time-step.

In addition to the truncation error discussed in Sec.~\ref{sec:basicsDMRG} present in all variational methods, the t-DMRG is also limited by the expansion of the time-evolution operator~\eqref{equ:trotter}. This second uncertainty can be controlled by the choice of the time-step $\delta t$ (see Ref.~\cite{Gobert_spindynamics} for a detailed discussion).

\subsubsection{Momentum-energy correlations}\label{sec:tDMRGcorrelation}

To obtain the spectral functions
($S^{\alpha\beta}_{q_y}(q,\omega)$ in~\eqref{equ:correlation1}), we first compute the correlations in
space and time
\begin{equation}\label{equ:DMRGcorrelation}
S^{\alpha\beta}_{l,k}(t_n) =\langle 0 |e^{it_nH} S^\alpha_{l+L/2,k}e^{-it_nH}S^\beta_{L/2,1}|0\rangle
\end{equation}
with $l=-L/2+1,-L/2+2,\ldots,L/2$, $k=1,2$, and $t_n=n\delta t$ ($n=0,1,\ldots,N_t$) is the discrete time used. These
correlations are calculated by time-evolving the ground state
$\ket{\psi_0}\equiv\ket{0}$ and the excited state
$\ket{\psi_1}=S^\beta_{L/2,1}|\psi_0\rangle$ using the t-DMRG
(see Sec.~\ref{sec:timeDMRG}). Afterwards the overlap of
$S^\alpha_{l+L/2,k}\ket{\psi_1(t)}$ and $\ket{\psi_0(t)}$ is
evaluated to obtain the correlation function~\eqref{equ:DMRGcorrelation}.

In an infinite system reflection symmetry would be fulfilled.
To minimize the finite system corrections and to recover the
reflection symmetry of the correlations, we average them
\begin{equation}
 \frac{1}{2}\left(S^{\alpha\beta}_{-l,k}(t_n)+S^{\alpha\beta}_{l,k}(t_n)\right)\rightarrow S^{\alpha\beta}_{l,k}(t_n).
\end{equation}
We then compute the symmetric (antisymmetric) correlations (upon leg exchange)
(see Sec.~\ref{par:zerocorrelations})
\begin{equation}
 S^{\alpha\beta}_{l,q_y}(t_n)=2(S^{\alpha\beta}_{l,1}(t_n)\pm S^{\alpha\beta}_{l,2}(t_n))
\end{equation}
with the rung momentum $q_y=0,\pi$, respectively. Finally, we perform a numerical Fourier transform\footnote{The negative time correlations in the sum for $n<0$ are deduced from their value at positive time
since $S^{\alpha\beta}_{q_y}(q,-t_n)={ S^{\alpha\beta}_{q_y}}^\dagger(q,t_n)$, with $S^{\alpha\beta}_{q_y}(q,t_n)=\sum_le^{-iql}S^{\alpha\beta}_{l,q_y}(t_n)
$, for translation invariant systems, and for correlations such as ${S^\alpha}^\dagger=S^\beta$ with $\alpha,\beta=z,+,-$. In order to delete the numerical artefacts appearing in the zero frequency component of $S^{zz}_0(q,\omega)$ due to the boundary effects and the limitation in the numerical precision, we compute the Fourier transform of $S^{zz}_{l,0}(t_n)$ only with the imaginary part that has no zero frequency component as proposed in Ref.~\cite{pereira_spectrum_fermion1D}.}
\begin{equation}\label{equ:numFourrier}
S^{\alpha\beta}_{q_y}(q,\omega)\approx\delta t\sum_{n=-N_t+1}^{N_t}\sum_{l=-L/2+1}^{L/2}e^{i(\omega t_n -ql)}S^{\alpha\beta}_{l,q_y}(t_n)
\end{equation}
for discrete momenta $q=2\pi k/L$ ($k=0,1,\ldots,L-1$) and
frequencies $\omega$. The momentum $q$ has the reciprocal units of the interrung
spacing $a$ ($a=1$ is used if not mentioned otherwise). Due to the finite time step $\delta t$, our
computed $S^{\alpha\beta}_{q_y}(q,\omega)$ are limited to the
frequencies from $-\pi/\delta t$ to $\pi/\delta t$.
The finite calculation time $t_f=N_t\delta t$ induces artificial
oscillations of frequency $2\pi/t_f$ in
$S^{\alpha\beta}_{q_y}(q,\omega)$. To eliminate these artefacts
and reduce the effects of the finite system length, we apply
a filter to the time-space correlations before the numerical
Fourrier transform~\eqref{equ:numFourrier}, i.e.
\begin{equation}
S^{\alpha\beta}_{l,q_y}(t_n)f(l,t_n)\rightarrow
S^{\alpha\beta}_{l,q_y}(t_n).
\end{equation}
We tried different functional
forms for the filter $f(l,t_n)$ (cf.
Ref.~\cite{white_time_DMRG} as well). In the following
the results are obtained by a Gaussian filter
\begin{equation}
 f(l,t_n)=e^{-\left(4l/L\right)^2-\left(2t_n/t_f\right)^2}
\end{equation}
if not stated
otherwise. As the effect of this filtering on the
momentum-energy correlations consists to convolve them by a
Gaussian function
\begin{equation} 
f(q,\omega)=t_f
L/(32\pi)e^{-\left(\omega
t_f/4\right)^2-\left(qL/8\right)^2},
\end{equation}
it minimizes the numerical artefacts but further reduces the momentum-frequency resolution.

After checking the convergence, typical values we used in the simulations of dynamic quantities are system lengths of up to $L=160$
sites while keeping a few hundred DMRG states $M$. We limited
the final time $t_f$ to be smaller than the time necessary
for the excitations to reach the boundaries ($t_f\sim L/2u$ with
$u$ the LL velocity in Fig.~\ref{fig:LLparameter}) in order to
minimize the boundary effects. The computations for the BPCB parameters, Eq.~\eqref{equ:couplings}, were typically done with
a time step of $\delta t =0.0355 ~{J_\parallel}^{-1}$ up to $t_f=71~{J_\parallel}^{-1}$ (but calculating
the correlations only every second time step).
The
momentum-frequency limitations are then
$\delta\omega\approx0.11 ~J_\parallel$ and $\delta q\approx
0.1~a^{-1}$. Concerning the other couplings and the spin chain
calculations, we used a time step $\delta t =0.1 ~{J_\parallel}^{-1}$
up to $t_f=100 ~{J_\parallel}^{-1}$ (also with the correlation
evaluations every second time steps) for a momentum-frequency
precision $\delta\omega\approx0.08 ~J_\parallel$ and $\delta
q\approx 0.1~a^{-1}$.

Different techniques of extrapolation in time (using linear prediction or fitting the long time evolution with a guessed asymptotic form cf.~Refs.~\cite{white_spectral_heisenberg_spin_1,Barthel_spectrum_1D}) were recently used to improve the frequency resolution of the computed correlations. Nevertheless, as none of them can be applied systematically for our ladder system due to the presence of the high energy triplet excitations (which result in a superposition of very high frequency oscillations), we decided not to use them.

\subsection{Finite temperature DMRG}\label{sec:temperatureDMRG}

The main idea of the finite temperature DMRG
~\cite{Verstraete_finiteT_DMRG,Zwolak_finiteT_DMRG,White_finT}
(T-DMRG) is to represent the density matrix of the physical
state as a pure state in an artificially enlarged Hilbert
space. The auxiliary system is constructed by simply doubling
the physical system. Doing so, we define the totally mixed Bell state on the auxiliary system as
\begin{equation}
|\psi(0)\rangle=\frac{1}{{N_\sigma}^{L/2}}\prod_{l=1}^L\sum_{\sigma_l}|\sigma_l^P\sigma_l^C\rangle 
\end{equation}
where $|\sigma_l^P\sigma_l^C\rangle$ is the state at the bond $l$
of the auxiliary system which has the same value $|\sigma_l^P\rangle=|\sigma_l^C\rangle=\ket{\sigma_l}$ on the
two sites of the bond (the {\it physical} ($P$) and its {\it copy} ($C$)). The sum
$\sum_{\sigma_l}$ is done on all these $N_\sigma$ states
$|\sigma_l\rangle$. Considering the property\footnote{$\mathbb{I}^P$ is the identity operator on the physical system and $\mathrm{Tr}_C$ is the partial trace on the copy system.}
\begin{equation}\label{equ:psiproperty}
\mathrm{Tr}_C\ket{\psi(0)}\bra{\psi(0)}=\frac{\mathbb{I}^P}{{N_\sigma}^L}\end{equation}
of the state $\ket{\psi(0)}$, it is possible to construct the Boltzmann distribution at finite temperature $T=1/\beta$.

Starting from the infinite temperature limit, we evolve down in imaginary time the physical part of
$|\psi(0)\rangle$ to obtain
\begin{equation}
|\psi(\beta)\rangle=e^{-\beta H/2}|\psi(0)\rangle
\end{equation}
using the t-DMRG algorithm presented in Sec.~\ref{sec:timeDMRG}
with imaginary time. In order to avoid an overflow error,
this state is renormalized at each step of the imaginary time
evolution:
\begin{equation}
 \frac{\ket{\psi(\beta)}}{\langle\psi(\beta)|\psi(\beta)\rangle}\rightarrow \ket{\psi(\beta)}.
\end{equation}
Hence, according to~\eqref{equ:psiproperty}, we get the Boltzmann distribution through
\begin{equation}
\mathrm{Tr}_C\ket{\psi(\beta)}\bra{\psi(\beta)}= \frac{e^{-\beta H}}{\mathrm{Tr}[e^{-\beta H}]}.
\end{equation}
Therefore, the expectation value of an operator $O$ acting in the physical system with respect to the normalized state $|\psi(\beta)\rangle$ is directly related to its thermodynamic average, i.e.~
\begin{equation}
\langle O\rangle_\beta=\frac{\mathrm{Tr}[Oe^{-\beta H}]}{\mathrm{Tr}[e^{-\beta H}]}=\langle\psi(\beta)|O |\psi(\beta)\rangle.
\end{equation}
We use this method to compute the average value of the local rung
magnetization $m^z(T)$ and energy per rung $E(T)$ in the center of
the system. Additionally we extract the
specific heat $c(T)$ by
\begin{equation}
c(\beta+\delta\beta/2) \approx -\frac{(\beta+\delta\beta/2)^2}{2\delta\beta} \left(\langle E\rangle_{\beta+\delta\beta}-\langle E\rangle_{\beta}\right)
\end{equation}
where $\delta\beta$ is the imaginary time-step used in the T-DMRG.

To reach very low temperatures $T\rightarrow 0$ for the specific heat,
we approximate the energy by its expansion in $T$
\begin{equation}\label{equ:lowTEexpansion}
E(T)\approx E_0 + \sum_{i=2}^{n}\alpha_iT^i
\end{equation}
up to $n=4$. The energy at zero temperature $E_0$ is
determined by the original DMRG in Sec.~\eqref{sec:basicsDMRG}. Since
$E(T)$ has a minimum at $T=0$ the linear term in the expansion~\eqref{equ:lowTEexpansion} does not exist. The numbers
$\alpha_i$ ($i=2,3,4$) are obtained fitting the expansion on the
low $T$ values of the numerically computed $E(T)$.

After checking the convergence, typical system lengths used for the finite temperature
calculations are $L=80$ ($L=100$ for the spin chain mapping)
keeping a few hundred DMRG states $M$ and choosing a temperature
step of $\delta\beta=0.02~{\rm K}^{-1}$ ($\delta\beta=0.01~{\rm K}^{-1}$ for the spin chain mapping).

Let us note that recently a new method has been developed to treat finite temperatures ~\cite{White_TDMRG_thermal_state,Stoudenmire_TDMRG_thermal_state,ScholwockDMRG_MPS} which is very promising to reach even lower temperatures.

\section{Luttinger Liquid (LL)}\label{sec:luttinger_liquid}

In quantum systems, the interactions between particles can lead to very different physics which depend strongly on the dimensionality of the system. For instance, in high dimensions many systems enter into the universality class of Fermi liquids. This theory describes systems in which the elementary excitations are quasiparticles~\cite{bruus_flensberg_manybody}. Contrarily in 1D systems the effects of the interactions is so strong that the excitations are generally collective. The Luttinger liquid theory describes such systems in which the collective excitations are free bosonic excitations with linear spectrum. In this situation, the physics is described by the Hamiltonian~\cite{giamarchi_book_1d,gogolin_1dbook}
\begin{equation}\label{equ:luttingerliquid}
H_{\textrm{LL}} = \frac{1}{2\pi}\int dr\left[uK\left(\partial_r\theta(r)\right)^2+\frac{u}{K}\left(\partial_r\phi(r) \right)^2\right],
\end{equation}
where $\phi$ and $\theta$ are canonically commuting bosonic fields,
\begin{equation}\label{equ:boscomrel}
[\phi(r),\partial_{r'}\theta(r')]=i\pi\delta(r-r'). 
\end{equation}
Many gapless 1D interacting quantum systems belong to the Luttinger Liquid (LL) universality class: the dynamics of their low-energy excitations is governed by the Hamiltonian~\eqref{equ:luttingerliquid} and the local operators of the underlying model are written through the free boson fields $\theta$ and $\phi$ (the latter procedure is often called bosonization). 

The dimensionless parameter $K$ entering Eq.~\eqref{equ:luttingerliquid} is customarily called the Luttinger parameter, and $u$ is the propagation velocity of the bosonic excitations (velocity of sound). These parameters are non-universal and depend strongly on the underlying model. Once computed all the time and space correlations can be determined asymptotically by the field theory corresponding to the Hamiltonian $H_{LL}$.

From an experimental point of view, theoretical predictions of the LL theory have been observed in a growing number of 1D systems such as the organic conductors~\cite{schwartz_electrodynamics}, quantum wires~\cite{auslaender_quantumwire_tunneling}, carbon nanotubes~\cite{ishii_LL_nanotube}, edge states of quantum Hall effect~\cite{grayson_LL_quantum_hall}, ultracold atoms~\cite{bloch_cold_atoms_optical_lattices_review}, antiferromagnetic spin chain~\cite{lake_LL_spinchain} or spin ladder systems~\cite{dagotto_review_ladders}. In these systems characteristic features of the LL theory such as the power law behavior of some correlation or spectral functions, discussed in Secs.~\ref{sec:LLparameters} and~\ref{sec:luttingerliquidcorr} for spin-$1/2$ ladders, have been observed. However, since the details of the interactions are rarely known, only a theoretical estimate of the power law exponents related to the parameter $K$ was usually possible. 

In this section, we present the LL predictions of the low energy physics of spin-$1/2$ ladders. Computing precisely the parameters of the LL theory for the spin-$1/2$ ladder model, we quantitatively describe its low energy properties. Compared to the measurements on the compound BPCB in chapter~\ref{sec:staticproperties}, this description provides the first {\it quantitative} test of the LL theory.

\subsection{Bosonization of the spin-$1/2$ chain and ladder}

It has been shown that the gapless regime of the spin-$1/2$ ladder model~\eqref{equ:spinladderhamiltonian} is described by the LL theory~\cite{giamarchi_ladder_coupled,furusaki_correlations_ladder, chitra_spinchains_field}. In particular the bosonization of the local spin operators performed for both the strong ($\gamma\rightarrow0$) and the weak ($\gamma\rightarrow\infty$) coupling limits are smoothly connected~\cite{giamarchi_ladder_coupled}. In the following, we perform the more straightforward strong coupling procedure based on the spin chain mapping (Sec.~\ref{sec:spinchainmap}) starting with a reminder of the bosonic description of the mapped spin-$1/2$ XXZ chain.

\begin{figure}[h]
\begin{center}
\includegraphics[width=3.5cm]{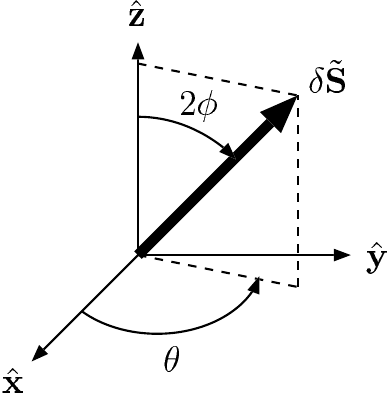}
\end{center}
\caption{Using a classical interpretation, the fields $2\phi$ and $\theta$ can be viewed as the two polar angles of the spin fluctuation $\delta \tilde {\bf S}$.\label{fig:Bosonization_rep}}
\end{figure}

The spin-$1/2$ XXZ chain, Eq.~\eqref{equ:strongcouplinghamiltonian}, in the gapless phase is a well-known example of a model belonging to the LL universality class. Its local operators are expressed through the boson fields as follows~\cite{giamarchi_book_1d}:
\begin{align}\label{equ:luttingeroperator1}
\tilde S^\pm(r) &= e^{\mp i\theta(r)}\left[\sqrt{2A_x}(-1)^r +2\sqrt{B_x}\cos(2\phi(r)-2\pi\tilde m^z r)\right],\\
\tilde S^z(r)&= \tilde m^z-\frac{\partial_r \phi(r)}{\pi}
+\sqrt{2A_z}(-1)^r\cos(2\phi(r)-2\pi \tilde m^zr).\label{equ:luttingeroperator2}
\end{align}
Here the continuous coordinate $r=la$ is given in units of the lattice spacing~$a,$ $\tilde m^z=\langle\tilde M^z\rangle/L$ is the magnetization per site of the spin chain, and $A_x$, $B_x$ and $A_z$ are coefficients which depend on the parameters of the model~\eqref{equ:strongcouplinghamiltonian}. How to calculate $K,$ $u,$ $A_x$, $B_x,$ and $A_z$ is described in Sec.~\ref{sec:LLparameters}. For the XXZ spin-$1/2$ chain, a geometrical representation of the two fields $\theta$ and $\phi$ in Eqs.~\eqref{equ:luttingeroperator1} and~\eqref{equ:luttingeroperator2} is easily obtained using their classical interpretation. As shown in Fig.~\ref{fig:Bosonization_rep}, they can be  seen as the two polar angles of the spin fluctuation $\delta \tilde {\bf S}=\tilde {\bf S}-\tilde m^z\hat {\bf z}$ which gives an intuition of the origin of the two terms in the LL Hamiltonian~\eqref{equ:luttingerliquid}. The first term $(\partial_r\phi(r))^2$ measuring the spatial fluctuations of $\phi(r)$ is related to the longitudinal ($z$) direction interaction term in~\eqref{equ:strongcouplinghamiltonian}. In contrast, the transverse ($xy$) interactions are responsible for the second term $(\partial_r\theta(r))^2$ in~\eqref{equ:luttingerliquid} related to the spatial fluctuations of $\theta(r)$. Similarly to the original spin commutation relation~\eqref{equ:spincomm},
the quantum nature of the two fields comes from their commutation relation~\eqref{equ:boscomrel}. The latter relation induces a competition between the ordering in the transverse and the longitudinal direction. Due to the strong effects of quantum fluctuations in 1D systems, the correlation functions in a LL decay algebraically with exponents depending on $K$ (see Secs.~\ref{sec:LLparameters} and~\ref{sec:zeroT_correlation}). 

As discussed in appendix~\ref{sec:tjmodelmapping}, the Hamiltonian~\eqref{equ:strongcouplinghamiltonian} is the leading term in the strong coupling expansion of the spin-$1/2$ ladder model~\eqref{equ:spinladderhamiltonian}. Using this strong coupling approach, local operators of the latter model are bosonized by combining Eqs.~\eqref{equ:spinchainmaping}, \eqref{equ:luttingeroperator1}, and \eqref{equ:luttingeroperator2}
\begin{align}\label{equ:pmluttingeroperator}
S^\pm_j(r)&= (-1)^j\ e^{\mp i\theta(r)+i\pi r}\left[\sqrt{A_x}+\sqrt{2B_x}\cos(2\phi(r)-2\pi m^zr)\right]\\
S^z_j(r)&= \frac{m^z}{2}-\frac{\partial_r \phi(r)}{2\pi}
+\sqrt{\frac{A_z}{2}}\cos(2\phi(r)-2\pi m^zr).\label{equ:zzluttingeroperator}
\end{align}
with $j=1,2$ is the number of the leg. We would like to stress that even for a small $\gamma$ some parameters out of $K,$ $u,$ $A_x$, $B_x,$ and $A_z$ show significant numerical differences if calculated within the spin chain~\eqref{equ:strongcouplinghamiltonian} compared to the spin ladder~\eqref{equ:spinladderhamiltonian} (see Fig.~\ref{fig:LLparameter}). We discuss this issue in Sec.~\ref{sec:LLparameters}.

In the following, we first  discuss the numerical determination and the properties of the parameters $K,$ $u,$ $A_x$, $B_x,$ and $A_z$.
Furthermore we recall some properties of the LL focusing on the finite and zero temperature correlations which are directly related to many experimental quantities (see chapters~\ref{sec:staticproperties} and~\ref{sec:dynamicalcorrelation}).

\subsection{Luttinger liquid parameter determination}\label{sec:LLparameters}

\begin{figure}[!h]
\begin{center}
\includegraphics[width=0.5\linewidth]{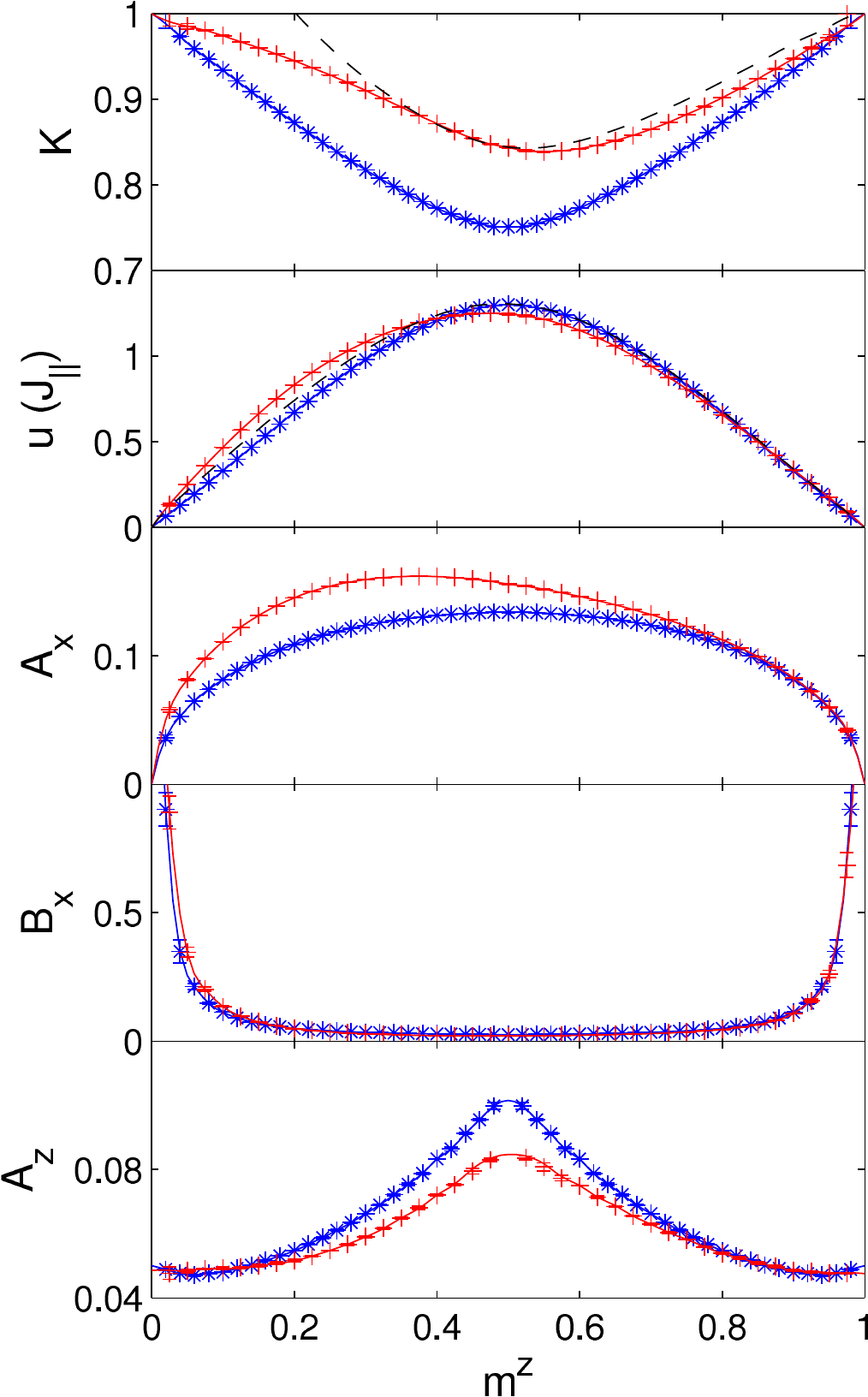}
\end{center}
\caption{LL parameters $u$, $K$ and the prefactors of the spin operators $A_x$, $B_x$, $A_z$ versus
the magnetization per rung, $m^z$, computed for a spin ladder with the BPCB couplings~\eqref{equ:couplings}~(red crosses) and for the spin chain mapping
(blue stars). The strong coupling expansion of $u$ and $K$ up to second order in $\gamma$ (discussed in appendix~\ref{sec:secondorder_LL}) is plotted in black dashed lines. [Taken from Ref.~\cite{Bouillot_ladder_statics_dynamics}]\label{fig:LLparameter}}
\end{figure}

In this paragraph we detail the determination of the LL parameters $u$, $K$ and the prefactors $A_x$, $B_x$ and $A_z$ (see Eqs.~\eqref{equ:luttingerliquid}, \eqref{equ:luttingeroperator1},~\eqref{equ:luttingeroperator2},~\eqref{equ:pmluttingeroperator} and~\eqref{equ:zzluttingeroperator}) using two main properties of the LL ground state i.e. the algebraic decay of the correlation functions\footnote{These relations are given for the spin chain~\eqref{equ:strongcouplinghamiltonian} since from these the relations for the spin ladders can be easily inferred using the spin chain mapping~\eqref{equ:spinchainmappingLL}.}~\eqref{equ:xxcorrelationsimplify} and~\eqref{equ:zzcorrelationsimplify} as well as the susceptibility\footnotemark[\value{footnote}]~\eqref{equ:uoverKcalculation}. These parameters are necessary for a {\it quantitative} use of the LL theory. The parameters $K$, $A_x$, $B_x$ and $A_z$ and their dependence on the magnetic field have been previously determined in Refs.~\cite{hikihara_LL_ladder_magneticfield,Usami_LL_parameter,furusaki_correlations_xxz_magneticfield} for different values of the couplings than those considered here.  We obtain these parameters in two steps~\cite{Klanjsek_NMR_3Dladder,Klanjsek_3Dladder_Ax}:
\begin{itemize}
  \item[(i)] We determine the ratio $u/K$ from its relation to the static LL susceptibility\footnotemark[\value{footnote}]~\cite{haldane_luttinger,giamarchi_book_1d}
  \begin{equation}\label{equ:uoverKcalculation}
  \frac{u}{K}=\frac{1}{\pi \frac{\partial \tilde m^z}{\partial \tilde h^z}}.
  \end{equation}
  We numerically compute the static susceptibility using DMRG and infer the ratio $u/K$ with a negligible error.

  \item[(ii)] The parameter $K$ and the prefactors $A_x$, $B_x$ and $A_z$ are extracted by fitting numerical results for the static correlation functions obtained by DMRG  with their analytical LL expression\footnotemark[\value{footnote}]~\cite{hikihara_LL_ladder_magneticfield}
  \begin{align}
  \label{equ:xxcorrelationsimplify}
  \langle \tilde S^x_l\tilde S^x_{l'}\rangle&=A_x\frac{(-1)^{l-l'}}{|l-l'|^{\frac{1}{2K}}}- B_x(-1)^{l-l'}\frac{\cos[q(l-l')]}{|l-l'|^{2K+\frac{1}{2K}}}\\
  \label{equ:zzcorrelationsimplify}
  \langle \tilde S^z_l\tilde S^z_{l'}\rangle&=\tilde m^{z\,2}+A_z(-1)^{l-l'}\frac{\cos[q(l-l')]}{|l-l'|^{2K}}-\frac{K}{2\pi^2|l-l'|^2}
  \end{align}
  These correlations computed for infinite systems decay algebraically with the distance $|l-l'|$ with a $K$ dependent exponent. In practice, we use the more sophisticated expressions 
  Eqs. \eqref{equ:xxcorrelationfurusaki}, \eqref{equ:zzcorrelationfurusaki} and~\eqref{equ:zmagnetizationfurusaki} (for $\langle \tilde S^z_l\rangle$) shown in appendix~\ref{sec:staticcorrfinite} and derived in Ref.~\cite{hikihara_LL_ladder_magneticfield}. These take into account the boundary effects which are present in the finite DMRG computations but neglected in Eqs.~\eqref{equ:xxcorrelationsimplify} and~\eqref{equ:zzcorrelationsimplify}.

  We first fit the transverse correlation ($xx$-correlation $\langle \tilde S^x_l\tilde S^x_{l'}\rangle$) to extract the parameters $K$, $A_x$, and $B_x$. Then we use the previously extracted value for $K$ to fit the longitudinal correlation ($zz$-correlation $\langle \tilde S^z_l\tilde S^z_{l'}\rangle$) and the magnetization, $\langle \tilde S^z_l\rangle$, which allow us to determine $A_z$. The values determined by both fits are very close and in Fig.~\ref{fig:LLparameter} the average value of both is shown.
\end{itemize}

All the results presented in Fig.~\ref{fig:LLparameter} were obtained for $L=200$ and several hundred DMRG states after an average on the four sets of used data points in the fit $10<l,l'<170$, $30<l,l'<170$,  $10<l,l'<190$, $30<l,l'<190$. The error bars correspond to the maximum discrepancy of these four fits from the average. We further checked that different system lengths lead to similar results.

The LL parameters of the ladder system~\eqref{equ:spinladderhamiltonian} for the BPCB couplings~\eqref{equ:couplings} are
presented in Fig.~\ref{fig:LLparameter} as a function of the
magnetization per rung. Additionally we show the parameters of the spin chain mapping (computed for the spin chain Hamiltonian~\eqref{equ:strongcouplinghamiltonian}) for comparison.
When the ladder is just getting magnetized, or when the ladder is almost fully polarized,
$K\rightarrow 1$ (free fermion limit) and $u\rightarrow 0$  (because of the low density of
triplons in the first case, and low density of singlets in the second case).
Between these two limits $K<1$ due to the triplet-triplet repulsion (see Eq.~\eqref{equ:H0} in the strong coupling expansion). For the spin chain mapping,
the reflection symmetry around $m^z=0.5$ arises from the
symmetry under $\pi$ rotation around the $x$ or $y$ axis of the spin chain.
This symmetry has no reason to be present in the original ladder model,
and is an artefact of the strong coupling limit, when truncated to the lowest order term as shown in appendix~\ref{sec:tjmodelmapping}.
The values for the spin ladder
with the compound BPCB parameters can deviate strongly from this
symmetry. The velocity $u$ and the prefactor $B_x$ remain very close to the values for the spin chain mapping. In contrast, the
prefactors $A_z$, $A_x$ and the exponent $K$ deviate
considerably and $A_x$ and $K$ become strongly asymmetric. The origin of the asymmetry lies in the contribution of the higher triplet states~\cite{giamarchi_ladder_coupled}, and can be understood using a strong coupling expansion of the Hamiltonian~\eqref{equ:spinladderhamiltonian} up to second order in $\gamma$ (see appendix~\ref{sec:secondorder_LL}). This asymmetry has
consequences for many experimentally relevant quantities and it
was found to cause for example strong asymmetries in the 3D
order parameter, its transition temperature and the NMR
relaxation rate as will be discussed in chapter~\ref{sec:staticproperties} (see Fig.~\ref{fig:T1},
Fig.~\ref{fig:criticaltemperature} and Fig.~\ref{fig:orderparameter}).

\subsection{Dynamical correlations}\label{sec:luttingerliquidcorr}

In this paragraph, we focus on the retarded correlations defined in the time-space as
\begin{equation}\label{equ:correlationdef}
 \chi^{\alpha\beta}_{ij}(r,t)=-i \Theta(t)\left\langle\left[S^\alpha_i(r,t), S^\beta_j(0,0)\right]\right\rangle
\end{equation}
with $\Theta(t)$ the Heaviside function, $\alpha,\beta=\pm,z$ such as $S^\alpha={S^\beta}^\dagger$ and $S^\alpha_i(r,t)=e^{iHt}S^\alpha_i(r)e^{-iHt}$ is the time evolution of the operator $S^\alpha_i(r)$. Their Fourier transform is computed as  $\chi^{\alpha\beta}_{ij}(q,\omega)=\int dr\ dt\ e^{i(\omega t- qr)}\chi^{\alpha\beta}_{ij}(r,t)$. These correlations are necessary for the mean field determination of the transition temperature $T_c$ to the 3D-ordered phase (see Sec.~\ref{sec:transitiontemperature}). They are also directly related to the NMR relaxation rate $T^{-1}_1$ (see Sec.~\ref{sec:relaxationtime}) and the INS cross-section (see Sec.~\ref{sec:INS}) through the spectral functions
\begin{equation}\label{equ:spectralfun}
S^{\alpha\beta}_{ij}(q,\omega)=\int_{-\infty}^\infty dr\ dt\ e^{i(\omega t- qr)}\langle S^\alpha_i(r,t) S^\beta_j(0,0)\rangle.
\end{equation}
As discussed in Sec.~\ref{sec:timeDMRG}, these spectral functions are also accessible numerically at zero temperature~\eqref{equ:numFourrier} and have the following properties
\begin{align}\label{equ:spectralfunlimit}
S^{\alpha\beta}_{ij}(q,\omega)=\frac{2}{e^{-\beta\omega}-1}\Im\left(\chi^{\alpha\beta}_{ij}(q,\omega)\right)&\xrightarrow[1\gg\beta\omega]{} -\frac{2}{\beta\omega}\Im\left(\chi^{\alpha\beta}_{ij}(q,\omega)\right)\notag\\
&\xrightarrow[\beta\rightarrow\infty]{} -2\Theta(\omega)\Im\left(\chi^{\alpha\beta}_{ij}(q,\omega)\right).
\end{align}
Hence, the spectral function $S^{\alpha\beta}_{ij}$ diverges in the low energy limit $1\gg\beta\omega$ unless the correlation $\chi^{\alpha\beta}_{ij}$ vanishes in this limit. At zero temperature, $S^{\alpha\beta}_{ij}$ vanish for all negative frequencies. As we will see in Sec.~\ref{par:zerocorrelations}, $S^{\alpha\beta}_{ij}$ measure the excitations of the system, their vanishing thus physically means that at zero temperature no excited state has an energy lower than the ground state.

\subsubsection{Finite temperature LL correlations}
Using the bosonization formalism~\eqref{equ:pmluttingeroperator} and~\eqref{equ:zzluttingeroperator}, and taking into account only the most relevant terms, we can compute the Fourrier transform of the  correlations~\eqref{equ:correlationdef} as described in Ref.~\cite{giamarchi_book_1d} for the LL Hamiltonian~\eqref{equ:luttingerliquid}:
\begin{align}\label{equ:pmcorrelationqo}
\chi^{\pm\mp}_{ij}(q,\omega)=&(-1)^{i+j}A_x \ f_{1/4K}(q-\pi,\omega,\beta)\\
\chi^{zz}_{ij}(q,\omega)=&\frac{A_z}{8}\left[f_{K}(1-2\pi m^z,\omega,\beta)+f_{K}(1+2\pi m^z,\omega,\beta)\right]+\frac{1}{4\pi^2}g_K(q,\omega)\label{equ:zzcorrelationqo}
\end{align}
with
\begin{align}
f_\nu(q,\omega,\beta) =&-\frac{\sin\left(\pi\nu\right)}{u}\left(\frac{2\pi}{\beta u}\right)^{2\nu-2}B\left(-i\frac{\beta(\omega-uq)}{4\pi}+\frac{\nu}{2},1-\nu\right)\\&\times B\left(-i\frac{\beta(\omega+uq)}{4\pi}+\frac{\nu}{2},1-\nu\right)\notag\\
g_\nu(q,\omega)=&\frac{\pi q^2u\nu}{(\omega+i0^+)^2-u^2q^2}
\end{align}
where $B(x,y)=\frac{\Gamma(x)\Gamma(y)}{\Gamma(x+y)}$. The correlations~\eqref{equ:pmcorrelationqo} and~\eqref{equ:zzcorrelationqo} are linear combinations of the functions $f_\nu$ and $g_\nu$ which depend only on the two LL parameters $u$ and $K$. The weight of the $f_\nu$ component is related to the prefactors $A_x$ and $A_z$ for $\chi^{\pm\mp}_{ij}$ and $\chi^{zz}_{ij}$, respectively. In contrast the component $g_\nu$ in~\eqref{equ:zzcorrelationqo}  has a constant prefactor and is invariant in temperature. As we will see in chapter~\ref{sec:staticproperties}, these correlations are necessary to compute the critical temperature $T_c$ of the 3D transition through a mean field treatment of the interladder coupling $J'$.

Thus using~\eqref{equ:pmcorrelationqo},~\eqref{equ:zzcorrelationqo} and~\eqref{equ:spectralfunlimit}, the local correlations (momentum average of the spectral functions~\eqref{equ:spectralfun})
\begin{equation}\label{equ:local_correlation}
S^{\alpha\beta}(\omega)=S^{\alpha\beta}_{ii}(r=0,\omega)=\frac{1}{2\pi}\int dq\ S^{\alpha\beta}_{11}(q,\omega)
\end{equation}
become in the low energy limit $1\gg\beta\omega$:
\begin{align}\label{equ:local_correlationLL}
S^{\pm\mp}(\omega\rightarrow0)=&\frac{2A_x\cos\left(\frac{\pi}{4K}\right)}{u}\left(\frac{2\pi}{\beta u}\right)^{\frac{1}{2K}-1}B\left(\frac{1}{4K},1-\frac{1}{2K}\right)\\
S^{zz}(\omega\rightarrow0)=&\ \frac{A_z\cos\left(\pi K\right)}{2u}\left(\frac{2\pi}{\beta u}\right)^{2K-1}B\left(q,1-2K\right)+\frac{K}{4\pi\beta u^2}.
\end{align}
In this low energy limit, the local correlations~\eqref{equ:local_correlationLL} are directly related to the NMR relaxation rate $T_1^{-1}$, Eq.~\eqref{equ:T1init}.

\subsubsection{Zero temperature correlations in the LL}\label{sec:zeroT_correlation}

\begin{figure}[!h]
\begin{center}
\includegraphics[width=0.7\linewidth]{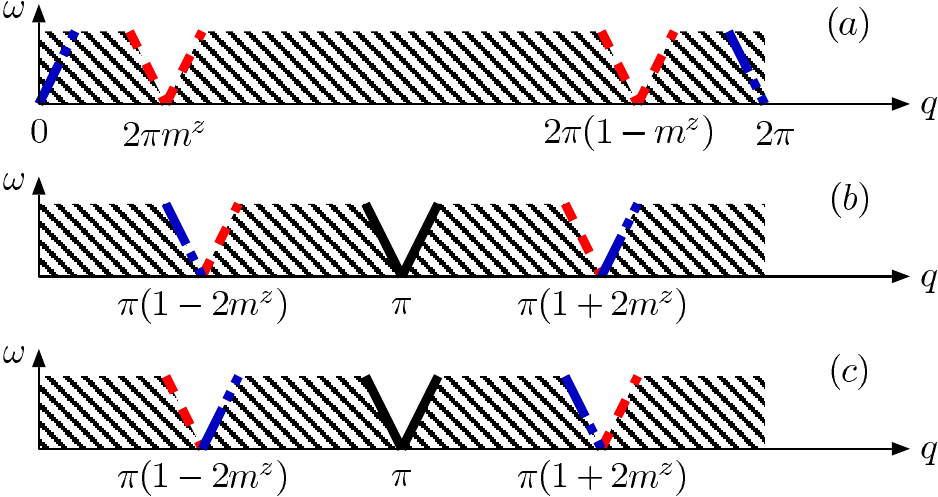}
\end{center}
\caption{\label{fig:LLcorrschema}Map of the low energy spectral functions of the LL model where the white areas represent the continuum of excitations. In the striped areas no excitations are possible. (a) $S^{zz}_0(q,\omega)$: the dash-dotted lines (blue) are the excitation peaks close to $q=0,2\pi$ and the dashed lines (red) are the continuum lower boundary with edge exponent $1-K$ close to $q=2\pi m^z,2\pi(1- m^z)$. (b) $S^{+-}_\pi(q,\omega)$, (c) $S^{-+}_\pi(q,\omega)$: the continuum lower boundary close to $q=\pi,\pi(1\pm 2m^z)$ is represented by solid lines (black) (edge exponent $1-1/4K$), dashed lines (red) (edge exponent $1-\eta_-=2-1/4K-K$) and dash-dotted lines (blue) (edge exponent $1-\eta_+=-1/4K-K$). [Taken from Ref.~\cite{Bouillot_ladder_statics_dynamics}]}
\end{figure}

At zero temperature the correlation functions in the LL have been computed in Ref.~\cite{chitra_spinchains_field,furusaki_correlations_ladder,Sato_NMR_frustrated_ladder}.   In the following, we give directly the expression of the symmetric ($+$) and antisymmetric ($-$) spectral functions $S^{\alpha\beta}_{q_y}=2(S^{\alpha\beta}_{11}\pm S^{\alpha\beta}_{12})$ with rung momentum $q_y=0,\pi$, respectively\footnote{Note that the edge exponents in the incommensurate branches of the correlations \eqref{equ:LLtimepmcorrelation} are inverted compared to their expression in Ref.~\cite{furusaki_correlations_ladder} and pictured in Fig.~\ref{fig:LLcorrschema}.b-c.}.  These are the relevant quantities for a comparison with INS measurements (see Sec.~\ref{sec:INS}). They are derived analogously to the finite temperature correlations\footnote{Note that the two last terms of the zero temperature spectral functions $S^{\pm\mp}_\pi$ originate from those which mix the two fields $\theta$ and $\phi$ in the bosonic description. These were neglected in the bosonic derivation of the finite temperature correlations~\eqref{equ:pmcorrelationqo}.}~\eqref{equ:pmcorrelationqo} and~\eqref{equ:zzcorrelationqo} in the limit $\beta\rightarrow\infty$ using Eq.~\eqref{equ:spectralfunlimit}:
\begin{align}
S^{zz}_0(q,\omega)&=(2\pi{m^z})^2\delta(q)\delta(\omega)+ \frac{K\omega}{u}\Theta(\omega)\left[\delta(\omega-uq)+\delta(\omega+uq)\right]\notag\\
&+\frac{\pi^2A_z}{u\Gamma(K)^2}
\left[\Theta(\omega-u|q-2\pi m^z|)\left(\frac{4u^2}{\omega^2-u^2(q-2\pi m^z)^2}\right)^{1-K}\right.\notag\\ 
&+\{m^z\rightarrow 1-m^z\}\Bigg]\label{equ:LLtimezzcorrelation}
\end{align}
\begin{align}
S^{+-}_\pi(q,\omega)&=\frac{8\pi^2A_x}{u\Gamma(1/4K)^2}\Theta(\omega-u|q-\pi|)\left(\frac{4u^2}{\omega^2-u^2(q-\pi)^2}\right)^{1-1/4K}\notag\\
&+\frac{4\pi^2B_x}{u\Gamma(\eta_+)\Gamma(\eta_-)}\Bigg[\Theta(\omega-u|q-\pi(1-2m^z)|)\notag\\
&\times\left(\frac{2u}{\omega-u[q-\pi(1-2m^z)]}\right)^{1-\eta_-}
\left(\frac{2u}{\omega+u[q-\pi(1-2m^z)]}\right)^{1-\eta_+}\notag\\
&+\{m^z\rightarrow -m^z\}\Bigg]\label{equ:LLtimepmcorrelation}
\end{align}
with $\eta_{\pm}=1/4K\pm1+K$. The spectral function $S^{-+}_\pi$ is obtained replacing $m^z\rightarrow-m^z$ in the $S^{+-}_\pi$ expression Eq.~\eqref{equ:LLtimepmcorrelation}.

\begin{figure}[!h]
\begin{center}
\includegraphics[width=0.5\linewidth]{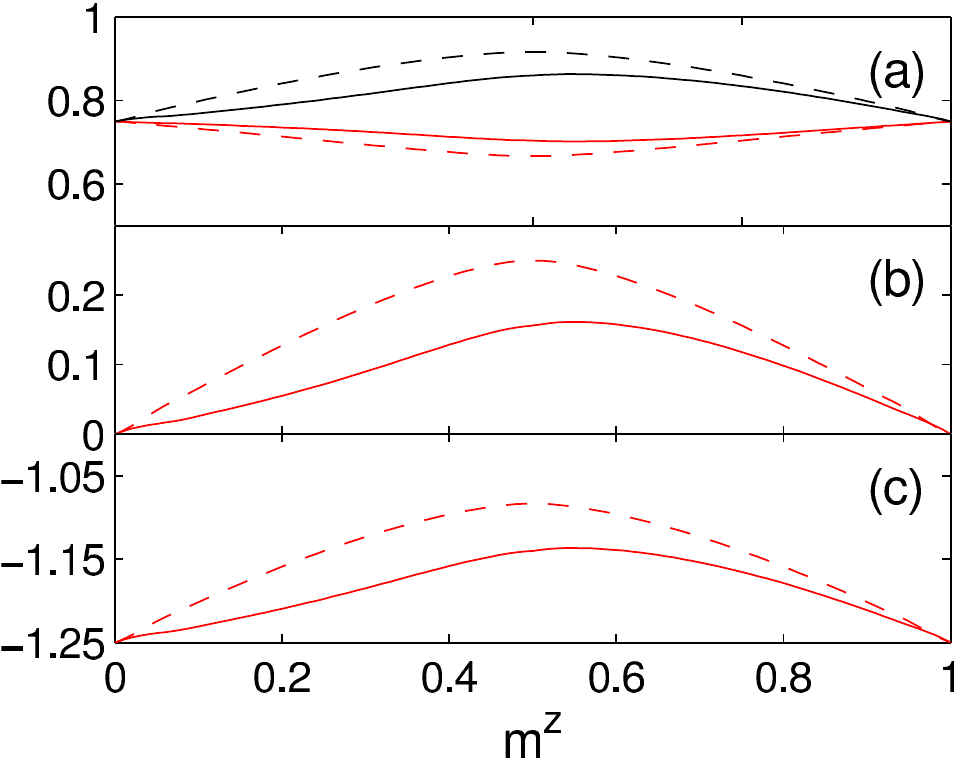}
\end{center}
\caption{Different exponents that appear in the LL correlation functions, Eqs.~\eqref{equ:LLtimezzcorrelation} and~\eqref{equ:LLtimepmcorrelation}, versus the magnetization $m^z$. The solid (dashed) lines are determined from the ladder (spin chain mapping) exponent $K$ in Fig.~\ref{fig:LLparameter}. The exponent $1-K$  of the $S^{zz}_0$ correlations is shown in (b), and the exponent  $1-1/4K$ of the $S^{\pm\mp}_\pi$ correlations at the $q=\pi$ branch in (a) (lower red curves). The exponents $1-\eta_-=2-1/4K-K$ (upper black curves) in (a) and $1-\eta_+=-1/4K-K$ in (c) correspond to both sides of the incommensurate branches of the $S^{\pm\mp}_\pi$ (see Fig.~\ref{fig:LLcorrschema}). [Data taken from Ref.~\cite{Bouillot_ladder_statics_dynamics}]\label{fig:decay_exp}}
\end{figure}

The expressions Eq.~\eqref{equ:LLtimezzcorrelation} and Eq.~\eqref{equ:LLtimepmcorrelation} exhibit the typical behavior of the frequency-momentum LL correlations: a continuum of low energy excitations exists with a linear dispersion with a slope given by the Luttinger velocity $\pm u$.
The spectral weight at the lower boundary of the continuum displays an algebraic singularity with the exponents related to the Luttinger parameter $K$. A summary of this behavior is sketched in Fig.~\ref{fig:LLcorrschema}.
For the considered system the longitudinal
correlation $S^{zz}_0$ is predicted to diverge with the
exponent $1-K$ at its lower edge. As shown in
Fig.~\ref{fig:decay_exp}.b the exponent of this divergence is
very weak $<~0.2$ for the parameters of BCPB. The transverse
correlations $S^{\pm\mp}_\pi$ exhibit a distinct behavior
depending on the considered soft mode. Close to $q=\pi$ the
weight diverges with an exponent given by $1-1/4K$. This
divergence is strong for the considered parameters
($1-1/4K\approx3/4\gg0$ in Fig.~\ref{fig:decay_exp}.a). In
contrast at the soft mode  $q=\pi(1-2m^z),\pi(1+2m^z)$ a
divergence (cusp) is predicted at the lower edge with the
exponent $2-1/4K-K\approx 3/4$ in
Fig.~\ref{fig:decay_exp}.a ($-1/4K-K\approx -5/4$ in
Fig.~\ref{fig:decay_exp}.c).

\section{Mean field approximation}\label{sec:mean-field}

Up to now, we have presented methods adapted to deal with one
dimensional systems. In real compounds, an interladder coupling
is often present. As discussed in
Sec.~\ref{sec:weakinterladderc}, in the incommensurate regime
this interladder coupling $J'$ (cf.~Eq.~\eqref{equ:coupledladdershamiltonian}) can lead to a new
three dimensional order (3D-ordered phase in
Fig.~\ref{fig:phasediagram}.b) at temperatures of the order of
the coupling $J'$. In the case of BPCB the interladder coupling is
much smaller than the coupling inside the ladders, i.e. $J'\ll
J_\perp,J_\parallel$ (Sec.~\ref{sec:bpcb}). Therefore, unless one is extremely close
to $h_{c1}$ or $h_{c2}$ one can treat the interladder coupling
within a mean field approximation. Let us emphasize that this approach
incorporates all the fluctuations inside a ladder. However, it
overestimates the effect of $J'$ by neglecting quantum
fluctuations between different ladders. Such effects can partly be
taken into account by a suitable change of the interladder
coupling~\cite{Thielemann_ND_3Dladder} to an effective value that will be discussed in Sec.~\ref{sec:coupledladderproperties}. Close to the critical
fields the interladder coupling $J'$ becomes larger than the
effective energy of the one dimensional system. This forces one to
consider a three dimensional approach from the start and brings
the physics of the system in the universality class of
Bose-Einstein condensation
~\cite{giamarchi_ladder_coupled,giamarchi_BEC_dimers_review}.
In the following we consider that we are far enough (i.e.~by an
energy of the order of $J'$) away from the critical points so
that we can use the mean field approximation.

The mean field approximation of the interladder interactions in the 3D Hamiltonian $H_{\textrm{3D}}$ (Eq.~\eqref{equ:coupledladdershamiltonian}) reads
\begin{equation}\label{equ:meanfield_rel}
{\bf S}_{l,k,\mu}\cdot{\bf S}_{l',k',\mu'}\cong {\bf S}_{l,k,\mu}\cdot\langle{\bf S}_{l',k',\mu'}\rangle + \langle{\bf S}_{l,k,\mu}\rangle\cdot{\bf S}_{l',k',\mu'}-\langle{\bf S}_{l,k,\mu}\rangle\langle{\bf S}_{l',k',\mu'}\rangle
\end{equation}
and the ladders decouple. 
Since the single ladder correlation functions along the magnetic field direction ($z$~axis) decay
faster than the staggered part of the ones in the perpendicular $xy$~plane (see Eqs.~\eqref{equ:xxcorrelationsimplify}~and~\eqref{equ:zzcorrelationsimplify}
for the LL exponent $K$ of the ladder shown in
Fig.~\ref{fig:LLparameter}), the three dimensional order will
first occur in this plane. Thus the dominant order parameter
is the $q=\pi$ staggered magnetization perpendicular to
the applied magnetic field. Focusing on one of the ladders $\mu$ of the system, we thus introduce the order parameters
\begin{align}\label{equ:mean-fieldvalue}
\langle S^x_{l,k}\rangle=-(-1)^{l+k} m^x_a
\quad\text{and}\quad\langle S^z_{l,k}\rangle=\frac{m^z}{2}-(-1)^{l+k}m^z_a
\end{align}
assuming the staggered $xy$~ordering to be along the $x$~axis. $m^z_a$ will be very small and therefore neglected. Hence
 $H_{\textrm{3D}}$~\eqref{equ:coupledladdershamiltonian} becomes
\begin{equation}\label{equ:mean-fieldhamiltonian}
H_{\textrm{MF}}=J_{\parallel}H_{\parallel} + J_{\perp} H_{\perp}
+\frac{n_cJ'm^z}{4}\sum_{l,k}S_{l,k}^z
+\frac{n_cJ'm^x_a}{2}\sum_{l,k}(-1)^{l+k}S_{l,k}^x.
\end{equation}
Here we suppose that the coupling is dominated by $n_c$ neighboring ladders which are antiferromagnetically ordered (along the $x$ axis) with respect to each other, where $n_c$ is the rung connectivity ($n_c=4$ for the case of BPCB, cf.~Fig. \ref{fig:structure}). This mean field Hamiltonian
corresponds to a single ladder in a site dependent magnetic
field with a uniform component in the $z$~direction and a staggered component in the $x$~direction. The ground state wave function of the Hamiltonian must be determined fulfilling the self-consistency condition for $m^z$ and $m^x_a$ using numerical or analytical methods.

\subsection{Numerical mean field}\label{sec:numericalmean-field}
The order parameters $m^z$ and $m^x_a$ can be computed
numerically by treating the mean field Hamiltonian $H_{\textrm{MF}}$
self consistently with DMRG. These parameters are evaluated
recursively in the center of the ladder (to minimize the
boundary effects) starting with $m^z=0$ and $m^x_a=0.5$. An
accuracy of $<10^{-3}$ on these quantities is quickly reached
after a few recursive iterations (typically $\sim 5$) of the DMRG keeping few hundred DMRG states and treating a system of length $L=150$. We
verified by keeping as well the alternating part of the $z$~order
parameter $m^z_a$ that this term is negligible ($<10^{-5}$).

\subsection{Analytical mean field}

Using the low energy LL description of our ladder system (see
Sec.~\ref{sec:luttinger_liquid}), it is possible to treat the
mean field Hamiltonian $H_{\textrm{MF}}$ within the bosonization
technique. Introducing the LL operators~\eqref{equ:pmluttingeroperator} and~\eqref{equ:zzluttingeroperator} in $H_{\textrm{MF}}$ \eqref{equ:mean-fieldhamiltonian} and keeping only the
most relevant terms leads to the Hamiltonian~\cite{schulz_coupled_spinchains,nagaosa_quasi1D_SC}
\begin{equation}\label{equ:sinegordonhamiltonian}
H_{\textrm{SG}} = \frac{1}{2\pi}\int dr\left[uK\left(\partial_r\theta(r)\right)^2+\frac{u}{K}\left(\partial_r\phi(r) \right)^2\right]\\+\sqrt{A_x}n_cJ'm_a^x\int dr\cos(\theta(r))
\end{equation}
where we neglected the mean field renormalization of $h^z$ in~\eqref{equ:mean-fieldhamiltonian}. This Hamiltonian differs from
the standard LL Hamiltonian $H_{\textrm{LL}}$~\eqref{equ:luttingerliquid} by a cosine term corresponding to
the $x$~staggered magnetic field in~\eqref{equ:mean-fieldhamiltonian}.  It is known as the sine-Gordon Hamiltonian~\cite{coleman_equivalence,luther_chaine_xyz,giamarchi_book_1d}. In the range of the typical $K$ values for BPCB (see Fig.~\ref{fig:LLparameter}) the cosine term in \eqref{equ:sinegordonhamiltonian} is relevant~\cite{giamarchi_book_1d} and orders the field $\theta(r)$. As pictured in Fig.~\ref{fig:Bosonization_rep} this ordering is responsible for the staggered transverse magnetization $m^x_a$. The expectation values of the fields can be derived exactly using integrability~\cite{lukyanov_sinegordon_correlations}. In particular $m^x_a$ can be determined self-consistently as shown in Sec.~\ref{sec:orderparameter}.

\chapter{Static properties and NMR relaxation rate}\label{sec:staticproperties}

In chapter~\ref{sec:coupledladder}, we have seen that the physics of weakly coupled spin-$1/2$ ladders is particularly rich. In the following, we explore the diversity of their phase diagram, pictured in Fig.~\ref{fig:phasediagram}, by computing several physical quantities such as the magnetization, the rung state density and the specific heat. In particular, we test the LL low energy prediction and evaluate the related crossover to the quantum critical regime. Furthermore we discuss the effect of the 3D interladder coupling computing the staggered magnetization in the 3D-ordered phase and its critical temperature. We finally discuss the NMR relaxation rate in the LL gapless regime related to the low energy dynamics. All of these physical quantities are computed for the BPCB parameters (see Sec.~\ref{sec:bpcb}). Hence they can be directly compared to the experiments on BPCB discussed in detail at the end of this chapter.

\section{Critical fields}\label{sec:criticalfields}

\begin{figure}[!h]
\begin{center}
\includegraphics[width=0.5\linewidth]{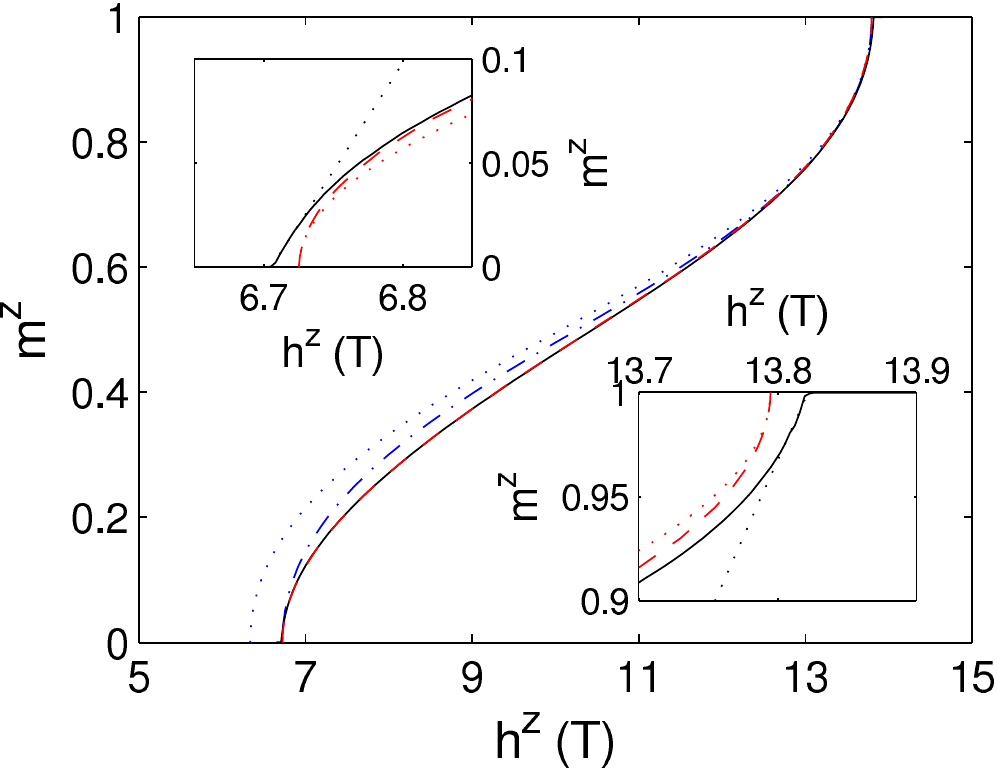}
\end{center}
\caption{Dependence of the magnetization per rung $m^z$ on the magnetic field $h^z$ at zero temperature for the single ladder computed by DMRG with the BPCB couplings (Sec.~\ref{sec:bpcb}) (dashed red line), the spin chain mapping (dotted blue line) rescaled to fit with the single ladder critical fields (dash-dotted blue line), and for the weakly coupled ladders treated by the mean field approximation (solid black line). The insets emphasize the different behavior of the magnetization curves for the single (dashed red line) and weakly coupled (solid black line) ladders close to the critical fields which are indistiguishable in the main part of the figure. The dotted lines in the insets correspond to the linear and square root like critical behavior.
See also Fig~\ref{fig:NMR_magz} for a comparison with the NMR measurements on BPCB. [Taken from Ref.~\cite{Bouillot_ladder_statics_dynamics}]\label{fig:mz_zeroT}}
\end{figure}

The zero temperature magnetization contains extremely useful
information. Its behavior directly gives the critical
values of the magnetic fields $h_{c1}$ and $h_{c2}$ at which
the system enters and leaves the gapless regime, respectively (Fig.~\ref{fig:phasediagram}.b). In
Fig.~\ref{fig:mz_zeroT} the dependence of the magnetization on the applied magnetic field is shown for a single
ladder and for weakly coupled ladders. At low magnetic field,
$h^z<h_{c1}$, the system is in the gapped spin liquid regime
with zero magnetization, and spin singlets on the rungs
dominate the behavior of the system\footnote{The perturbative expression of the ground state in the spin liquid regime and the corresponding singlet and triplet densities are given in appendix~\ref{sec:tjmodelmapping}.}, see Fig.~\ref{fig:tripletdensity}.
At $h^z=h_{c1}$, the
Zeemann interaction closes the spin gap to the rung triplet band
$\ket{t^+}$ (Fig.~\ref{fig:phasediagram}).
Above $h^z>h_{c1}$ the triplet $\ket{t^+}$ band starts to be
populated leading to an increase of the magnetization with
$h^z$.  The lower
critical field in a 13th order expansion~\cite{Weihong_spin_ladder} in $\gamma$ is $h_{c1} \approx 6.73~{\rm
T}$ for the BPCB parameters. At the same time the singlet and the high energy triplets occupation decreases as shown in	 Fig.~\ref{fig:tripletdensity}. For $h^z>h_{c2}= J_\perp+2J_\parallel\approx13.79~{\rm
T}$ (for the compound BPCB), the $\ket{t^+}$ band is completely filled and the other bands are depopulated. The system becomes
fully polarized ($m^z=1$) and gapped. The two critical fields,
$h_{c1}$ and $h_{c2}$, are closely related to the two ladder
exchange couplings, $J_\perp$ and $J_\parallel$. As they are experimentally easily accessible, assuming that a ladder Hamiltonian is an accurate description of the experimental system, these critical fields can be used to
determine the ladder couplings~\cite{Klanjsek_NMR_3Dladder}.

Such a general behavior of the magnetization is seen
for both the single ladder and the weakly coupled ladders in
Fig.~\ref{fig:mz_zeroT}. In particular, the effect of a small
coupling $J'$ between the ladders is completely negligible in the
central part of the curve. Only in the vicinity of the critical
fields, the single ladder and the coupled ladders show a
distinct behavior. The single ladder behaves like an empty
(filled) one-dimensional system of non-interacting fermions
which leads to a square-root behavior $m^z\propto(h^z-h_{c1})^{1/2}$ close to the lower critical field and  $1-m^z \propto(h_{c2}-h^z)^{1/2}$ close to the upper critical field. In contrast, in the system of weakly
coupled ladders, a 3D-ordered phase appears at low enough temperatures
in the gapless regime (see Secs.~\ref{sec:weakinterladderc} and~\ref{sec:mean-field}). The
magnetization dependence close to the critical fields becomes linear, $m^z\propto h^z-h_{c1}^{3D}$, and $1-m^z \propto h_{c2}^{3D}-h^z$, respectively~\cite{giamarchi_ladder_coupled,sachdev_qaf_magfield}. In comparison with the single ladder, the critical
fields $h_{c1}^{3D}$ and $h_{c2}^{3D}$ are shifted by a value of the order of $J'$ in comparison with $h_{c1}$ and $h_{c2}$. This
behavior is in the universality class of the Bose-Einstein
condensation~\cite{giamarchi_BEC_dimers_review,giamarchi_ladder_coupled}. Appearing very close to the critical fields these 3D effects are at the limit of validity of the mean field approximation. Nevertheless they are qualitatively reproduced by this approximation as shown in the insets of Fig.~\ref{fig:mz_zeroT}.

For comparison, the magnetization of a single ladder in the spin chain mapping is also plotted in
Fig.~\ref{fig:mz_zeroT}. This approximation reproduces well the
general behavior of the ladder magnetization discussed above.
However, note that for the exchange coupling constants considered here the
lower critical field in this approximation is different from the ladder one.
The lower critical field is
$h_{c1}^{\rm XXZ}=J_\perp-J_\parallel\approx 6.34 ~{\rm
T}<h_{c1}$. The upper critical field
$h_{c2}^{\rm XXZ}=J_\perp+2J_\parallel=h_{c2}$ is the same as for the ladder. If we rescale $h_{c1}^{\textrm{XXZ}}$ and $h_{c2}^{\textrm{XXZ}}$ to match the critical fields $h_{c1}$ and $h_{c2}$ ($\tilde h^z\rightarrow
\frac{(\tilde h^z-h_{c1}^{\textrm{XXZ}})(h_{c2}-h_{c1})}{h_{c2}^{\textrm{XXZ}}-h_{c1}^{\textrm{XXZ}}}+h_{c1}$), the magnetization curve gets very close to the one calculated for a ladder.
However, in contrast to the magnetization curve for the ladder, the corresponding curve in the spin chain mapping is symmetric with respect to its center at $h_m^{\textrm{XXZ}}=\frac{h_{c1}^{\textrm{XXZ}}+h_{c2}^{\textrm{XXZ}}}{2}=J_\perp+J_\parallel/2$ due to the absence of the high energy triplets.

\begin{figure}[!h]
\begin{center}
\includegraphics[width=0.5\linewidth]{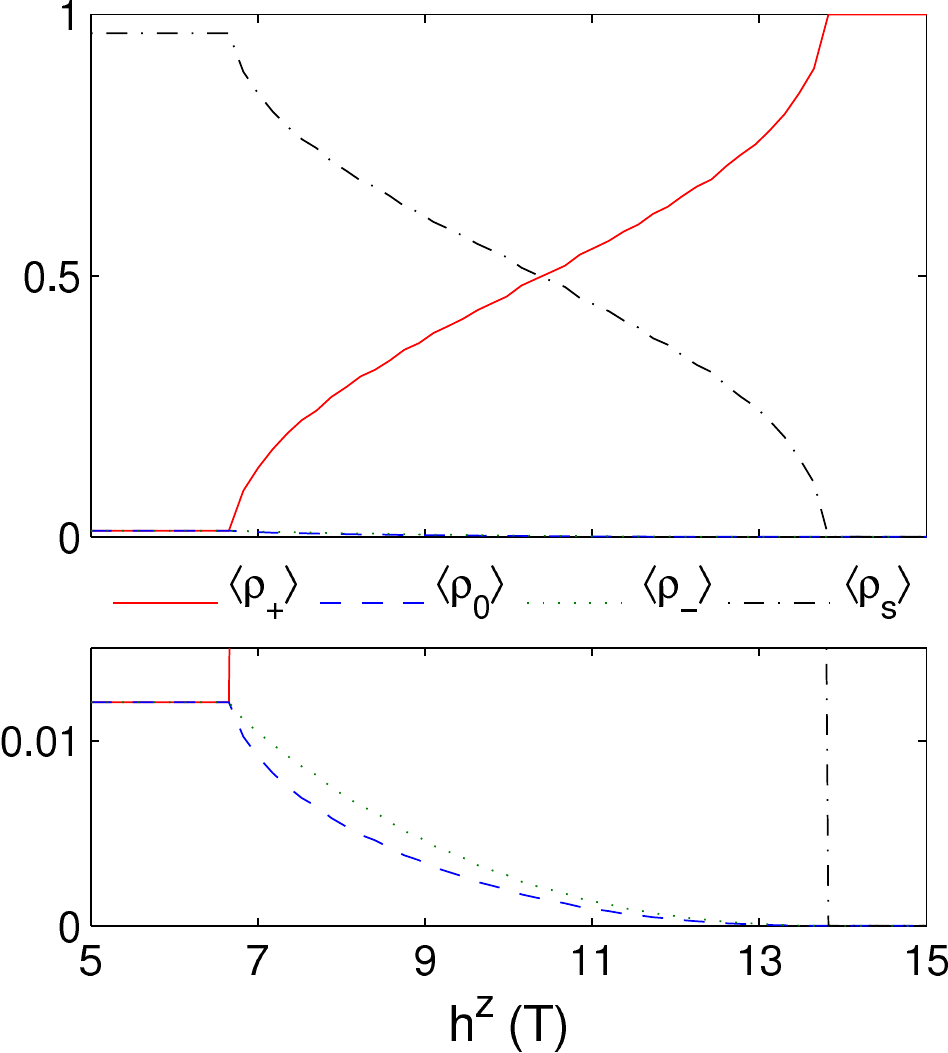}
\end{center}
\caption{Rung state density versus the applied magnetic field $h^z$ at zero temperature for the single ladder computed by DMRG with the BPCB couplings. The dash-dotted (black) lines correspond to the singlet density $\langle \rho_s\rangle$. The triplet densities are represented by the solid (red) lines for $\langle \rho_+\rangle$, the dashed (blue) lines for $\langle \rho_0 \rangle$ and the dotted (green) lines for $\langle\rho_-\rangle$. [Taken from Ref.~\cite{Bouillot_ladder_statics_dynamics}]
\label{fig:tripletdensity}}
\end{figure}

\section{The Luttinger liquid regime and its crossover to the critical regime}

The thermodynamics of the spin-$1/2$ ladders has been studied in the
past for different regimes and coupling ratio $\gamma$~\cite{Troyer_thermo_ladder,Wang_thermo_ladder,wessel01_spinliquid_bec,Ruegg_thermo_ladder,normand_bond_spinladder}.
We here summarize the main interesting features of the
magnetization and the specific heat focusing on the crossover between
the LL regime and the quantum critical region using the BPCB
parameters (Sec.~\ref{sec:bpcb}).
As the interladder exchange coupling $J'$ is supposed very small compared to the
ladder exchange couplings $J_\parallel$ and $J_\perp$, it is reasonable
to neglect $J'$ in the regime far from the
3D phase. Therefore we focus on a single ladder in the
following.

\subsection{Finite temperature magnetization}\label{sec:finiteTmagnetization}

\begin{figure}[!h]
\begin{center}
\includegraphics[width=\linewidth]{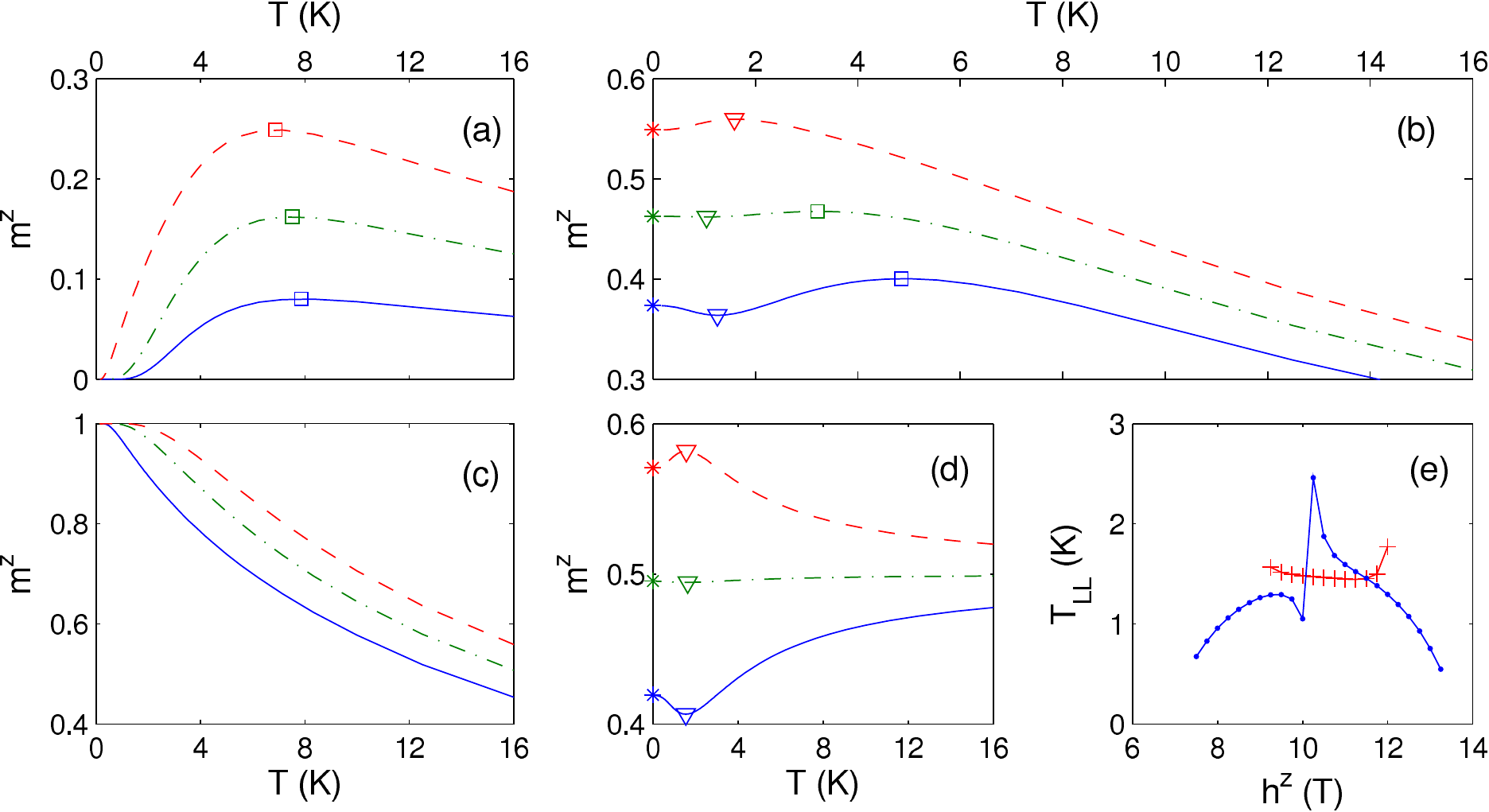}
\end{center}
\caption{Temperature dependence of the magnetization per rung, $m^z(T)$, for the ladder with the BPCB couplings~\eqref{equ:couplings} (a) in the spin liquid regime for $h^z=2~{\rm T}$ (solid blue lines), $h^z=4~{\rm T}$ (dash-dotted green lines) and $h^z=6~{\rm T}$ (dashed red lines), (b) (and (d) for the spin chain mapping) in the gapless LL regime for $h^z=9~{\rm T}$ (solid blue lines), $h^z=10~{\rm T}$ (dash-dotted green lines), $h^z=11~{\rm T}$ (dashed red lines) and (c) in the fully polarized regime for $h^z=15~{\rm T}$ (solid blue lines), $h^z=17~{\rm T}$ (dash-dotted green lines) and $h^z=19~{\rm T}$ (dashed red lines). The results were obtained using T-DMRG. The stars at $T=0~\mathrm{K}$ are the ground state magnetization per rung determined by zero temperature DMRG. The triangles (squares) mark the low (high) temperature extrema. (e) Crossover temperature $T_{\rm LL}$ of the LL to the quantum critical regime versus the applied magnetic field (blue circles for the extremum in $m^z(T)|_{h^z}$ criterium and red crosses for the maximum in $c(T)|_{h^z}$ criterium). See also Fig.~\ref{fig:comp_crossover} for a comparison with magnetocaloric effect measurements on BPCB. [Data taken from Ref.~\cite{Bouillot_ladder_statics_dynamics}] \label{fig:Tmagnetization}}
\end{figure}

We start the description of the temperature dependence of the magnetization, $m^z$, in the two gapped regimes: the spin liquid phase and the fully polarized phase. For small magnetic fields $h^z<h_{c1}$, the magnetization
vanishes exponentially, $m^z\propto\sqrt{T}e^{- (h_{c1}-h^z)/T}$, at low temperature. As shown in Fig.~\ref{fig:Tmagnetization}.a, this decay slowly  disappears while the gap $h_{c1}-h^z$ is closed ($h^z\rightarrow h_{c1}$). After a maximum
at intermediate temperatures $m^z$ decreases to zero for large
temperatures due to strong thermal fluctuations. Similar features appear for large magnetic fields $h^z>h_{c2}$. As shown in Fig.~\ref{fig:Tmagnetization}.c, the
magnetization increases exponentially up to $m^z=1$ at
low temperature, $1-m^z\propto \sqrt{T}e^{-(h^z-h_{c2})/T}$, and decreases monotonously in the limit of
infinite temperature. As in the spin liquid phase, the low temperature exponential behavior becomes more pronounced while the gap  $h^z-h_{c2}$ increases.

In the gapless regime, the magnetization at low temperature has
a non-trivial behavior that strongly depends on the applied
magnetic field. As shown in Fig.~\ref{fig:Tmagnetization}.b, in this regime ($h_{c1}<h^z<h_{c2}$) new extrema appear in the magnetization at low temperature. This behavior can be
understood close to the critical fields where the ladder can be
described by a one-dimensional fermion model with negligible interaction between fermions. Indeed, in
this simplified picture~\cite{giamarchi_book_1d} and in more
refined calculations~\cite{Maeda_spinchain_magnetization,Wang_thermo_ladder,wessel01_spinliquid_bec}
the magnetization has an extremum where the temperature reaches
the chemical potential, i.e., at the temperature at which the energy of
excitations starts to feel the curvature of the energy band. The type of the low temperature extrema depends on the magnetic field derivative of the LL velocity~\cite{Maeda_spinchain_magnetization}  ($\partial u/\partial h^z$). Thus a maximum (minimum) is expected if $\partial u/\partial h^z<0$ ($\partial u/\partial h^z>0$). This
specific behavior is illustrated in
Fig.~\ref{fig:Tmagnetization}.b with the curve for $h^z=11~{\rm T}$
($h_m=\frac{h_{c1}+h_{c2}}{2}<h^z<h_{c2}$) with $(\partial u/\partial h^z)|_{h^z=11{\rm T}}<0$ (see Fig.~\ref{fig:LLparameter}). The low temperature maximum moves to higher temperature for $h^z<h_m$ and crosses over to the already
discussed maximum for $h^z<h_{c1}$ (see Fig.~\ref{fig:Tmagnetization}.a). Symmetrically with respect
to $h_m$, a low temperature minimum appears in the curve for $h^z=9~{\rm T}$
($h_{c1}<h^z<h_m$) with  $(\partial u/\partial h^z)|_{h^z=9{\rm T}}>0$ (see Fig.~\ref{fig:LLparameter}). This minimum slowly disappears when
$h^z\to h_m$ for which $(\partial u/\partial h^z)|_{h_m}\approx 0$ (the curve for $h^z=10~{\rm T}$ is close to that).

The location of the lowest extremum is a reasonable
criterion to characterize the crossover temperature between the
LL and the quantum critical regime~\cite{Maeda_spinchain_magnetization,Wang_thermo_ladder,wessel01_spinliquid_bec},
since the extremum occurs at temperatures of the order of the
chemical potential. A plot of this crossover temperature versus
the magnetic field is presented in
Fig.~\ref{fig:Tmagnetization}.e. Following this criterium, the
crossover has a continuous shape far from $h_m$. Nevertheless, close to $h_m$  we have $\partial u/\partial h^z\approx 0$ and the low energy extremum disapears. The criterium is thus not well defined and presents a discontinuity at $h_m$ which is obviously an artefact. In the vicinity of $h_m$, we thus
use another crossover criterium based on the specific heat (see
Sec.~\ref{sec:specificheat}) that seems to give a more accurate
description. 

The temperature dependence of the magnetization of the spin chain mapping, Fig.~\ref{fig:Tmagnetization}.d, exhibits a single low temperature
maximum if $h_m^{\textrm{XXZ}}<h^z<h_{c2}^{\textrm{XXZ}}$ (minimum
if $h_{c1}^{\textrm{XXZ}}<h^z<h_m^{\textrm{XXZ}}$). The appearance of a single extremum and its convergence to $m^z\rightarrow 0.5$ when $T\rightarrow\infty$
is due to the exact
symmetry with respect to the magnetic field $h_m^{\textrm{XXZ}}$. This approximation reproduces the main low energy features of the ladder but fails to describe the high energy behavior which strongly depends on the high energy triplets.

\subsection{Specific heat}\label{sec:specificheat}

\begin{figure}[!h]
\begin{center}
\includegraphics[width=\linewidth]{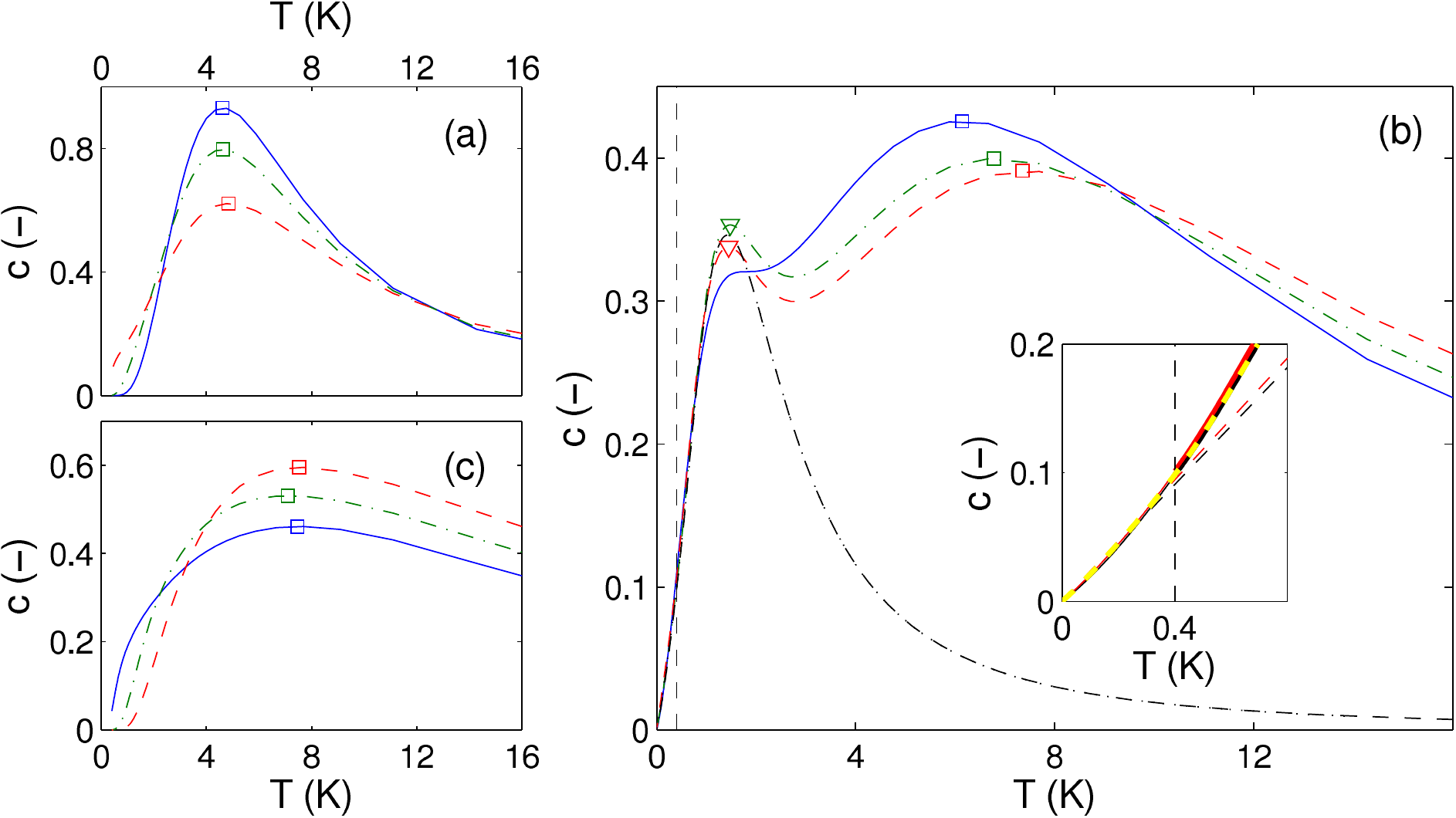}
\end{center}
\caption{Specific heat per rung $c$ versus temperature $T$ computed with T-DMRG for the BPCB couplings~\eqref{equ:couplings} (a) in the spin liquid regime for $h^z=2~{\rm T}$ (solid blue line), $h^z=4~{\rm T}$ (dash-dotted green line), and $h^z=6~{\rm T}$ (dashed red line), (b) in the gapless LL regime for $h^z=9~{\rm T}$ (solid blue line), $h^z=10~{\rm T}$ (dash-dotted green line), and $h^z=11~{\rm T}$ (dashed red line) and (c) in the fully polarized regime  for $h^z=15~{\rm T}$ (solid blue line), $h^z=17~{\rm T}$ (dash-dotted green line), and $h^z=19~{\rm T}$ (dashed red line). 
In (b) the spin chain mapping at $h^z=10~{\rm T}$ solved by T-DMRG (Bethe ansatz from Ref.~\cite{Bouillot_ladder_statics_dynamics}) are plotted in dashed (dotted) black line. Note that the two lines are hardly distinguishable. The triangles (squares) mark the low (high) temperature maxima of the specific heat versus temperature. The vertical dashed line marks the temperature $T=0.4~{\rm K}$ below which the DMRG results are extrapolated (see Sec.~\ref{sec:temperatureDMRG}). The inset in (b) shows the low temperature dependence of the specific heat per rung for $h^z=10~\mathrm{T}$. The T-DMRG calculations are in red thick lines for the ladder with the BPCB couplings~\eqref{equ:couplings} (black thick lines for the spin chain mapping). The two curves can hardly be distinguished. Their low temperature polynomial extrapolation is plotted in thin lines below $T=0.4~{\rm K}$ (represented by a vertical dashed line). The linear low temperature behavior of the LL is represented by dashed lines (red for the ladder, black for the spin chain mapping). The dashed yellow lines correspond to the Bethe ansatz~\cite{Bouillot_ladder_statics_dynamics} computation for the spin chain mapping. See also Fig.~\ref{fig:chalspe_comp} for a comparison with measurements on BPCB. [Data taken from Ref.~\cite{Bouillot_ladder_statics_dynamics}]\label{fig:cvsT}}
\end{figure}

Similarly to the magnetization discussed in Sec.~\ref{sec:finiteTmagnetization}, the specific heat of spin-$1/2$ ladders shows in the spin liquid and fully polarized phases the typical behavior of gapped regimes. At low temperature the specific heat grows exponentially: $c\propto T^{-3/2}e^{-(h_{c1}-h^z)/T}$ and $c\propto T^{-3/2}e^{-(h^z-h_{c2})/T}$ for both phases respectively. After reaching a maximum when the gapped excitations start to be thermally populated, in the quantum critical regime (see Fig.~\ref{fig:phasediagram}.b), it slowly decreases to zero at high temperature due to the strong temperature fluctuations. These specific features are shown in Figs.~\ref{fig:cvsT}.a and~\ref{fig:cvsT}.c where $c(T)$ is plotted for various applied magnetic fields in both gapped regimes.

As presented in Fig.~\ref{fig:cvsT}.b, the behavior of the specific heat becomes more subtle in the gapless regime in which the contribution due to the gapless spinon excitations appear at low temperature. This
results in a peak around $T\sim 1.5~{\rm K}$. This peak is most
pronounced for the magnetic field values lying mid value between
the two critical fields. At higher temperatures the
contribution of the gapped triplet excitations leads to a second
peak which exists also in the gapped regimes as discussed above and shown in  Figs.~\ref{fig:cvsT}.a and~\ref{fig:cvsT}.c. Its position depends on the magnetic field (see Ref.~\cite{Wang_thermo_ladder} for a detailed discussion). To separate
out the contribution from the low lying spinon excitations, we
compare the specific heat of the ladder to the results obtained
by the spin chain mapping in which we just keep the lowest two modes of the ladder (see Sec.~\ref{sec:spinchainmap} and appendix~\ref{sec:tjmodelmapping}).
The resulting effective chain model is solved using
Bethe ansatz~\cite{Bouillot_ladder_statics_dynamics} and
T-DMRG methods. The
agreement between these methods is excellent and the corresponding curves in Fig.~\ref{fig:cvsT}.b can
hardly be distinguished. However, a clear difference with the
full spin ladder result is revealed. While at low
temperatures the curves are very close, the first peak
in the spin chain mapping already lacks some weight, which stems
from higher modes of the ladder.

In the inset of Fig.~\ref{fig:cvsT}.b, the low temperature region is
analyzed in more detail. At very low temperatures the spinon
modes of the ladder can be described by the LL theory (see Sec.~\ref{sec:luttinger_liquid}) which
predicts a linear rise with temperature~\cite{giamarchi_book_1d,Suga_LL_spinchain} inversely proportional
to the spinon velocity $u$ (shown in Fig.~\ref{fig:LLparameter} versus the applied magnetic field for the BPCB couplings)
\begin{equation}\label{equ:LLspec_heat}
c_{\textrm{LL}}(T)=\frac{T\pi}{3u}.
\end{equation}
In the inset of Fig.~\ref{fig:cvsT}.b we compare the results of the LL,
the Bethe ansatz~\cite{Bouillot_ladder_statics_dynamics} and the DMRG results for the effective spin chain
and the numerical DMRG results taking the full ladder into
account. The numerical results for the adaptive T-DMRG at finite
temperature are extrapolated to zero temperature by connecting
algebraically to zero
temperature DMRG results (see Sec.~\ref{sec:temperatureDMRG}). A very good agreement between~\eqref{equ:LLspec_heat} and numerics is found for
low temperatures. However, at higher
temperatures, the slope of the $T\to 0$ LL
description slightly changes with respect to the curves calculated with other methods. This change of slope reflects the fact that the
curvature of the energy dispersion must be taken into account
when computing the finite temperature specific heat, and this
even when the temperature is quite small compared to the
effective energy bandwidth of the system. The effective
spin chain and the numerical results for the ladder agree 
for higher temperatures (depending on the magnetic field),
before the higher modes of the ladder cause deviations.

As for the magnetization (Sec.~\ref{sec:finiteTmagnetization}),
the location of the low temperature peak can be interpreted as the
crossover of the LL to the quantum critical regime.  Indeed, in
a free fermion description which is accurate close to the critical fields, this
peak appears at the temperature for which the excitations stem
from the bottom of the energy band. The corresponding temperature
crossover is compared in Fig.~\ref{fig:Tmagnetization}.e to the
crossover temperature extracted from the first magnetization
extremum (Sec.~\ref{sec:finiteTmagnetization}). The two
crossover criteria are complementary due to their domain of
validity. The first specific heat maximum is well pronounced
only in the center of the gapless phase. In contrast in this regime the disappearance of low energy extrema renders the magnetization
criterium very imprecise (cf.~Sec.~\ref{sec:finiteTmagnetization}). In Fig.~\ref{fig:comp_crossover}, both criteria have been
applied on the magnetocaloric effect and specific heat measurements on BPCB~\cite{Ruegg_thermo_ladder}. These experimentally extracted crossovers are in perfect agreement with the ones computed by T-DMRG.

More generally comparisons of the computed specific heat with actual
experimental data~\cite{Ruegg_thermo_ladder} for BPCB are excellent (see Fig.~\ref{fig:chalspe_comp} in the experimental Sec.~\ref{sec:thermo_BPCB}). For these comparisons the theoretical data are computed with $g=2.06$ related to the experimental orientation of the sample with respect to the magnetic field~\cite{Ruegg_thermo_ladder} (see Sec.~\ref{sec:bpcb}) and rescaled by a factor $0.98$ in agreement with the global experimental uncertainties\footnote{An additional scaling factor $7.47\ {\rm mJ/gK}$ has to be applied on the theoretical specific heat (per rung) to convert to the experimental units.}.

\section{Spin-lattice relaxation rate}\label{sec:relaxationtime}

The NMR spin-lattice relaxation in quantum spin systems is mainly due to the pure magnetic coupling 
$$
H_{e-n}=\gamma_n A_{\alpha\beta}I^\alpha S^\beta
$$
between electronic and nuclear spins ${\bf S} =(S^x,S^y,S^z)$ and ${\bf I}=(I^x,I^y,I^z)$, respectively. $A_{\alpha\beta}$ with $\alpha,\beta=x,y,z$ is the hyperfine tensor related to the dipolar interaction between the spins ${\bf S}$ and ${\bf I}$. $\gamma_n$ is the nuclear gyromagnetic ratio of the measured nuclear spin ${\bf I}$ ($\gamma_n=19.3~\mathrm{MHz/T}$ for the measurements on BPCB done on the $\text{N}(1)$ atoms in Fig.~\ref{fig:chemical_BPCB} (see
Ref.~\cite{Klanjsek_NMR_3Dladder})). Due to this form of the coupling the NMR spin-lattice relaxation rate
$T_1^{-1}$ is directly related through the Redfield equations \cite{slichter_NMR_book} to the local correlations 
$S^{\alpha\beta}(\omega_0)$ defined in Eq.~\eqref{equ:local_correlation}
\begin{equation}\label{equ:T1init}
T_1^{-1}=\gamma_n^2 A_\perp^2 S^{+-}(\omega_0)+\gamma_n^2 A_\parallel^2 S^{zz}(\omega_0).
\end{equation}
$\omega_0=h^z\gamma_n$ is the Larmor frequency.  $A_\parallel$ and $A_\perp$ are the longitudinal and transverse components of the hyperfine tensor which have the same order of magnitude than the components $A_{\alpha\beta}$.

\begin{figure}[!h]
\begin{center}
\includegraphics[width=0.5\linewidth]{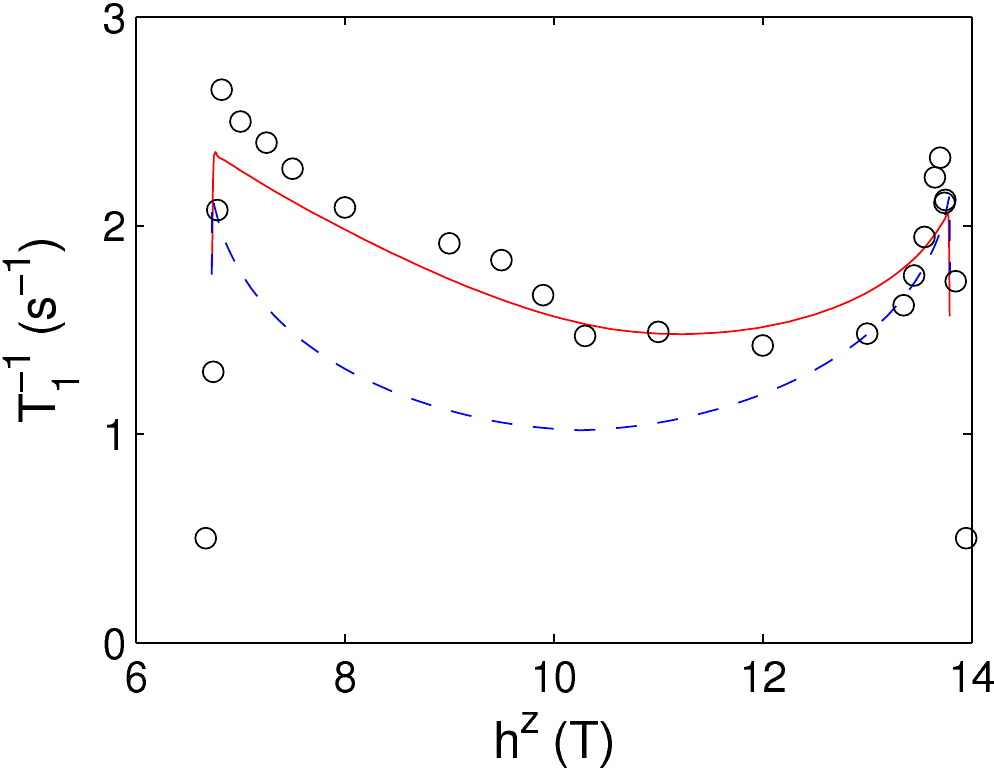}
\end{center}
\caption{Magnetic field dependence of the NMR relaxation rate, $T_1^{-1}(h^z)$, at $T=250~\mathrm{mK}$. The solid red line is the bosonization determination using the ladder LL parameters for the BPCB couplings shown in Fig.~\ref{fig:LLparameter} (the dashed blue line uses the LL parameters of the spin chain mapping). The black circles are the measured NMR relaxation rate on BPCB done on the $\text{N}(1)$ atoms in Fig.~\ref{fig:chemical_BPCB} from Ref.~\cite{Klanjsek_NMR_3Dladder}. [Taken from Ref.~\cite{Bouillot_ladder_statics_dynamics}]\label{fig:T1}}
\end{figure}

Assuming $J_\parallel\gg T$, the relaxation rate $T^{-1}_1$ can be computed in the gapless regime using the LL low energy description  (Sec.~\ref{sec:luttinger_liquid}) of the electronic spin dynamics. We introduce only the most relevant local LL correlation~\eqref{equ:local_correlationLL} (in the limit $T\gg\omega_0$)   into Eq.~\eqref{equ:T1init} (i.e. the transverse component $S^{+-}(\omega\rightarrow0)\gg S^{zz}(\omega\rightarrow0)$ for the LL parameters shown in Fig.~\ref{fig:LLparameter}). We obtain
\begin{equation}\label{equ:T1}
T_1^{-1}=\frac{2\gamma_n^2A_\perp^2A_x\cos\left(\frac{\pi }{4K}\right)}{u}\left(\frac{2\pi T}{ u}\right)^{\frac{1}{2K}-1}B\left(\frac{1}{4K},1-\frac{1}{2K}\right).
\end{equation}
The theoretical shape of
$T^{-1}_1(h_z)$ plotted in Fig. \ref{fig:T1} at
$T=250~\mathrm{mK}\gg T_c\approx J'$ is totally determined by the LL
parameters (Fig.~\ref{fig:LLparameter}). Thus, similarly to the LL parameters, it shows a strong asymmetry with respect to the center of the gapless phase which is perfectly reproduced by the NMR measurements on BPCB (see  Fig.~\ref{fig:T1}). The only free (scaling) parameter, $A_\perp=0.041~\mathrm{T}$, is deduced
from the fit of Eq.~\eqref{equ:T1} to the experimental data~\cite{Klanjsek_NMR_3Dladder}. The fitted parameter agrees with that obtained from direct N$^{14}$ NMR determination~\cite{Klanjsek_NMR_3Dladder} and rescale globally the theoretical curves without modifying its shape. For
comparison, the $T^{-1}_1$ obtained in the spin chain mapping
approximation is also plotted in  Fig.~\ref{fig:T1}. As for
other physical quantities, this description fails to
reproduce the non-symmetric shape.

\section{Properties of weakly coupled ladders}\label{sec:coupledladderproperties}

As discussed in Sec.~\ref{sec:weakinterladderc}, the interladder coupling $J'$ induces a low temperature ordered phase (the 3D-ordered phase
in Fig.~\ref{fig:phasediagram}.b). Using the mean field
approximation presented in Sec.~\ref{sec:mean-field} we
characterize the ordering and compute the critical temperature
and the order parameter related to this phase.

\subsection{3D order transition temperature}\label{sec:transitiontemperature}

\begin{figure}[!h]
\begin{center}
\includegraphics[width=0.5\linewidth]{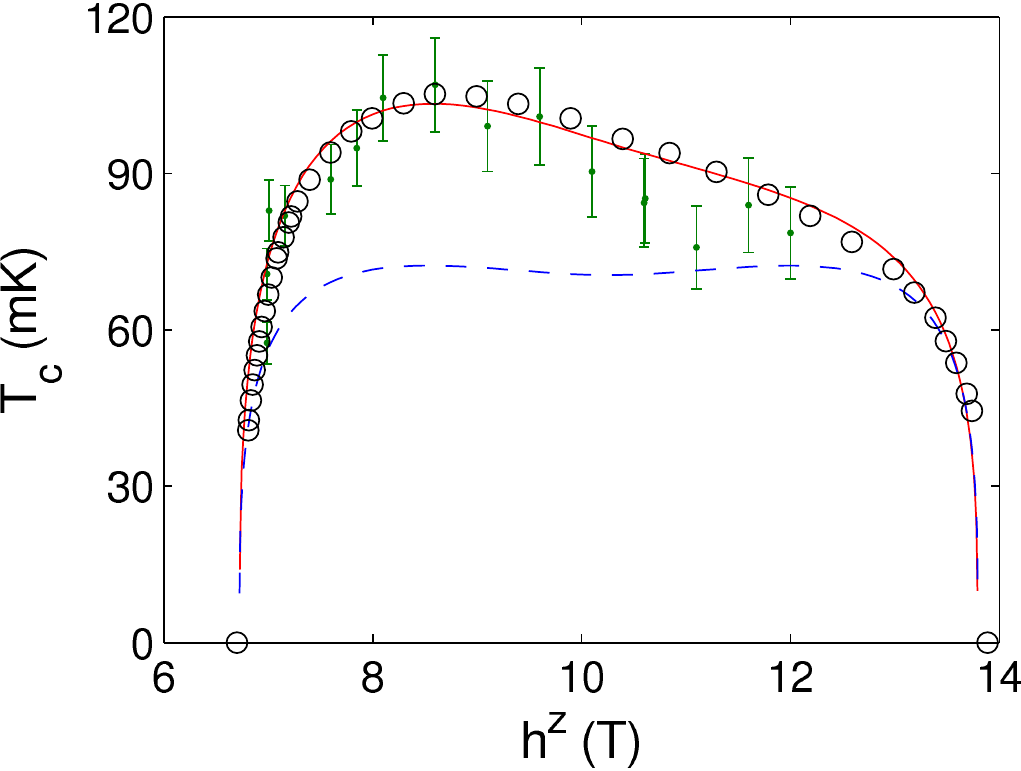}
\end{center}
\caption{Magnetic field dependence of the transition temperature between the gapless regime and the 3D-ordered phase, $T_c(h^z)$, is plotted in solid red line for the ladder LL parameters of BPCB shown in Fig.~\ref{fig:LLparameter} (in dashed blue line for the LL parameters of the spin chain mapping). The NMR measurements from Ref.~\cite{Klanjsek_NMR_3Dladder} are represented by black circles and the neutron diffraction measurements from Ref.~\cite{Thielemann_ND_3Dladder} by green dots. [Taken from Ref.~\cite{Bouillot_ladder_statics_dynamics}]
\label{fig:criticaltemperature}}
\end{figure}

In order to compute the critical temperature of the 3D
transition, we follow
Ref.~\cite{giamarchi_ladder_coupled} and treat the
staggered part of the mean field Hamiltonian $H_{\textrm{MF}}$~\eqref{equ:mean-fieldhamiltonian} perturbatively using linear
response. The instability of the resulting mean field transverse
susceptibility for an order with momentum $q=\pi$ (staggered order), due to the 3D transition, appears at $T_c$
when~\cite{scalapino_q1d}
\begin{equation}\label{equ:instabilitycondition}
\left.\chi_{11}^{+-}(q=\pi,\omega=0)\right|_{T_c}=-\frac{1}{n_cJ'}.
\end{equation}
Where $\chi_{11}^{+-}$ is the transverse correlation
function of an isolated single ladder system defined in Eq.~\eqref{equ:correlationdef}. As $T_c\approx J'\ll J_\parallel$, this correlation can be computed analytically (see
Eq.~\eqref{equ:pmcorrelationqo}) using the LL low energy description
of the isolated ladder (Eq.~\eqref{equ:luttingerliquid}) in the gapless regime. Applying the condition~\eqref{equ:instabilitycondition} to the LL correlation~\eqref{equ:pmcorrelationqo} leads to the critical temperature
\begin{equation}\label{equ:criticaltemperature}
T_c=\frac{u}{2\pi}\left(\frac{A_xJ'n_c\sin\left(\frac{\pi}{4K}\right)B^2\left(\frac{1}{8K},1-\frac{1}{4K}\right)}{2u}\right)^{\frac{2K}{4K-1}}.
\end{equation}
Introducing the computed LL parameters $u$, $K$ and $A_x$ (see
Fig.~\ref{fig:LLparameter}) in this expression, we get
the critical temperature~\cite{Klanjsek_NMR_3Dladder} as a function of the magnetic field. Only $J'$ remains as a free parameter. Fitting our results for the transition temperature to the experimental data (Fig.~\ref{fig:criticaltemperature}) allows us to extract the {\it mean field} interladder coupling $J'_{\textrm{MF}}\approx20~\mathrm{mK}$ for the experimental compound
BPCB. The asymmetry of the LL
parameters induces a strong asymmetry of $T_c$ with respect
to the middle of the 3D phase which is in very good agreement with the experimentally observed asymmetry.

As the mean field approximation
neglects the quantum fluctuations between the ladders, the
critical temperature $T_c$ is overestimated for a given
$J'_{\textrm{MF}}$. In order to fully take the fluctuations into account a Quantum Monte Carlo (QMC) determination of $T_c$ based on the same 3D lattice structure is performed in Ref.~\cite{Bouillot_ladder_statics_dynamics}.
Let us note that QMC simulations of the coupled spin ladder Hamiltonian~\eqref{equ:coupledladdershamiltonian} are possible since the 3D lattice structure, Fig.~\ref{fig:structure}, is unfrustrated. Currently, this determination is only accessible for larger interladder couplings $J'$ and shows~\cite{Thielemann_ND_3Dladder} that the real critical
temperature is well approximated by the mean field
approximation, but with a rescaling of the real interladder
coupling $J'\approx27~{\rm mK}=\alpha^{-1} J'_{\textrm{MF}}$ with
$\alpha\approx0.74$. The rescaling factor $\alpha$ is essentially magnetic field independent~\cite{Thielemann_ND_3Dladder} and similar to the values obtained for other quasi one-dimensional antiferromagnets~\cite{yasuda_neelorder,Todo_chain_MFT}.

\subsection{Zero temperature 3D order parameter}\label{sec:orderparameter}

\begin{figure}[!h]
\begin{center}
\includegraphics[width=0.5\linewidth]{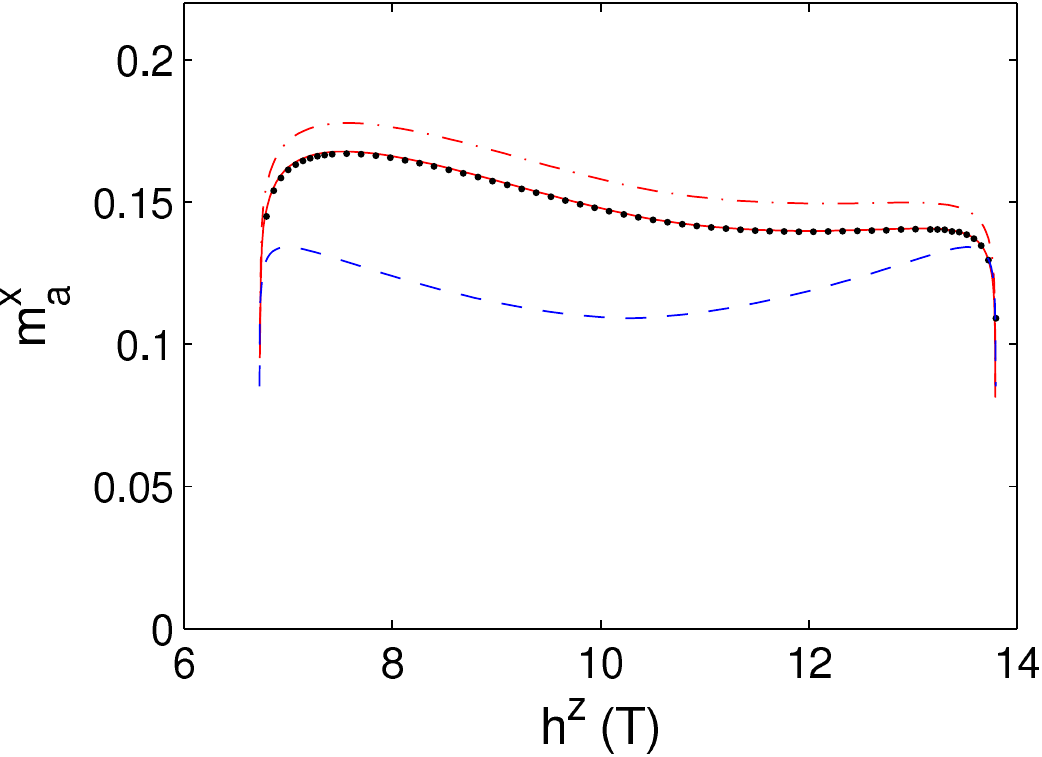}
\end{center}
\caption{Magnetic field dependence of the transverse staggered
  magnetization per spin, $m^x_a(h^z)$, at zero temperature in the 3D-ordered phase. Results of the analytical bosonization technique for the LL parameters of the compound BPCB shown in Fig.~\ref{fig:LLparameter} and $J'=27~{\rm mK}$ is represented by the dash-dotted red line (dashed blue line for the LL parameters of
  the spin chain mapping). The DMRG result for $J'=20~{\rm
    mK}(=J'_{\textrm{MF}})$ is represented by black dots (as a comparison
  the bosonization result for $J'=20~{\rm mK}$ is plotted in solid red
  line). Note, that these two curves are almost indistinguishable. A comparison of $m^x_a(h^z)$ with the experimentally determined value using ND and NMR is shown in Fig.~\ref{fig:orderparameter_exp}.
[Taken from Ref.~\cite{Bouillot_ladder_statics_dynamics}]\label{fig:orderparameter}}
\end{figure}

The staggered order parameter in the 3D-ordered phase, $m^x_a$, can  be
analytically determined at zero temperature using the mean field approximation for the interladder coupling and the bosonization technique (see Sec.~\ref{sec:mean-field}). As
$m^x_a=\sqrt{A_x}\langle\cos(\theta(r))\rangle$ in the
bosonization description (Sec.~\ref{sec:luttinger_liquid}) and the expectation value~\cite{lukyanov_sinegordon_correlations} of the
operator $e^{i\theta(r)}$ is
\begin{equation}\label{equ:expsinegordon} \left\langle
e^{i\theta(r)}\right\rangle = F(K)\left(\frac{\pi
\sqrt{A_x} n_cJ'm^x_a}{2u}\right)^{\frac{1}{8K-1}}
\end{equation}
for the sine-Gordon Hamiltonian $H_{\textrm{SG}}$~\eqref{equ:sinegordonhamiltonian} with
\begin{equation}
F(K)=\frac{\frac{\pi^2}{\sin\left(\frac{\pi}{8K-1}\right)}
\frac{8K}{8K-1}  \left[\frac {\Gamma\left(1-\frac 1{8K}\right)}
{\Gamma\left(\frac{1}{8K}\right)}\right]^{\frac{8K}
{8K-1}}}{\left[\Gamma \left(\frac{4K}{8K-1} \right)  \Gamma
\left(\frac{16K-3}{16K-2} \right)\right]^2},
\end{equation}
we can extract
\begin{equation}\label{equ:orderparameter}
m_a^x=\sqrt{A_x}F(K)^{\frac{8K-1}{8K-2}}\left(\frac{\pi n_c
A_xJ'}{2u}\right)^{\frac{1}{8K-2}}.
\end{equation}
This can be evaluated in the 3D-ordered phase by introducing into~\eqref{equ:orderparameter} the LL parameters $u$, $K$ and $A_x$ from Fig.~\ref{fig:LLparameter}. Fig.~\ref{fig:orderparameter} shows the order parameter
versus the magnetic field determined analytically and
numerically by DMRG (see Sec.~\ref{sec:numericalmean-field}). The two
curves are almost indistinguishable and exhibit a strongly
asymmetric camel-like shape~\cite{Klanjsek_NMR_3Dladder} with
two maxima close to the critical fields. The asymmetry of the curve
is again due to the
presence of the additional triplet states. This asymmetry disappears in the spin chain mapping. 

\section{Comparison with experimental results on BPCB}\label{sec:thermo_BPCB}

Many experimental measurements related to the theoretical results presented at the beggining of this chapter have been performed on BPCB. In order to characterize this compound and understand its physical behavior, we discuss, in this section, several experiments performed on BPCB and compare these to the theoretical predictions.

\begin{figure}[!h]
\begin{center}
\includegraphics[width=0.5\linewidth]{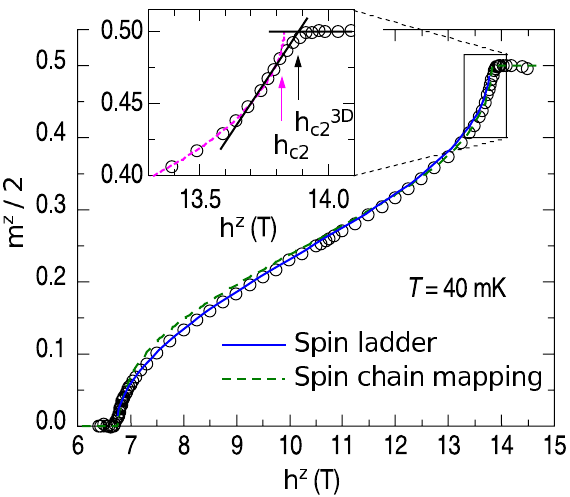}
\end{center}
\caption{Magnetic field dependence of the magnetization per Cu$^{2+}$ ion in BPCB ($m^z/2$) measured by NMR at $T=40\ \text{mK}$. The data are compared to the result of the DMRG calculation for a single ladder with the BPCB couplings (solid blue line) and for the spin chain mapping rescaled to fit with the single ladder critical fields (dashed green line), both at $T = 0$. Inset shows the critical linear dependence characteristic for weakly coupled ladders very close
to $h_{c2}^{3D}$ (solid black line) and the determination of $h_{c2}$ using the assumption of a square root critical behavior for a single ladder (dashed pink line). See also Fig~\ref{fig:mz_zeroT} for a comparison of the theoretical computations. [Taken from Ref.~\cite{Klanjsek_NMR_3Dladder}]\label{fig:NMR_magz}}
\end{figure}

\begin{figure}[!h]
\begin{center}
\includegraphics[width=0.45\linewidth]{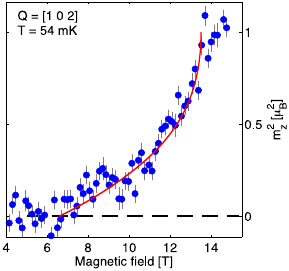}
\end{center}
\caption{Magnetic field dependence of the square of the magnetization per rung in BPCB, ${m^z}^2(h^z)$, measured by ND at $T=54\ \text{mK}$ (blue circles). The solid red line represents the theoretical prediction computed by DMRG with the BPCB couplings. [Taken from Ref.~\cite{Thielemann_ND_3Dladder}]\label{fig:mz2_ND}}
\end{figure}

The longitudinal magnetization that can be measured very
precisely by NMR at $T=40\ \text{mK}$  (see Ref.~\cite{Klanjsek_NMR_3Dladder}) agrees remarkably well with the one computed using the weakly coupled ladder model (see Fig.~\ref{fig:NMR_magz}). In particular the linear growth close to the critical fields due to the small interladder coupling $J'$ (Sec.~\ref{sec:criticalfields}) is highlighted (in the inset of Fig.~\ref{fig:NMR_magz}). Nevertheless, the main shape of the magnetization is not very sensitive to the underlying model (see Fig.~\ref{fig:mz_zeroT} and~\ref{fig:NMR_magz}). Thus it cannot be used to distinguish between various models. However once the model is chosen e.g. a spin ladder, it can
be used to fix precisely the parameters given the high accuracy
of the experimental data. In particular, the position of the critical fields are very sensitive to the values of the intraladder couplings (Sec.~\ref{sec:criticalfields}). The couplings determined by the magnetization are
$J_{\perp}\approx12.6~\mathrm{K}$ and $J_\parallel\approx3.55~\mathrm{K}$. Note that the magnetization curve has been also measured by ND~\cite{Thielemann_ND_3Dladder} (see Fig.~\ref{fig:mz2_ND}) and agrees perfectly with the NMR experiments\footnote{ $g=2.17$ for the experimental settings of ND measurements.}.

\begin{figure}[!h]
\begin{center}
\includegraphics[width=0.6\linewidth]{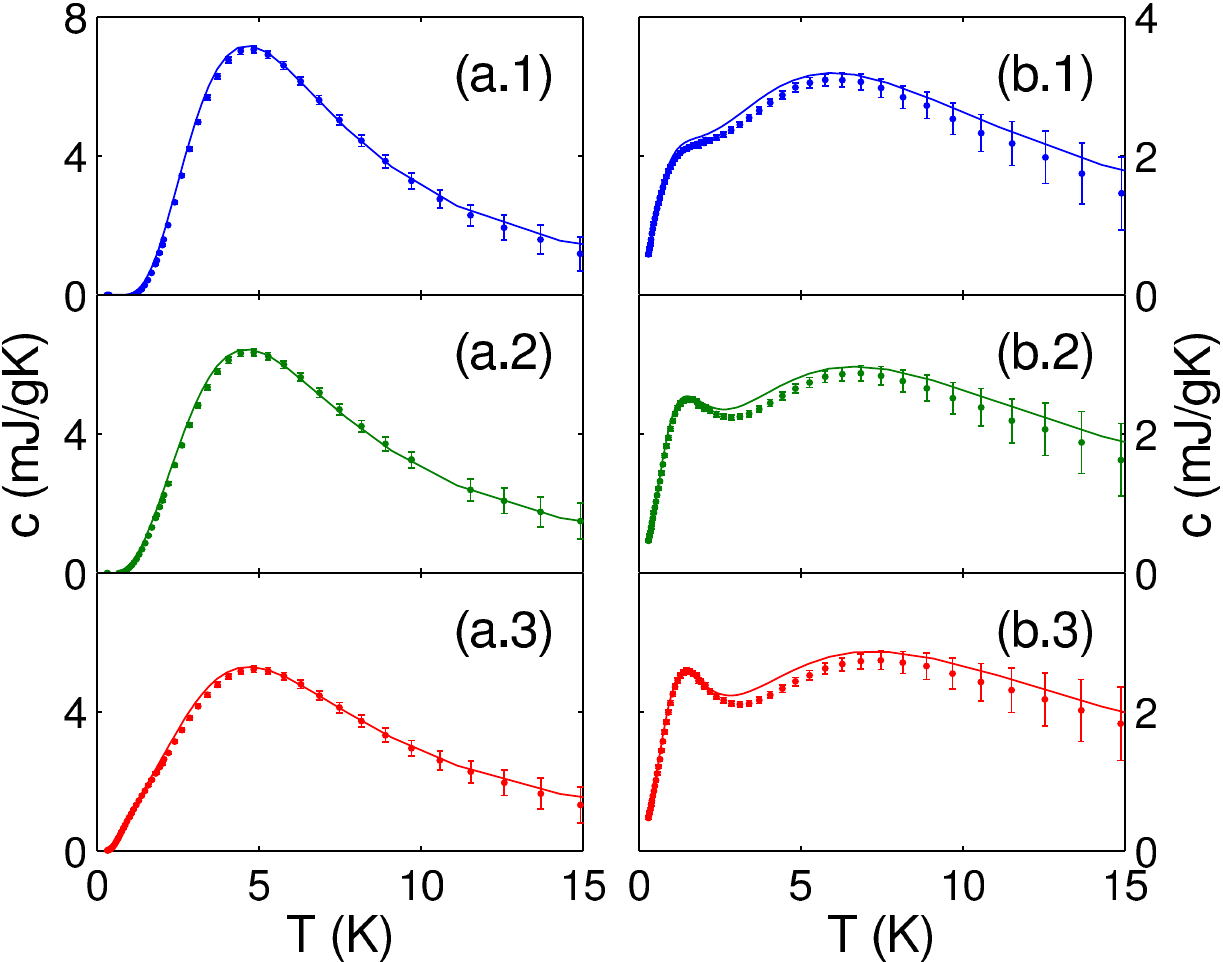}
\end{center}
\caption{Temperature dependence of the specific heat measurements $c(T)$ on the compound BPCB from Ref.~\cite{Ruegg_thermo_ladder} (dots) and the T-DMRG calculations (solid lines) in the spin liquid regime at (a.1) $h^z=0~{\rm T}$, (a.2) $h^z=3~{\rm T}$, (a.3) $h^z=5~{\rm T}$ and in the gapless LL regime at (b.1) $h^z=9~{\rm T}$, (b.2) $h^z=10~{\rm T}$, (b.3) $h^z=11~{\rm T}$. See also Fig.~\ref{fig:cvsT} for a comparison of the theoretical computations and Fig.~\ref{fig:chalspe_BPCB} for a false color picture of $c/T$ in a full range of temperature and magnetic field. [Data taken from Ref.~\cite{Bouillot_ladder_statics_dynamics}]\label{fig:chalspe_comp}}
\end{figure}

\begin{figure}[!h]
\begin{center}
\includegraphics[width=0.7\linewidth]{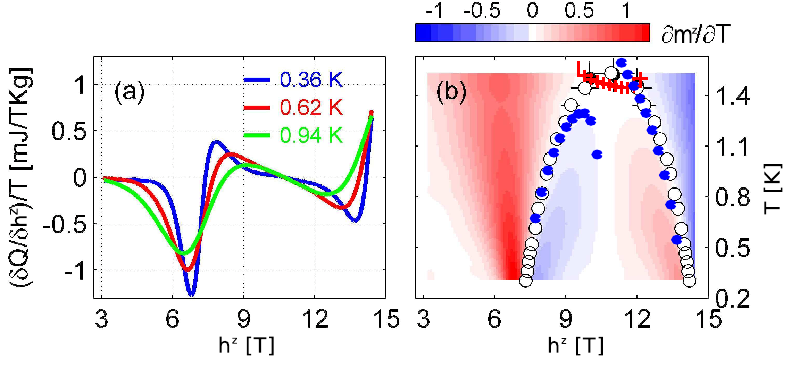}
\end{center}
\caption{Magnetocaloric effect measured on BPCB. (a) Heat-flow $\delta Q$ to and from the sample as a function of magnetic field divided by temperature, ($\delta Q/\delta h^z)/T=-(\partial M^z/\partial T)|_{h^z}$. (b) Crossover temperature $T_{LL}$ of the LL to the quantum critical regime versus the applied magnetic field $h^z$. White circles (black circles) denote the phase boundary derived from the $(\partial m^z/\partial T)|_{h^z}=0$ criterium computed from BPCB measurements shown in (a) ($(\partial c/\partial T)|_{h^z}=0$ criterium computed from BPCB measurements shown in Figs.~\ref{fig:chalspe_comp} and~\ref{fig:chalspe_BPCB}). As a comparison the blue circles and red crosses show the T-DMRG computations of this crossover presented in Fig.~\ref{fig:Tmagnetization} for both criteria respectively. [Taken from Ref.~\cite{Ruegg_thermo_ladder}]\label{fig:comp_crossover}}
\end{figure}

A more selective test to distinguish between various models is provided by
the specific heat. This is due to the fact that the specific heat contains information on high energy excitations which are characteristic for the underlying model.  As shown in Fig.~\ref{fig:chalspe_comp} the
experimental data from~\cite{Ruegg_thermo_ladder} are remarkably described, up to an accuracy of
a few percent, by a simple Heisenberg ladder Hamiltonian with the parameters extracted from the magnetization. In particular, not only the low temperature behavior and the crossover from the LL regime to the quantum critical regime (Fig.~\ref{fig:comp_crossover}.b) are covered by the ladder description, but also the higher maxima. This indicates that the ladder
Hamiltonian is an adequate description of the compound and insure that no other large magnetic exchanges are forgotten. The small discrepancies between the specific heat data and the calculation
which is essentially exact can have various sources. First of all,
the substraction of the non-magnetic term in the experimental data
can account for some of the deviations.
Furthermore the interladder coupling and the coupling anisotropies can induce slight changes in the behavior of the specific heat.

The quality of the determination of the model and its intraladder parameters becomes more evident in the
comparison of the NMR data for the relaxation rate $T_1^{-1}$ with the theoretical results of the Luttinger liquid theory as
shown in Fig.~\ref{fig:T1}. Only \emph{one}
adjustable parameter is left, namely the hyperfine coupling constant (see Sec.~\ref{sec:relaxationtime}).
This parameter allows one for a global rescaling of the
theoretical curve, but not for a change of its shape which is totally determined by the LL parameters (Fig.~\ref{fig:LLparameter}).
The agreement between the theory and the experimental data is very good over
the whole range of the magnetic field and only small deviations can be
seen. As discussed further, two other correlations included in the LL description of the low temperature 3D order and its critical temperature are tested with the same LL parameters. This compound thus allows us to \emph{quantitatively} test the Luttinger liquid
universality class. Even though the Luttinger liquid description is restricted to low energies,
in BPCB its range of validity is rather large. Indeed
at high energy, its breakdown is approximately signaled by the first peak of the specific
heat~\cite{Ruegg_thermo_ladder} (see Sec.~\ref{sec:specificheat}) or the cancellation of the magnetocaloric effect (equivalent to the first extrema of the magnetization versus the temperature discussed in Sec.~\ref{sec:finiteTmagnetization}). Here the experimentally determined crossover is located 
about $T \sim 1.5~{\rm K}$ at midpoint between $h_{c1}$ and $h_{c2}$, and agrees totally with its numerically computed value   (see Fig.~\ref{fig:comp_crossover}.b, respectively).
Given the low ordering temperature which has a maximum at about
$T \sim 100~{\rm mK}$ this leaves a rather large Luttinger regime for
this compound.

Finally, deviations from the simple ladder Hamiltonian can be present.
Small anisotropy of the couplings
can exist and indeed are necessary to interpret recent ESR experiments ~\cite{cizmar_esr_bpcb}. Other
terms such as longer range exchanges or Dzyaloshinskii-Moryia
(DM) terms might occur along the legs even if the latter is forbidden
by symmetry along the dominant rung coupling. Clearly all these
deviations from the Heisenberg model cannot be larger than a
few percents. They will not lead to any sizeable deviation
for the Luttinger parameters (Fig.~\ref{fig:LLparameter}) in the one dimensional regime.
Close to the critical points they can, however, play a more important role. It
would thus be interesting in subsequent studies to refine the model
to take such deviations into account.

Taking now the coupling between ladders into account, one can induce a transition to a three-dimensional ordered phase.
The transition temperature is shown in
Fig.~\ref{fig:criticaltemperature}. Experimentally it is determined by
NMR~\cite{Klanjsek_NMR_3Dladder} and neutron diffraction
measurements ~\cite{Thielemann_ND_3Dladder}.
Theoretically the ladders are described by Luttinger liquid theory and their interladder coupling is treated in a mean field approximation (Secs.~\ref{sec:luttinger_liquid} and~\ref{sec:mean-field}).
As shown in Fig.~\ref{fig:criticaltemperature}, the Luttinger liquid theory provides a remarkable description
of the transition to the transverse antiferromagnetic order at low
temperatures. The shape of $T_c(h^z)$ is almost perfectly reproduced, in agreement with both the
NMR~\cite{Klanjsek_NMR_3Dladder} and the ND data~\cite{Thielemann_ND_3Dladder}.
The comparison with the experiments determines the interladder coupling $J'$, the only adjustable parameter.
The simple mean field approximation would
give a value of $J' \sim 20~{\rm mK}$. As discussed in Sec.~\ref{sec:transitiontemperature}, mean field tends to
underestimate the coupling and it should be corrected by
an essentially field independent factor. Taking this into account we obtain a coupling
of the order of $J' = 27~{\rm mK}$.

\begin{figure}[!h]
\begin{center}
\includegraphics[width=0.5\linewidth]{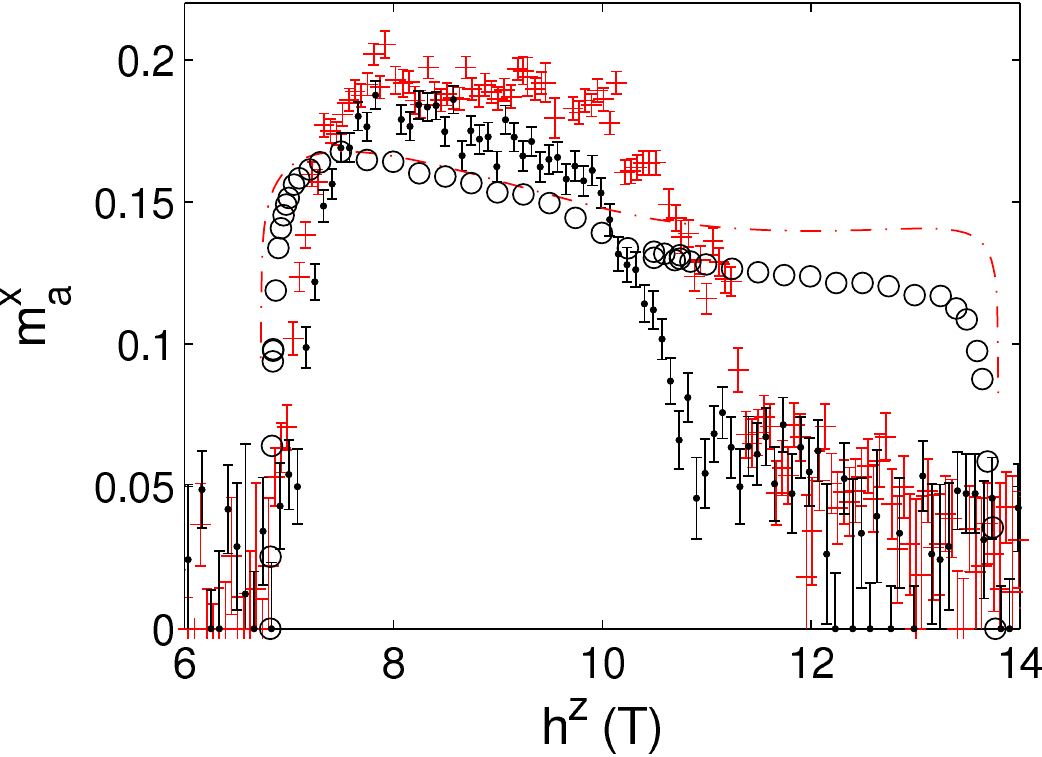}
\end{center}
\caption{Magnetic field dependence of the transverse staggered
  magnetization per spin, $m^x_a(h^z)$, in the 3D-ordered phase. Comparison between NMR measurements (black circles) done at $T=40~\mathrm{mK}$ from Ref.~\cite{Klanjsek_NMR_3Dladder} and
  scaled to the theoretical results for $J' =27~{\rm mK}$ (dash-dotted red line),  neutron
  diffraction measurements on an absolute scale from Ref.~\cite{Thielemann_ND_3Dladder} at
  $T=54~\mathrm{mK}$ ($T=75~\mathrm{mK}$) (red crosses (black dots)).  Recent neutron diffraction measurements as a function of temperature suggest that the data of Ref.~\cite{Thielemann_ND_3Dladder} was taken at temperatures approximately $10~{\rm mK}$ higher than the nominal indicated temperature. See also Fig~\ref{fig:orderparameter} for a comparison of the theoretical computations. [Taken from Ref.~\cite{Bouillot_ladder_statics_dynamics}]\label{fig:orderparameter_exp}}
\end{figure}

The order parameter in the antiferromagnetic phase can also be observed by
experiments. As discussed in Sec.~\ref{sec:transitiontemperature}, it shows a very interesting shape. At a pure experimental level
neutron diffraction and
NMR have some discrepancies as shown in
Fig.~\ref{fig:orderparameter_exp}. These discrepancies can be attributed to the different temperatures at which the data have been taken,
and a probable underestimation of the temperature in the neutron diffraction experiments~\cite{Thielemann_ND_3Dladder}. Indeed
the order parameter close to the critical magnetic field $h_{c2}$ is very sensitive to temperature, since the transition temperature drops steeply in this regime. Note that although
the NMR allows clearly for a more precise measurement of the
transverse staggered magnetization it cannot give its absolute
value. Thus the amplitude of the order parameter is fixed from the
neutron diffraction measurement. Even though a good agreement between the theoretical results and the experimental results is obtained, several questions concerning the deviations remain to be addressed.

First, the theoretical curve does not fully follow the shape of the experimental data. Particularly at high fields the experimental data shows a stronger decrease. A simple
explanation for this effect most likely comes from the fact that the
calculation is performed at zero temperature, while the
measurement is done at $40~{\rm mK}$. This is not a negligible temperature with
respect to $T_c$, in particular at magnetic fields close to $h_{c2}$. Extrapolation of the experimental data
to zero temperature~\cite{Klanjsek_NMR_3Dladder} improves the
agreement. Nevertheless, for a detailed comparison either lower temperature measurements or a calculation of the transverse staggered magnetization at finite temperature would be required. From a theoretical point of view, including the 3D coupling in the model~\eqref{equ:coupledladdershamiltonian} make the finite temperature computations difficult to perform (analytically and numerically). Such an investigation thus clearly require the development of more powerful techniques.

The second question comes from the amplitude of the
staggered magnetization. Indeed the experimental data seem to be slightly
above the theoretical curve, even if one uses the value $J' =
27~{\rm mK}$ for the interladder coupling. This is surprising since
one would expect that going beyond the mean field approximation
could only reduce the order parameter. Naively, one would thus need
a larger coupling, perhaps of the order of $J' \sim
60-80~{\rm mK}$ to explain the amplitude of the order parameter. This is a much
larger value than the one extracted from the comparison of $T_c$. How to reconcile
these two values remains open. The various anisotropies and additional
small perturbations in the ladder Hamiltonian could
resolve part of this discrepancy. However, it seems unlikely that they result in a correction of $J'$ by a factor of about 2-3.
Another origin might be the presence of some level of frustration present in the
interladder coupling. Clearly more experimental and theoretical
studies are needed on that point.



\chapter{Dynamical correlations of a spin ladder}\label{sec:dynamicalcorrelation}

In this chapter, we focus on the zero temperature spectral functions of a single spin-$1/2$ ladder computed with t-DMRG (see Sec.~\ref{sec:timeDMRG}). As we will see in the following these dynamical quantities
are direct probes of the excitations in the system and are experimentally accessible through INS measurements.

After an introduction of the relevant spectral functions related to spin-$1/2$ ladders, we discuss the possible rung excitations created by the spin operators. Next, we analyze in detail the computed spectra for the parameters of the compound BPCB (see Sec.~\ref{sec:bpcb}) separately in the gapped spin liquid and the gapless regime. In addition, these spectra are compared to analytical results when such results exist. In particular, we check the agreement with the LL description at low energy and use a strong coupling expansion (appendix~\ref{sec:tjmodelmapping}) to qualitatively characterize the different excitations occurring. Then we generalize our analysis  for different coupling ratios $\gamma$ from the weak ($\gamma\rightarrow \infty$) to strong
coupling ($\gamma\approx0$), and briefly discuss the influence of the weak interladder coupling on the excitations of the system. Finally, in Sec.~\ref{sec:INS}, we compare the low energy part of the computed spectra to the INS measurements on BPCB and provide a quantitative prediction for the high energy part of the INS spectra. Furthermore we give a short overview of the ND experiments for the measurement of static quantities.

\section{Zero temperature spectral functions}\label{par:zerocorrelations}

As discussed in Sec.~\ref{sec:luttingerliquidcorr}, in a ladder system different types of correlations are
possible. We focus here on the quantities
\begin{equation}\label{equ:correlation1}
S^{\alpha\beta}_{q_y}(q,\omega)=\sum_l\int_{-\infty}^\infty dt\langle S^\alpha_{l,q_y}(t)S^\beta_{0,q_y}\rangle e^{i(\omega t-ql)}
\end{equation}
where $S^\alpha_{l,q_y}=S^\alpha_{l,1}\pm S^\alpha_{l,2}$ are
the symmetric ($+$) and antisymmetric ($-$) operators with rung momentum\footnote{The rung momentum $q_y$ is a good quantum number.}
$q_y=0,\pi$ and parity in the rung direction $P=+1,-1$, respectively. The type of the correlation is denoted by $\alpha,\beta=z,+,-$. The time evolution $S^\alpha_{l,q_y}(t)=e^{iHt}S^\alpha_{l,q_y}e^{-iHt}$ is with respect to the Hamiltonian $H$~\eqref{equ:spinladderhamiltonian} of a single spin ladder.

These correlations are lattice versions of the spectral functions introduced in Sec.~\ref{sec:luttingerliquidcorr} with $q_y$ and $q$ the momenta in the rung and along the leg lying in the first Brillouin zone. The latter is given in reciprocal lattice units $a^{-1}$. 

Using the
reflection and translation invariance of an infinite size system
($L\rightarrow\infty$), we can
rewrite the considered correlations~\eqref{equ:correlation1} with ${S^\alpha}^\dagger=S^\beta$ in a spectral decomposition (at
zero temperature), i.e.~
\begin{equation}\label{equ:correlation2}
S^{\alpha\beta}_{q_y}(q,\omega)=\frac{2\pi}{L}\sum_\lambda|\langle\lambda|S_{q_y}^\beta(q)|0\rangle|^2\delta(\omega+E_0-E_\lambda)
\end{equation}
where $|0\rangle$ denotes the ground state of $H$ with energy
$E_0$, $S_{q_y}^\beta(q)=\sum_le^{-iql}S^\beta_{l,q_y}$,
$\sum_\lambda$ the sum over all eigenstates
$|\lambda\rangle$ of $H$, and $E_\lambda$ their eigenenergy. The form of
Eq.~\eqref{equ:correlation2} clearly shows that
$S^{\alpha\beta}_{q_y}(q,\omega)$ is non-zero if the operator
$S^\beta_{q_y}$ can create an excitation $|\lambda\rangle$ of
energy $E_0+\omega$ and momentum $q$ from the ground state. The
correlations $S_{q_y}^{\alpha\beta}$ are then direct probes of
the excitations $|\lambda\rangle$ in the system.

Since the experimentally relevant case (compound BPCB) corresponds to a relatively
strong coupling situation ($\gamma\ll 1$, Eq. \ref{equ:couplingratio}), we
use the decoupled bond limit introduced in
Sec.~\ref{sec:spinchainmap} to represent the expected
excitations on a single rung $|t^+\rangle$, $|t^0\rangle$, $|t^-\rangle$ or
$|s\rangle$. In table \ref{tab:excitations}, we summarize
the rung excitations created by the operators
$S^\beta_{q_y}$ and their properties. For example the operator $S^z_\pi$ applied on the singlet $\ket{s}$ excites the triplet $\ket{t^0}$. Typically the rung
parity $P$ is changed by applying an operator with rung momentum
$q_y=\pi$ and the $z$-magnetization is modified by $\Delta
M^z=\pm1$ by applying the operators $S^\pm_{q_y}$, respectively. 
\begin{table}[h!]
\begin{center}
$$
\begin{array}{l||cccccc}
&S_0^z&S_\pi^z&S_0^+&S_\pi^+&S_0^-&S_\pi^-\\
\hline
\hline
 |s\rangle&0&|t^0\rangle&0&-\sqrt{2}|t^+\rangle&0&\sqrt{2}|t^-\rangle\\
 |t^+\rangle&|t^+\rangle&0&0&0&\sqrt{2}|t^0\rangle&-\sqrt{2}|s\rangle\\
 |t^0\rangle&0&|s\rangle&\sqrt{2}|t^+\rangle&0&\sqrt{2}|t^-\rangle&0\\
 |t^-\rangle&-|t^-\rangle&0&\sqrt{2}|t^0\rangle&\sqrt{2}|s\rangle&0&0\\
\hline
P&+1&-1&+1&-1&+1&-1\\
\Delta M^z&0&0&+1&+1&-1&-1\\
\end{array}
$$
\end{center}
\caption{Rung excitations created by the symmetric and
antisymmetric operators in the decoupled bond limit. The elements of the first column represents the initial rung states on which the rung operators written in the first line apply. The effect of of these operators on the parity $P$ and the magnetization $M^z$ of the system is also summarized in the two last lines.\label{tab:excitations}}
\end{table}

\section{Excitations in the spin liquid}\label{sec:spinliquidexcitations}

\begin{figure}[h!]
\begin{center}
\includegraphics[width=0.7\linewidth]{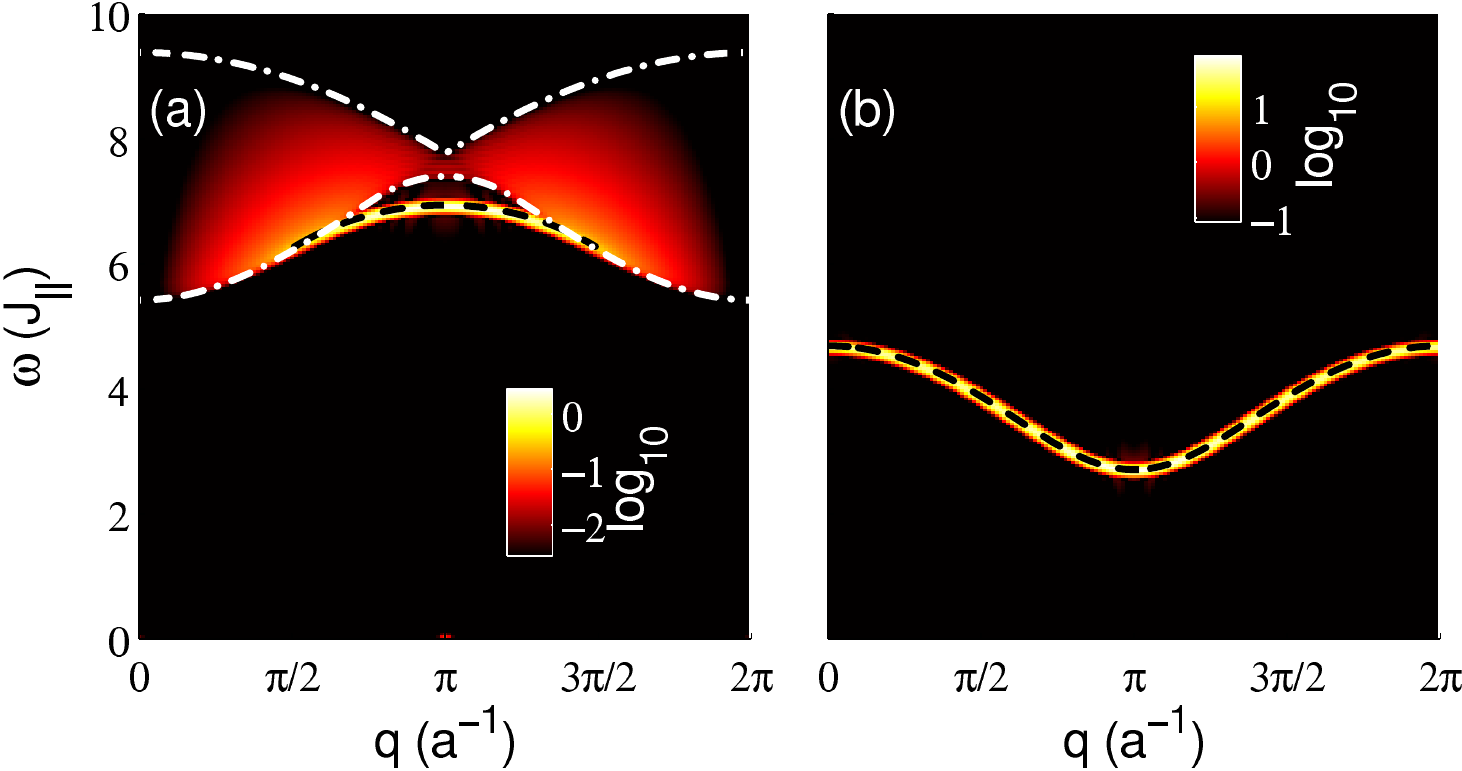}
\end{center}
\caption{Momentum-energy dependent correlation functions $S_{q_y}^{\alpha\alpha}(q,\omega)$  with $\alpha=x,y,z$ at $h^z=0$. Numerical results are shown with the color coding. (a) Symmetric part $S_0^{\alpha\alpha}(q,\omega)$. The dashed (black) line marks the $(q,\omega)$ position of the two-triplet bound state with dispersion relation $\omega_{tB}(q)$, Eq.~\eqref{equ:tripletboundstate}, in its existence interval \eqref{equ:qlimitbound}.
The dash-dotted (white) lines correspond to the boundaries of the two-triplet continuum.
(b) Antisymmetric part $S_\pi^{\alpha\alpha}(q,\omega)$.
The dashed (black) line corresponds to the predicted dispersion relation $\omega_t(q)$ of a single triplet excitation, Eq.~\eqref{equ:singletripletenergy}. [Taken from Ref.~\cite{Bouillot_ladder_statics_dynamics}]\label{fig:spectrummz0}}
\end{figure}

Using the decoupled bond limit in the spin liquid phase, the excitations in the system
can be pictured as the excitation of rung singlets to rung
triplets. At zero magnetic field $h^z=0$, the system is spin
rotational symmetric and the different triplet excitations have
the same energy $\sim J_\perp$. It has been seen
previously that in the spin liquid both single triplet excitations
and two-triplet excitations play an important role ~\cite{barnes_ladder,reigrotzki_ladder_field,Zheng_bound_state_ladder,sushkov_ladder_boundstates,knetter_ladder}.
We discuss these excitations in the following focusing on the
ones that can be created by the symmetric
$S^{\alpha\alpha}_0=2S^{\pm\mp}_0$ and the antisymmetric
$S^{\alpha\alpha}_\pi=2S^{\pm\mp}_\pi$ correlations (see
Fig.~\ref{fig:spectrummz0}) for the
BPCB parameters~\eqref{equ:couplings}, where $\alpha=x,y,z$ . Note that these correlations are
independent of the direction due to the spin
rotation symmetry.

\subsection{Single triplet excitation}\label{sec:singletripletspinliquid}

At zero magnetic field $h^z=0$, the system is in a global spin singlet state~($S=0$) \cite{Auerbach_book_magnetism}. The $q_y=\pi$ correlation
couples this ground state to states with an odd number of triplet excitations with rung parity $P=-1$ and
total spin $S=1,\ M^z=\pm1,0$ (see table~\ref{tab:excitations}). Nevertheless, only single triplet excitations are numerically resolved as shown in Fig.~\ref{fig:spectrummz0}.b.
Their spectral weight is concentrated in a very sharp peak whose dispersion relation, $\omega_t(q)$, can be approximated using a \emph{strong coupling expansion} in
$\gamma$ similar to the one described in appendix~\ref{sec:tjmodelmapping}. Up to third\footnote{
Note that an expression up to seventh order in $\gamma$ has been determined in
Ref.~\cite{oitmaa_ladderserieexp}.} order~\cite{reigrotzki_ladder_field} the dispersion is given by 
\begin{equation}\label{equ:singletripletenergy}
 \frac{\omega_t(q)}{J_\perp}=1+\gamma\cos q +\frac{\gamma^2}{4}(3-\cos 2q)
 +\frac{\gamma^3}{8}(3-2\cos q-2\cos 2q+\cos3q)+\mathcal{O}(\gamma^4).
\end{equation}
At the first order, it is simply a cosine dispersion~\cite{barnes_ladder}, i.e. $\omega_t(q)/J_\perp \approx
1+\gamma \cos q$. The lowest energy single triplet excitation has a momentum $q=\pi$ and corresponds to the spin liquid gap\footnote{Note that the spin gap has been determined up to 13th order in $\gamma$ ~\cite{Weihong_spin_ladder}.} or the first critical field $h_{c1}$ (see Secs.~\ref{sec:spinchainmap} and~\ref{sec:criticalfields}). In Fig.~\ref{fig:spectrummz0}.b we compare the numerical results
for the BPCB parameters~\eqref{equ:couplings} to the expression~\eqref{equ:singletripletenergy}. The strong coupling expansion describes very well the position of the excitations found numerically. The comparison with the known solutions serves as a check of the quality of our numerical results.

\subsection{Two-triplet excitations}\label{sec:zerofieldtwotriplets}

The structure of the $q_y=0$ correlation
is more complex (Fig.~\ref{fig:spectrummz0}.a). Due to the rung parity $P=1$ of the operators~$S^\alpha_0$, the
excitations correspond to an even number of
triplet excitations with total spin $S=1,\ M^z=\pm1,0$ excited from rung triplets already present in the ground state. We focus here on the two-triplet excitations that can be resolved numerically (Fig.~\ref{fig:spectrummz0}.a). These can be divided into a broad continuum and a very sharp triplet ($S=1$) bound state of
a pair of rung triplets. Since these excitations stem from the
coupling to triplets already present in the ground state (Fig.~\ref{fig:tripletdensity}), their
amplitude for the considered BPCB parameters~\eqref{equ:couplings} is
considerably smaller than the weight of the single triplet
excitations~\cite{knetter_ladder}.

The dispersion relation of the bound states, $\omega_{tB}(q)$, has been calculated
using a linked cluster series
expansion~\cite{Zheng_bound_state_ladder} up to third order in $\gamma$
\begin{multline}\label{equ:tripletboundstate}
\frac{\omega_{tB}(q)}{J_\perp}=2+\frac{\gamma}{2}(-3-2\cos q)+\frac{\gamma^2}{8}(11-2\cos q-4\cos2q)\\
+\frac{\gamma^3}{16}(17+9\cos q-8\cos 2q-5\cos 3q)+\mathcal{O}(\gamma^4).
\end{multline}
The first terms of
the expansion have an inverse cosine form and the bound state
only exists in an interval
\begin{equation}\label{equ:qlimitbound}
q_c< q<2\pi-q_c\quad\text{with}\quad q_c=\frac{2\pi}{3}-\frac{5\gamma}{2\sqrt{3}}-\frac{109\gamma^2}{48\sqrt{3}}+\mathcal{O}(\gamma^3)
\end{equation}
around $q=\pi$ (cf.~Ref.~\cite{Zheng_bound_state_ladder,sushkov_ladder_boundstates}).
The numerical
results for the BPCB parameters~\eqref{equ:couplings} agree very well with
the analytic form of the dispersion (Fig.~\ref{fig:spectrummz0}.a). The upper and lower limits of the continuum
can be determined by considering the boundary of the two non-interacting triplet continuum consisting of two excitations with the single triplet dispersion~\eqref{equ:singletripletenergy}. They agree very well with the results found numerically (Fig.~\ref{fig:spectrummz0}.a).

\section{Excitations in the gapless regime}\label{sec:gapless_dynamics}

A small applied magnetic field ($h^z<h_{c1}$), at first order,
 smoothly translates the excitations shown in
Fig.~\ref{fig:spectrummz0} by an energy $-h^z M^z$, with $M^z$ the magnetization of each excitation, due to the
Zeeman effect. However, if the magnetic field exceeds
$h_{c1}$, the system enters into the gapless regime and the structure of the excitations spectrum changes drastically. A
continuum of excitations at low energy arises. For small values of $\gamma$
most features of this low energy continuum are qualitatively
well described by considering the lowest two modes of the
ladder only. Beside the low energy continuum, a complex
structure of high energy excitations exist. Contrarily to the
low energy sector, this structure crucially depends on
the high energy triplet modes. In the following, we give a simple picture for
the excitations starting from the decoupled bond limit.
\begin{figure}
\begin{center}
\includegraphics[width=0.7\linewidth]{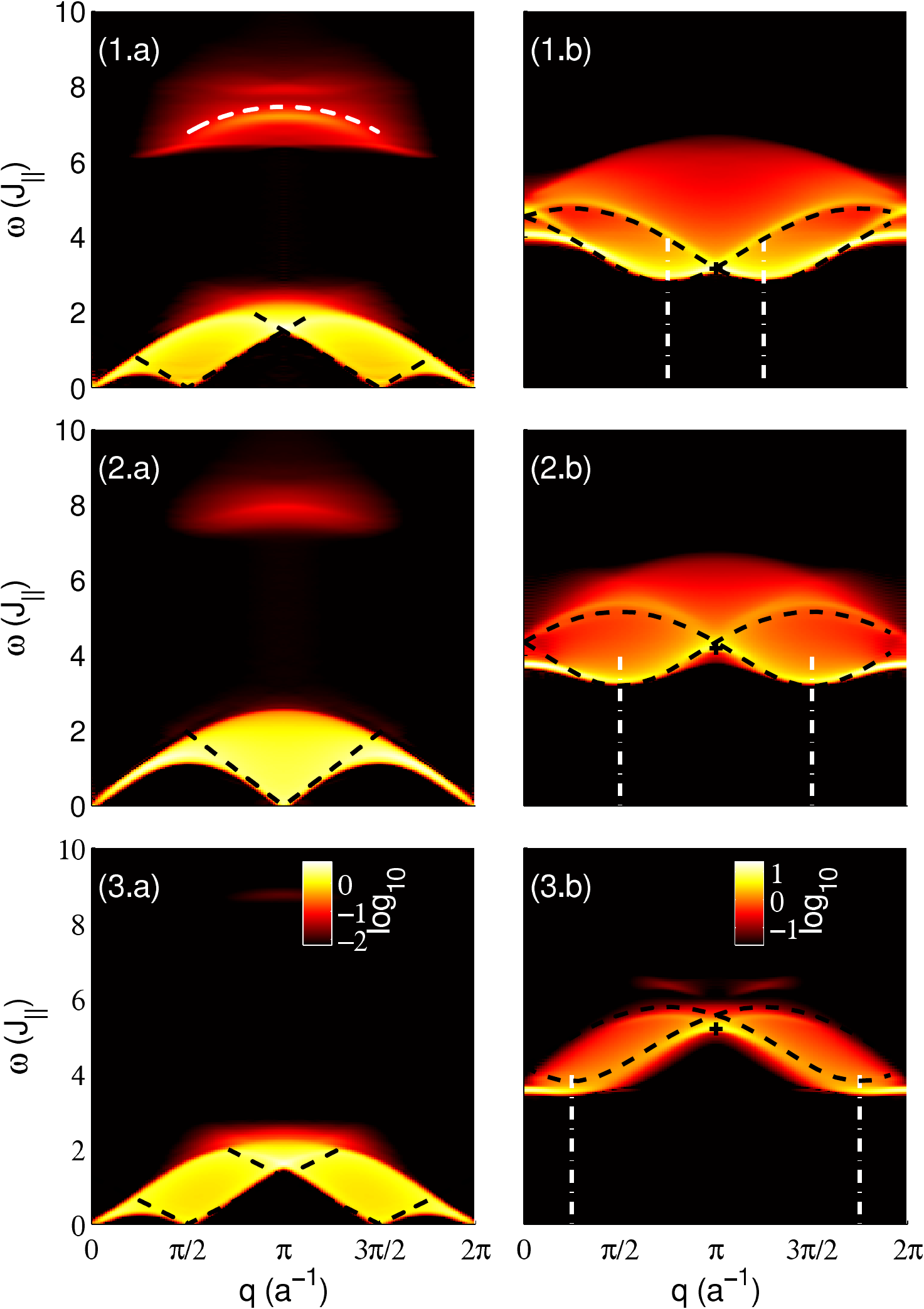}
\end{center}
\caption{Momentum-energy dependent $zz$-correlation function (1)~at $m^z=0.25$ ($h^z=3.153~J_\parallel$), (2)~at $m^z=0.5$ ($h^z=4.194~J_\parallel$), and (3)~at $m^z=0.75$ ($h^z=5.192~J_\parallel$). (a)~Symmetric part $S_0^{zz}(q,\omega)$ without Bragg peak at $q=0$. The dashed black lines correspond to the location of the slow divergences at the lower edge of the continuum predicted by the LL theory. The dashed white curve corresponds to the predicted two-triplet bound state location. (b)~Antisymmetric part $S_\pi^{zz}(q,\omega)$. The dashed black lines correspond to the position of the high energy divergences or cusps predicted by the approximate mapping onto the t-J model. The vertical white dash-dotted lines mark the momenta of the minimum energy of the high energy continuum and the black cross is the energy of its lower edge~\cite{furusaki_correlations_ladder} at $q=\pi$. [Taken from Ref.~\cite{Bouillot_ladder_statics_dynamics}]\label{fig:zzcorrelationmz}}
\end{figure}

\begin{figure}
\begin{center}
\includegraphics[width=0.7\linewidth]{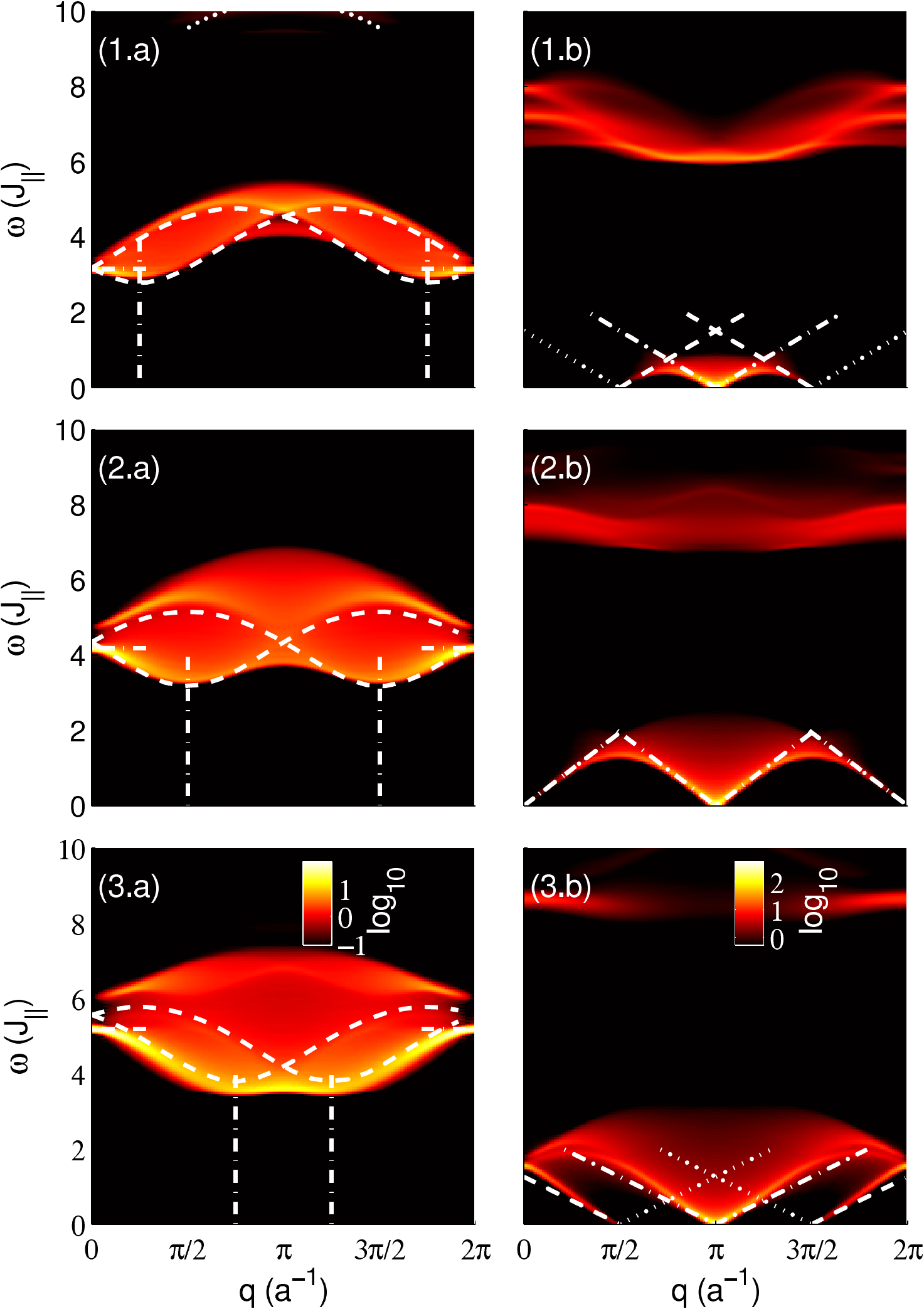}
\end{center}
\caption{Momentum-energy dependent $+-$-correlation function (1)~at $m^z=0.25$ ($h^z=3.153~J_\parallel$), (2)~at $m^z=0.5$ ($h^z=4.194~J_\parallel$), and (3)~at $m^z=0.75$ ($h^z=5.192~J_\parallel$). (a)~Symmetric part $S_0^{+-}(q,\omega)$. The vertical dash-dotted white lines mark the momenta of the minimum energy of the high energy continuum and the horizontal ones the frequency of its lower edge~\cite{furusaki_correlations_ladder} at $q=0,2\pi$. 
The dashed white lines correspond to the position of the high energy divergences or cusps predicted by the approximate mapping onto the t-J model.
The dotted white curve corresponds to the predicted two-triplet bound state location at $\omega\approx3h_z$ which is hardly visible.
(b)~Antisymmetric part $S_\pi^{+-}(q,\omega)$. The dashed and dash-dotted (dotted) white lines correspond to the location of the strong divergences (cusps) at the lower edge of the continuum predicted by the LL theory. [Taken from Ref.~\cite{Bouillot_ladder_statics_dynamics}]
\label{fig:pmcorrelationmz}}
\end{figure}

\begin{figure}
\begin{center}
\includegraphics[width=0.7\linewidth]{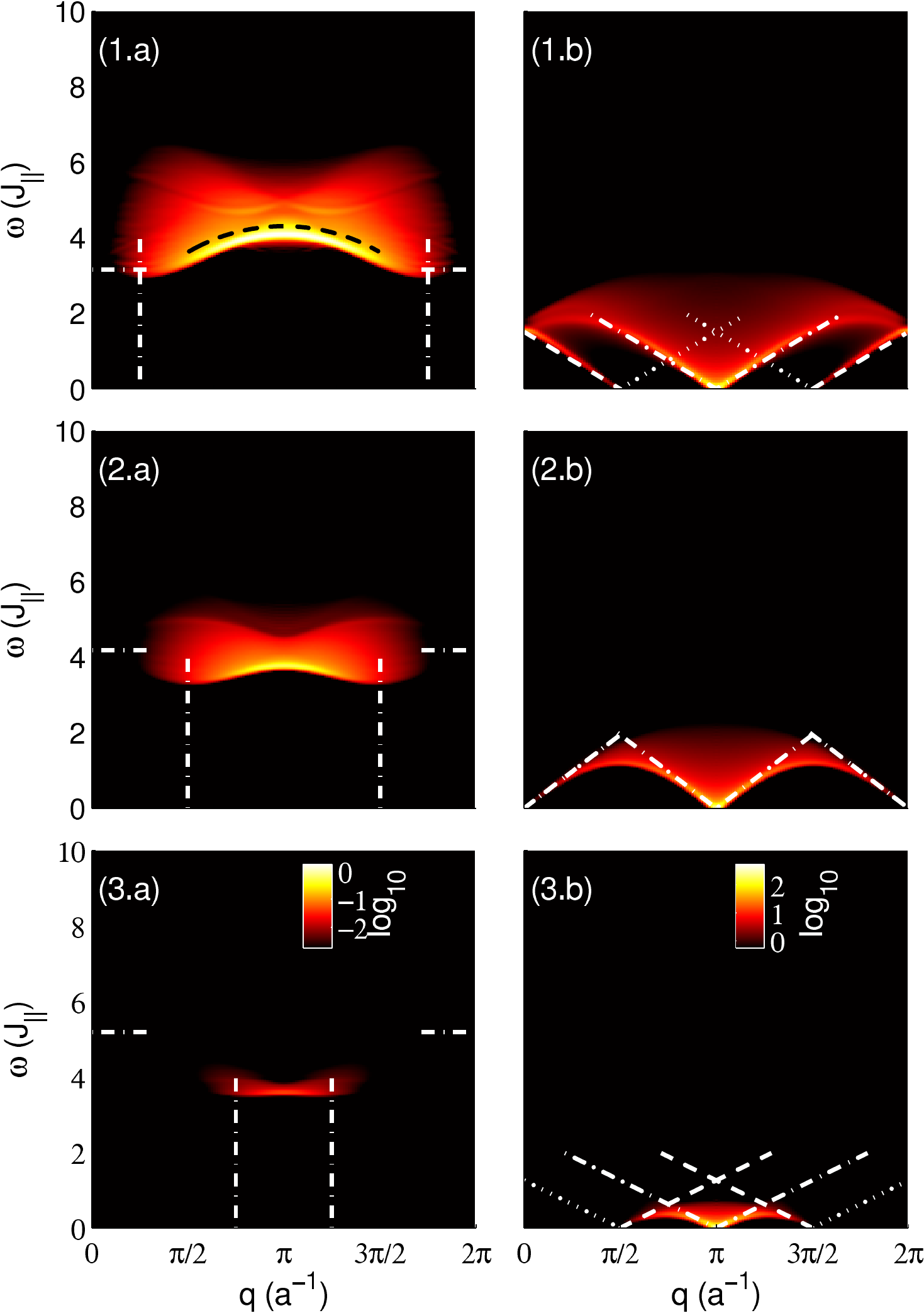}
\end{center}
\caption{Momentum-energy dependent $-+$-correlation function (1)~at $m^z=0.25$ ($h^z=3.153~J_\parallel$), (2)~at $m^z=0.5$ ($h^z=4.194~J_\parallel$), and (3)~at $m^z=0.75$ ($h^z=5.192~J_\parallel$). (a)~Symmetric part $S_0^{-+}(q,\omega)$. The vertical dash-dotted white lines correspond to the momenta at which the minimum energy of the high energy continuum occurs and the horizontal line to the frequency of its lower edge~\cite{furusaki_correlations_ladder} at $q=0,2\pi$. The dashed black curve corresponds to the predicted two-triplet bound state location. (b)~Antisymmetric part $S_\pi^{-+}(q,\omega)$. The dashed and dash-dotted (dotted) white lines correspond to the location of the strong divergences (cusps) at the lower edge of the continuum predicted by the LL theory. [Taken from Ref.~\cite{Bouillot_ladder_statics_dynamics}]\label{fig:mpcorrelationmz}}
\end{figure}

\subsection{Characterization of the excitations in the decoupled bond limit}\label{sec:excitationcharacterization}
The evolution of the spectra for the BPCB parameters with
increasing magnetic field are presented in
Fig.~\ref{fig:zzcorrelationmz} for $S^{zz}_{q_y}$, in
Fig.~\ref{fig:pmcorrelationmz} for $S^{+-}_{q_y}$, and in
Fig.~\ref{fig:mpcorrelationmz} for $S^{-+}_{q_y}$.
Three different classes of excitations occur:
\begin{itemize}
\item[(i)] a continuum of excitations at low energy for $S^{zz}_0$ and $S^{\pm\mp}_\pi$
\item[(ii)] single triplet excitations at higher energy with a clear substructure for $S^{zz}_\pi$, $S^{+-}_0$, and $S^{+-}_\pi$
\item[(iii)] excitations at higher energy for $S^{zz}_0$ and $S^{+-}_0$ and $S^{-+}_0$ stemming from two-triplet excitations which have their main weight around $q\approx \pi$.
\end{itemize}
In the following we summarize some of the characteristic
features of these excitations, before we study them in more
detail in Secs.~\ref{sec:lowenergyexcitations} and~\ref{sec:highenergyexcitation}.

\begin{itemize}
\item[(i)]  The continuum at low energy which does not exist in the
spin liquid is a characteristic signature of the gapless
regime. It stems from excitations within the low energy band which
corresponds to the $\ket{s}$ and $\ket{t_+}$ states in the decoupled bond limit (cf.~Fig.~\ref{fig:phasediagram}.a and table \ref{tab:excitations}):
\begin{itemize}
\item[$S^{zz}_0$ :] excitations within the triplet $\ket{t_+}$ mode
\item[$S^{\mp\pm}_\pi$ :] excitations between the singlet $\ket{s}$ and the triplet $\ket{t_+}$ mode.
\end{itemize}
This continuum is smoothly connected to the spin liquid spectrum in the case of $S^{-+}_\pi$. It originates from the single triplet $\ket{t^+}$ branch (Sec.~\ref{sec:singletripletspinliquid}) when the latter reaches the ground state energy due to the Zeeman effect. Since two modes play the main role in the description of these low energy features,
many of them can already be explained qualitatively by the spin chain mapping. The excitations in the chain have been
studied previously using a Bethe ansatz description and exact
diagonalization calculations in
Ref.~\cite{Muller_spinchain_dyncor}. More recently they
were computed in Ref.~\cite{caux_heisenbergchaindyn} to a high accuracy due
to recent progress in the Bethe ansatz method.
In particular, the boundary of the spectrum at very low energy is well described by this approach, since the LL velocity determining it is hardly influenced by the higher modes (cf.~Fig.~\ref{fig:LLparameter}).
However, a more quantitative description requires to take into account the higher modes of the system as well. In Sec.~\ref{sec:lowenergyexcitations} we compare in detail our results with the LL theory and the spin chain mapping pointing out their corresponding ranges of validity.

\item[(ii)] The single high energy triplet excitations form a continuum with a clear substructure.
In the decoupled bond limit, these excitations correspond to
\begin{itemize}
\item[$S^{zz}_\pi$ :] Single triplet excitations $|t^0\rangle$ at energy $\sim h^z$
\item[$S^{+-}_0$ :] Single triplet excitations $|t^0\rangle$ at energy $\sim h^z$
\item[$S^{+-}_\pi$ :] Single triplet excitations $|t^-\rangle$ at energy $\sim 2h^z$.
\end{itemize}
Many of the features of these continua can be understood by
mapping the problem onto a mobile hole in a spin chain, as pointed out first in Ref.~\cite{LaeuchliManep2007}.
We detail in Sec.~\ref{sec:singletripletLL} and appendix~\ref{sec:tjmodelmapping} this mapping. It opens the possibility to investigate the behavior of a single hole in a t-J like model using experiments in pure spin ladder compounds.

\item[(iii)] The high energy continuum, which has almost no weight close
to the Brillouin zone boundary ($q=0,2\pi$), is related to two-triplet excitations of the spin liquid (Sec.~\ref{sec:zerofieldtwotriplets}). They are
generated from high energy triplet components of the ground
state. Their weight therefore vanishes for $\gamma \to 0$ and the excitations correspond to
\begin{itemize}
\item[$S^{-+}_0$ :] Two-triplet excitations $\frac{1}{\sqrt{2}}(|t^0\rangle|t^+\rangle-|t^+\rangle|t^0\rangle)$ at energy $\sim h^z$
\item[$S^{zz}_0$ :] Two-triplet excitations  $\frac{1}{\sqrt{2}}(|t^+\rangle|t^-\rangle-|t^-\rangle|t^+\rangle)$ at energy $\sim2h^z$
\item[$S^{+-}_0$ :] Two-triplet excitations $\frac{1}{\sqrt{2}}(|t^0\rangle|t^-\rangle-|t^-\rangle|t^0\rangle)$ at energy $\sim 3h^z$.
\end{itemize}
\end{itemize}

\subsection{Low energy continuum}\label{sec:lowenergyexcitations}

In this section we concentrate on the low energy excitations of type (i) discussing first their support and then comparing their spectral weight to the LL prediction.

\subsubsection{Support of the low energy excitations}

\begin{figure}[h!]
\begin{center}
\includegraphics[width=0.6\linewidth]{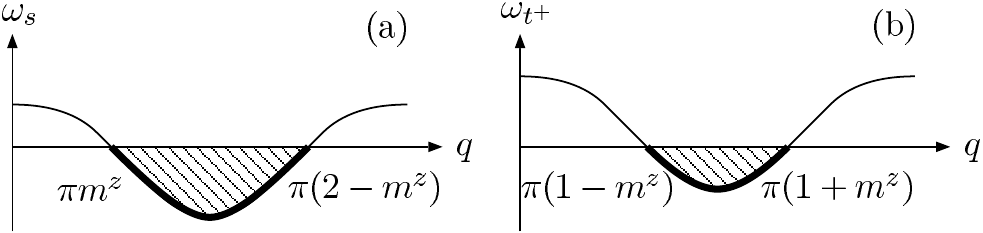}
\end{center}
\caption{Fermionic picture for the effect of the magnetic field from : Filling of (a) the singlet band $|s\rangle$, (b) the triplet band $|t^+\rangle$ in the gapless phase for a given magnetization $m^z$. [Taken from Ref.~\cite{Bouillot_ladder_statics_dynamics}]\label{fig:bandfilling}}
\end{figure}

The position of the soft modes in the low energy continuum can
be directly obtained from the bosonization representation
~\cite{chitra_spinchains_field,giamarchi_ladder_coupled,furusaki_correlations_ladder} (see Sec.~\ref{sec:luttingerliquidcorr}).
They can also be understood in a simple picture which we
outline in the following. The distribution of the rung state
population in the ground state depends on the magnetic field
$h^z$ (see Fig.~\ref{fig:tripletdensity}). Taking a fermionic point of view, the magnetic field
acts as a chemical potential that fixes the occupation of the
singlet and triplet rung states. Increasing the magnetic field
reduces the number of singlets, whereas at the same time the
number of triplets increases (see sketch in
Fig.~\ref{fig:bandfilling}). The Fermi level lies at the
momenta $q= \pi m^z, \pi(2-m^z)$ for the singlet states and at
the momenta $q= \pi(1-m^z), \pi(1+m^z)$ for the triplet states.
In this picture the soft modes correspond to excitations at the
Fermi levels. For transitions
$|t^+\rangle\leftrightarrow|t^+\rangle$ the transferred momenta
of these zero energy excitations are $q=0,2\pi m^z,2\pi(1-
m^z)$. In contrast, the interspecies transitions
$|t^+\rangle\leftrightarrow|s\rangle$ allow the transfer of
$q=\pi(1-2m^z),\pi,\pi(1+2m^z)$. Therefore, the positions of the
soft modes in the longitudinal correlation $S^{zz}_0$ which
correspond to transitions within the triplet states shift from the
boundaries of the Brillouin zone inwards towards $q=\pi$ when $m^z$ increases
(Fig.~\ref{fig:zzcorrelationmz}.a). In contrast, the positions
of the soft modes in the transverse correlations
$S^{\pm\mp}_\pi$ which induce transitions between the singlets
and the triplets move with increasing magnetic field outwards
(Figs.~\ref{fig:pmcorrelationmz}.b and~\ref{fig:mpcorrelationmz}.b).

The top of these low energy continua are reached when the
excitations reach the boundaries of the energy band. In
particular, the maximum of the higher boundary lies at the
momentum $q=\pi$ which is easily understood within the simple
picture drawn above (cf.~Fig.~\ref{fig:bandfilling}). A more
detailed description of different parts of these low energy
continua is given in Ref.~\cite{Muller_spinchain_dyncor}.

Let us compare the above findings with the predictions of the
LL theory for the dynamical correlations ~\cite{chitra_spinchains_field,furusaki_correlations_ladder,giamarchi_ladder_coupled}. Details on the LL description of the correlations
are given in Sec.~\ref{sec:luttingerliquidcorr}. The LL theory predicts
a linear momentum-frequency dependence of the lower continuum
edges with a slope given by the LL velocity $\pm
u$ (Fig.~\ref{fig:LLparameter}). The position of
the soft modes are given by the ones outlined in Sec.~\ref{sec:zeroT_correlation} (see
Fig.~\ref{fig:LLcorrschema}).  The predicted support at low
energy agrees very well with the numerical results (Fig.~\ref{fig:zzcorrelationmz}.a,~\ref{fig:pmcorrelationmz}.b, and~\ref{fig:mpcorrelationmz}.b). Of course
when one reaches energies of order $J_\parallel$ in the spectra one
cannot rely on the LL theory anymore. This is true in
particular for the upper limit of the spectra.

\subsubsection{Spectral weight of the low energy excitations}\label{sec:strongcouplingLLcomp}
Let us now focus on the distribution of the spectral
weight in the low energy continuum. In particular, we compare our numerical findings to the Luttinger liquid description. Qualitatively, the LL theory predictions for the low energy spectra are well reproduced by the DMRG computations.

The Luttinger liquid predicts typically an
algebraic behavior of the correlations at the low energy
boundaries which can be a divergence or a cusp.
\begin{itemize}
\item[$S^{zz}_0$ :] The Luttinger liquid predicts peaks at the $q=0,2\pi$ branches and a slow divergence at the lower edge of the incommensurate branches $q=2\pi m^z,2\pi(1-m^z)$ (with exponent $1-K\approx0.2\ll1$ (Fig.~\ref{fig:decay_exp}.b)). In the numerical results (Fig.~\ref{fig:zzcorrelationmz}.a) a slight increase of the weight towards the lower edge of the incommensurate branches can be seen.

\item[$S^{+-}_\pi$ :] A strong divergence at the lower edge of the $q=\pi$ branch (with exponent $1-1/4K\approx3/4\gg 0$ (Fig.~\ref{fig:decay_exp}.a)) is obtained within the Luttinger liquid description. This is in good qualitative agreement with the strong increase of the spectral weight observed in the numerical data (Fig.~\ref{fig:pmcorrelationmz}). A more interesting behavior is found close to the momenta $q=\pi(1\pm2m^z)$ in the incommensurate branches. Here a strong divergence is predicted for momenta higher (lower) than the soft mode $q=\pi(1- 2m^z)$ ($q=\pi(1+ 2m^z)$) with exponent $1-\eta_-\approx3/4\gg 0$ (Fig.~\ref{fig:decay_exp}.a). In contrast for momenta lower (higher) than the soft mode $q=\pi(1- 2m^z)$ ($q=\pi(1+ 2m^z)$) a cusp with exponent $1-\eta_+\approx-5/4\ll 0$ (Fig.~\ref{fig:decay_exp}.c) is expected. In the numerical results (Fig.~\ref{fig:pmcorrelationmz}.b) this very different behavior below and above the soft modes is evident. The divergence and cusp correspond to a large and invisible weight, respectively.

\item[$S^{-+}_\pi$ :] The same behavior as for $S^{+-}_\pi$ replacing $m^z\rightarrow -m^z$ can be observed in Fig.~\ref{fig:mpcorrelationmz}.b.
\end{itemize}

\begin{figure}[h!]
\begin{center}
\includegraphics[width=0.5\linewidth]{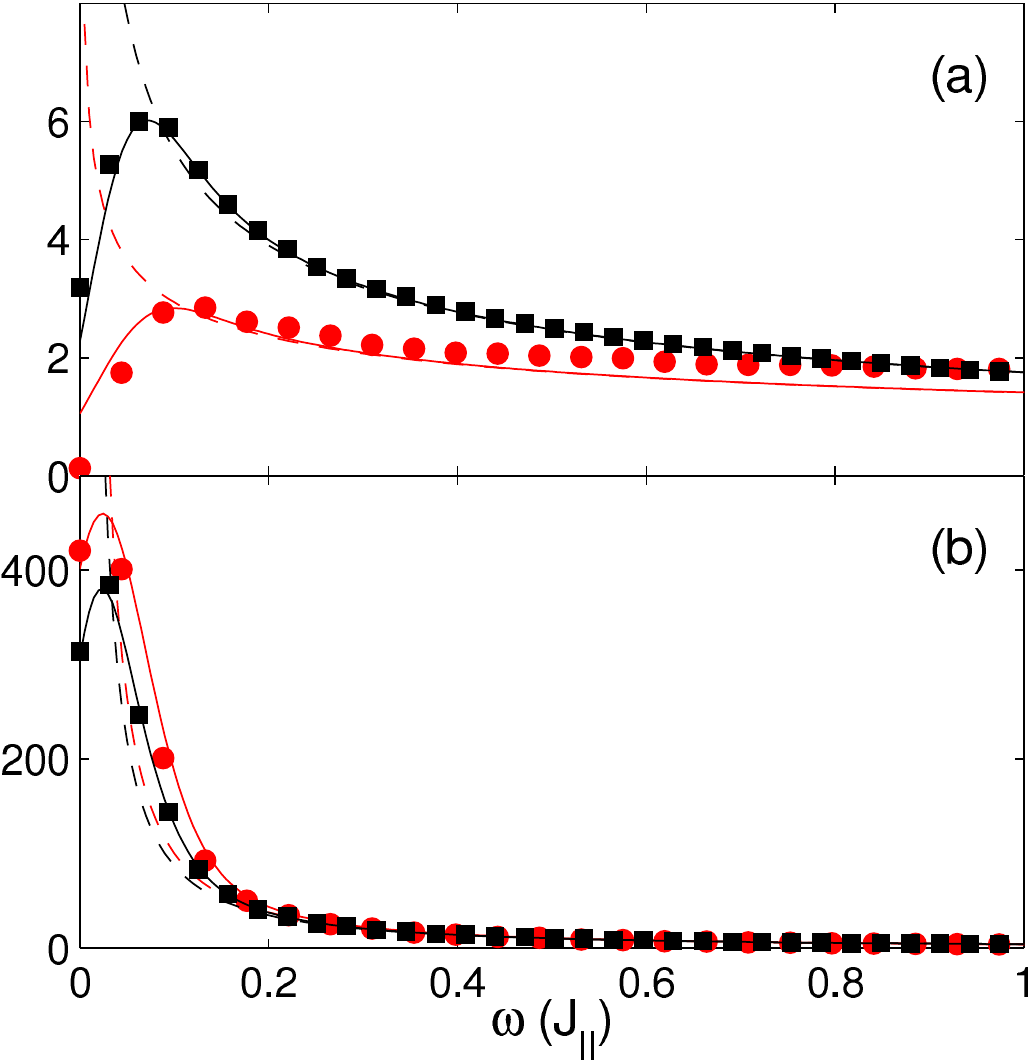}
\end{center}
\caption{Cuts at fixed momentum $q=\pi$ and magnetization $m^z=0.5$ of the low energy spectrum (a) $S^{zz}_0(q=\pi,\omega)$, and (b) $S^{-+}_\pi(q=\pi,\omega)$.  The (red) circles and the (black) squares are the numerical results for the ladder and its spin chain mapping, respectively. The dashed lines correspond to the LL predictions and the solid lines are the latter convolved with the same Gaussian filter than the numerical data. The DMRG frequency numerical
limitation is of the order of the peak broadening of width $\delta\omega\approx
0.1~J_\parallel$ (see Sec.~\ref{sec:tDMRGcorrelation}). [Taken from Ref.~\cite{Bouillot_ladder_statics_dynamics}]
\label{fig:LLcomparisons}}
\end{figure}

To compare quantitatively the predictions of the LL to the numerical results
we show in Fig.~\ref{fig:LLcomparisons} different cuts of the
correlations at fixed momentum $q=\pi$ and magnetization $m^z=0.5$ for
the ladder and the spin chain mapping.
These plots show the DMRG results, the LL
description, and the latter convolved with the
Gaussian filter. The filter had been used in the numerical data to avoid effects due to the finite time-interval simulated (see
Sec.~\ref{sec:tDMRGcorrelation}). Note that the amplitude of the LL results are inferred from the static correlation functions, such that the LL curve is fully determined and \emph{no fitting parameter} is left. Therefore, the convolved LL results can directly be compared to the numerical results. Even though the actual numerical resolution might not be good enough to resolve the behavior close to the divergences (cusps), interesting information as the arising differences between the spin chain mapping and the full ladder calculations can already be extracted. In Fig.~\ref{fig:LLcomparisons}.a, we show a cut through the correlation $S^{zz}_0(q=\pi,\omega)$.  The convolved LL and the numerical results compare very well. The difference between the
real ladder calculations and the spin chain mapping that neglect the effects of the higher triplet states
$|t^-\rangle$, $|t^0\rangle$ is obvious. From the LL description point of view, the shift of the spin chain correlation compared to the real ladder curve comes mainly from the prefactor $A_z$ and the algebraic exponent which are clearly modified by the effects of the high energy triplets (see Fig.~\ref{fig:LLparameter}).

For the transverse correlations, the LL theory predicts a strong
divergence (with an exponent $1-1/4K\approx3/4\gg0$ at the lower boundary of the
continuum branch at $q=\pi$.
A cut through the low energy continuum $S^{-+}_\pi(q=\pi,\omega)$ is shown in Fig.~\ref{fig:LLcomparisons}.b. The convolved Luttinger liquid reproduces well the numerical results.

From the comparisons in Fig.~\ref{fig:LLcomparisons} we show that the numerical computations allowing one to access the middle and high energy excitations and the analytical LL theory describing the low energy physics are complementary. Due to their large overlap, the combination of the two approaches provide a quantitative description of the correlations in a full range of energy. Nevertheless, a refined numerical investigation of the effects close to the continuum boundaries would be required in order to investigate the non LL edge singularities recently pointed out in Ref.~\cite{pustilnik_edge_exponent,Pereira_edge_singularities,Cheianov_treshold_singularities}.

\subsection{High energy excitations}\label{sec:highenergyexcitation}

Before looking in detail at the two kinds of high energy excitations presented in Sec.~\ref{sec:excitationcharacterization} we compare briefly our computed high energy spectra with the weak coupling
description ($\gamma\gg1$). 

\subsubsection{Weak coupling description of the high energy excitations}\label{sec:weakcouplinghighenergy}
In the weak coupling limit, information on the spectrum
can be extracted from the bosonization
description~\cite{chitra_spinchains_field,giamarchi_ladder_coupled,furusaki_correlations_ladder}.
In particular one expects a power law singularity at the lower
edge continuum with a minimal position at $q=\pi(1\pm m^z)$
(for $S^{zz}_\pi$), $q=\pi m^z, \pi(2-m^z)$ (for
$S^{\pm\mp}_0$) and an energy $h^z$ at momentum $q=\pi$ (for
$S^{zz}_\pi$), $q=0$ (for $S^{\pm\mp}_0$).
Except for $S^{-+}_0$ in which the spectral weight is
too low for a good visualization, our computed spectra
reproduces well the predictions for the minimal positions even though the coupling strength considered is not in the weak coupling limit (cf.~Figs.~\ref{fig:zzcorrelationmz}.b,~\ref{fig:pmcorrelationmz}.a and~\ref{fig:mpcorrelationmz}.a).

\subsubsection{High energy single triplet excitations}\label{sec:singletripletLL}

The high energy single triplet continua originate from the transition of the low energy rung states $\ket{s}$ and $\ket{t^+}$ to the high energy triplets $\ket{t^0}$ and $\ket{t^-}$. The excitations coming from the singlets $\ket{s}$ (in $S^{zz}_\pi$ and $S^{+-}_\pi$) are already present in the spin liquid phase (cf.~Sec.~\ref{sec:singletripletspinliquid}) in which they have the shape of a sharp peak centered on the triplet dispersion. The transition between the gapped spin liquid and the gapless regime is smooth and consists in a splitting and a broadening of the triplet branch that generates a broad continuum of new excitations. Contrarily to the latter the excitations coming from the low energy triplets $\ket{t^+}$ (in $S^{+-}_0$) are not present in the spin liquid phase. The corresponding spectral weight appears when $h^z>h_{c1}$.

An interpretation of the complex structure of these high energy continua can be obtained in terms of itinerant quantum chains.
 Using a strong coupling expansion of the Hamiltonian~\eqref{equ:spinladderhamiltonian} (appendix~\ref{sec:tjmodelmapping}) one can map the high energy single triplet excitations $\ket{t^0}$ to a single hole in a system populated by two types of particles with
pseudo spin $|\tilde\uparrow\rangle=|t^+\rangle$,
$|\tilde\downarrow\rangle=|s\rangle$ (with the notation of 
Sec.~\ref{sec:spinchainmap}).

In this picture the effective Hamiltonian of the $J_\perp$ energy sector is approximately equivalent to the half
filled anisotropic 1D t-J model with one hole (see appendix~\ref{sec:sec_sector_Jperp}). The effective Hamiltonian is given by
\begin{equation}\label{equ:t-J}
H_\text{t-J}= H_\text{XXZ} + H_\text{t} + H_\text{s-h}+\epsilon.
\end{equation}
where $\epsilon=(J_\perp+h^z)/2 $ is an energy shift, $H_\text{XXZ}$ is the XXZ spin-$1/2$ chain Hamiltonian~\eqref{equ:strongcouplinghamiltonian} and $H_\text{t}=J_\parallel/2\sum_{l,\sigma}(c_{l,\sigma}^\dagger c_{l+1,\sigma}^\phd +h.c.) 
$ is the usual hopping term. Here $c^\dagger_{l,\sigma}$ ($c_{l,\sigma}$) is the creation (annihilation) operator of a fermion with pseudo spin $\sigma=\tilde \uparrow,\tilde\downarrow$ at the site $l$. Note that although we are dealing here with spin states, it is possible to 
faithfully represent the three states of each site's Hilbert space ($\ket{s}$, $\ket{t^+}$, $\ket{t^0}$) using a fermionic representation. 

In addition to the usual terms of the t-J model, a nearest neighbor interaction term between one of the spins and the hole arises
\begin{equation}
H_\text{s-h}=-\frac{J_\parallel}{4}\sum_l\left[n_{l,h}n_{l+1,\tilde\uparrow}+n_{l,\tilde\uparrow}n_{l+1,h}\right].
\end{equation}
Here $n_{l,h}$ is the density operator of the hole on the site $l$.
In this language the spectral weight of $S^{zz}_\pi$ and $S^{+-}_0$ 
corresponding to the single high energy triplet
excitations is equivalent to the single particle spectral functions of the spin up and spin down particle, respectively:
\begin{equation}
\begin{array}{lllll}
S^{zz}_\pi&\propto&\langle c^\dagger_{\tilde\downarrow} c_{\tilde\downarrow}^\phd\rangle&\mbox{ with hole of type }&|s\rangle\rightarrow|t^0\rangle\\
S^{+-}_0&\propto&\langle c^\dagger_{\tilde\uparrow} c_{\tilde\uparrow}^\phd\rangle&\mbox{ with hole of type }&|t^+\rangle\rightarrow|t^0\rangle.
\end{array}
\end{equation}
Here $\langle c^\dagger_\sigma
c_\sigma^\phd\rangle(q,\omega)=\sum_\lambda|\langle\lambda|c_{q,\sigma}|0\rangle|^2\delta(\omega+E_0-E_\lambda)$.

\begin{figure}[!h]
\begin{center}
\includegraphics[width=0.4\linewidth]{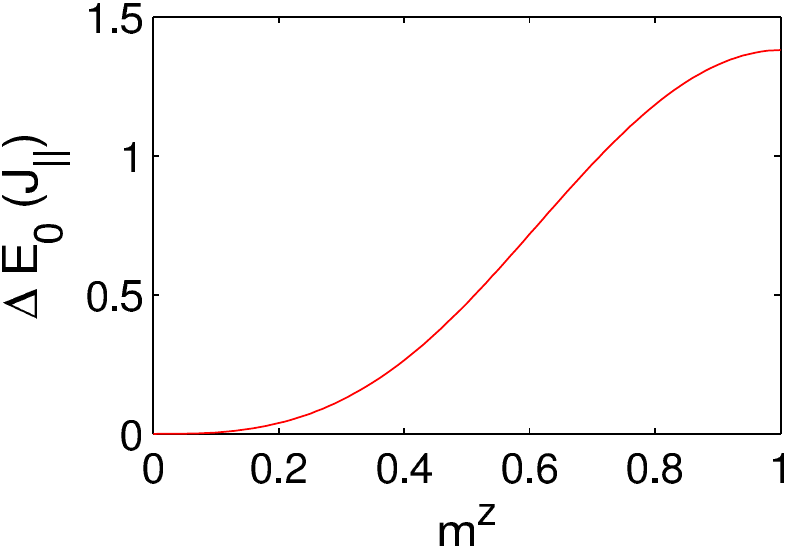}
\end{center}
\caption{Shift of the ground state energy per rung versus the magnetization $\Delta E_0(m^z)=E_0(0)-E_0(m^z)$. [Taken from Ref.~\cite{Bouillot_ladder_statics_dynamics}]\label{fig:groundE}}
\end{figure}

For the standard t-J model (for $SU(2)$ invariant XXX spin chain background and without the anisotropic term $H_\text{s-h}$ in Eq.~\eqref{equ:t-J}), these spectral functions have been studied in
Refs.~\cite{sorella_spectrum_hubbard_1D,sorella_spectrum_hubbard_1D_long}. The
presence of singularities of the form
\begin{equation}\label{equ:highenergysingularity}
\langle c^\dagger_\sigma c_\sigma^\phd\rangle(q,\omega)\propto[\omega-\omega_{t^0}(q-q_\nu)]^{2X_\nu(q)-1}
\end{equation}
were found. Here $\omega_{t^0}(q)$ is the $\ket{t^0}$ triplet
dispersion relation, $q_\nu$ the spinon momentum at the Fermi
level and $X_\nu$ the algebraic decay exponent at the
singularity. This exponent is not known in our case and depends
on the magnetization $m^z$ and the momentum $q$. Eq.~\eqref{equ:highenergysingularity} generates a
peak or a cusp at the energy
$\omega=\omega_{t^0}(q-q_\nu)$. The spinon momentum $q_\nu$ depends on the type of the rung state before excitation
($\nu=s,t^+$). For an excitation created from a singlet state one obtains
$q_s=\pm\pi m^z$ (for $S^{zz}_\pi$) and 
from the triplet state $q_{t^+}=\pi(1\pm m^z)$ (for $S^{+-}_0$) (Fig.~\ref{fig:bandfilling}). At $h^z=0$, a series expansion of
$\omega_{t^0}(q)$ can be performed (cf.~$\omega_t(q)$ in Eq.~\eqref{equ:singletripletenergy}). To extend this expression into the gapless phase ($h_{c1}<h^z<h_{c2}$), we approximate
$\omega_{t^0}(q)$ by shifting the value $\omega_t(q)$ at $h^z=0$ by the Zeeman shift, i.e.
\begin{equation}
\omega_{t^0}(q)=\omega_t(q)+\Delta E_0(h^z).
\end{equation}
Here we used the shift of the
ground state energy per rung $\Delta
E_0(h^z)=E_0(h^z)-E_0(0)$. $\Delta E_0$ was determined by
DMRG calculations (Fig.~\ref{fig:groundE} for the BPCB
parameters). The resulting momentum-frequency positions
$\omega=\omega_{t^0}(q-q_\nu)$ of the high energy
singularities (cusps or divergencies) are plotted 
on the spectrum
Fig.~\ref{fig:zzcorrelationmz}.b and
Fig.~\ref{fig:pmcorrelationmz}.a. They agree remarkably well with
the shape of the computed spectra, in particular, for small magnetic field \footnote{For the correlations
$S^{zz}_\pi$ and $S^{+-}_0$, some of these singularities
correspond to the lower edge description in
Ref.~\cite{furusaki_correlations_ladder} and discussed in Sec.~\ref{sec:weakcouplinghighenergy}.}.  Neglecting the additional interaction term $H_\text{s-h}$, the t-J model Hamiltonian would lead to a symmetry of these excitations with respect to half magnetization. 
However, in the numerical spectra the effect of the interaction shows up in a clear asymmetry of these excitations (compare Figs.~\ref{fig:zzcorrelationmz}.1.b  and~\ref{fig:pmcorrelationmz}.3.a).
In particular, in the $S^{+-}_0$ correlation some of the weight is seemingly detaching and pushed towards the upper boundary of the continuum (Fig.~\ref{fig:pmcorrelationmz}.3.a) for large magnetization. 

A similar mapping can be performed for the single $\ket{t^-}$ excitation. 
In contrast to the $J_\perp$ sector, in the $2J_\perp$ sector not only the $\ket{t^-}$ excitation occurs, but the effective Hamiltonian mixes also $\ket{t^0}$ triplets into the description. Therefore, the description by a single hole in a spin-$1/2$ chain breaks down and more local degrees of freedom are required.
This results in a more complex structure as seen in Fig.~\ref{fig:pmcorrelationmz}.1.b. Previously high-energy excitations in dimerized antiferromagnets have been described by a rather general mapping to an X-ray edge singularity problem~\cite{Kolezhuk_ESR_high_fields,Kolezhuk_response_high_fields,Friedrich_edge_haldane}. It is interesting though that in the present setup these excitations
can be understood as  t-J hole spectral functions, which display a much richer structure than anticipated.

\subsubsection{High energy two-triplet excitations}\label{sec:twotripletLL}

The two-triplet continua and bound states already discussed in the spin liquid phase (cf.~Sec.~\ref{sec:zerofieldtwotriplets}) are still visible in the gapless regime in the symmetric
correlations ($S^{zz}_0$ and $S^{\pm\mp}_0$). At low magnetic field the location of their maximal spectral weight can
be approximated by the expression of the bound state dispersion at zero field, $\omega_{tB}(q)$ in Eq.~\eqref{equ:tripletboundstate}, shifted by the Zeeman energy\footnote{The Zeeman shift includes both the shift of the ground state (Fig.~\ref{fig:groundE}) and the shift of the excited state.}.
The two-triplet excitation location obtained in this way agrees to a good extent with the location found in the numerical calculations
(cf.~Figs.~\ref{fig:zzcorrelationmz}.1.a,~\ref{fig:pmcorrelationmz}.1.a and~\ref{fig:mpcorrelationmz}.1.a). Since these excitations
are generated from the high energy triplet components in the
ground state and these vanish with increasing magnetic field (cf.~Fig.~\ref{fig:tripletdensity}),
their residual spectral weight slowly disappears with
increasing magnetization.

\section{Weak to strong coupling evolution}

\begin{figure}[H]
\begin{center}
\includegraphics[width=0.7\linewidth]{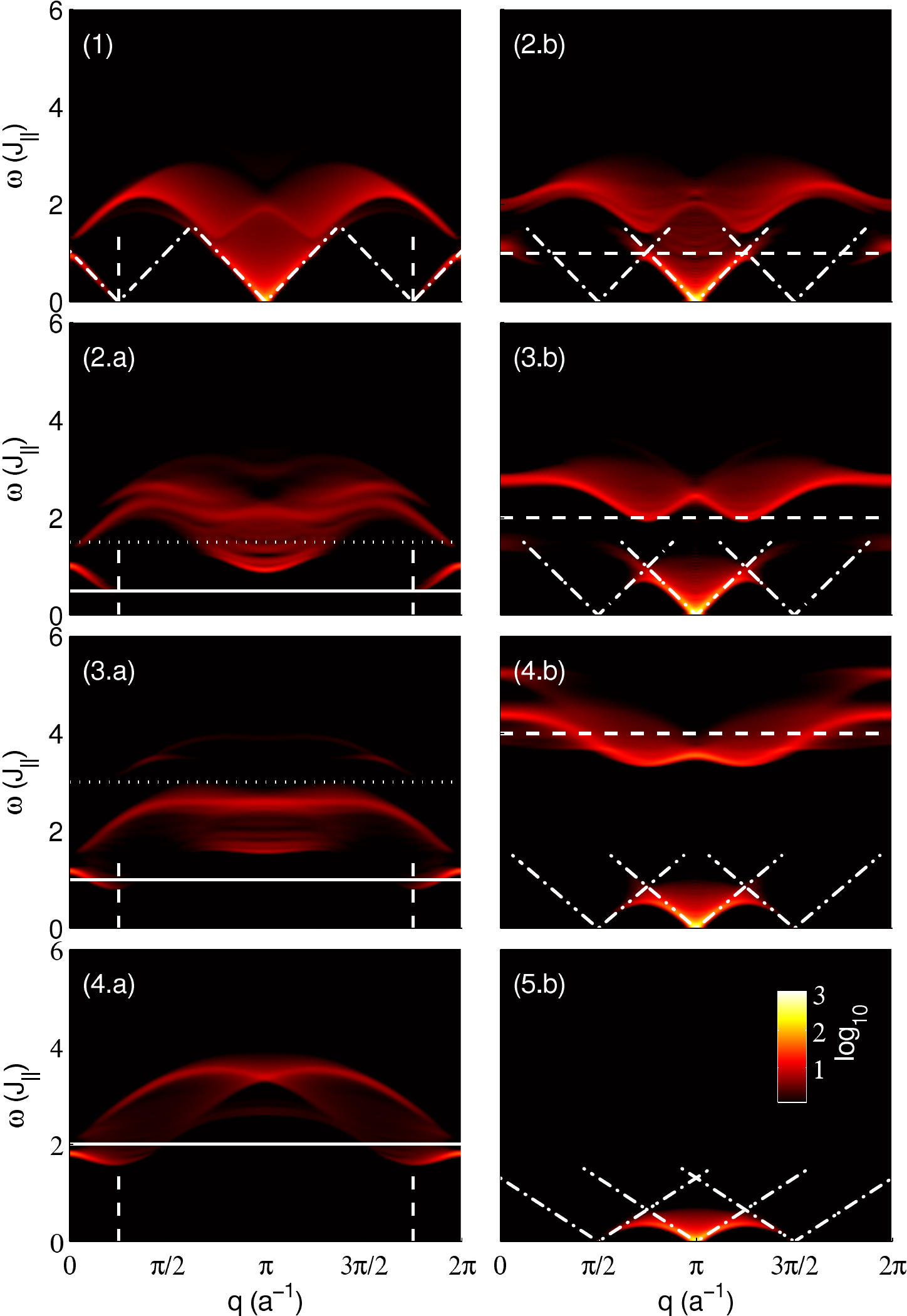}
\end{center}
\caption{Momentum-energy dependent $+-$-correlations ($S_{q_y}^{+-}(q,\omega)$) at $m^z=0.25$ for different ladder couplings ($\gamma=J_{\parallel}/J_{\perp}$) (1) $\gamma\rightarrow\infty$, (2) $\gamma=2$, (3) $\gamma=1$, (4) $\gamma=0.5$, (5) $\gamma\rightarrow0$. The symmetric (antisymmetric) correlations with $q_y=0$ ($q_y=\pi$) are presented in the figures labeled by a (b). In (1,2-4.a) the vertical dashed lines represent the incommensurate momenta of the low energy branches of the single spin chain at $q=\pm\pi m^z=\pm\pi/4$ (they also correspond to the predicted momenta of the lowest energy excitations of the symmetric correlations~\cite{furusaki_correlations_ladder}). The horizontal solid (dotted) horizontal lines in (2-4.a) correspond to the approximate energy $J_\perp$ ($3J_\perp$) of the single triplet excitations of type (ii) (two-triplet excitations of type (iii)). The horizontal dashed lines in (2-4.b) correspond to the approximate energy $2J_\perp$ of the excitations of type (ii). The dash-dotted lines in (1) correspond to the linear low energy boundaries of the continuum of excitations given by the LL theory applied to a single Heisenberg chains $\gamma\rightarrow \infty$. The dash-dotted lines in (2-5.b) correspond to the linear low energy boundaries of the continuum of excitations given by the LL theory for a spin ladder with finite $\gamma$ (Sec.~\ref{sec:luttingerliquidcorr}). [Taken from Ref.~\cite{Bouillot_ladder_statics_dynamics}]
\label{fig:weaktostrong}}
\end{figure}

For all the excitation spectra presented above the intrachain coupling ratio of BPCB $\gamma =J_\parallel/J_\perp\approx1/3.55\ll 1$ was taken. For this chosen value of $\gamma$, a strong coupling approach gives a reasonable
description of the physics. In this section we discuss the evolution of the spectra from weak ($\gamma\rightarrow\infty$) to strong coupling ($\gamma\rightarrow0$).
To illustrate this behavior, we show in
Fig.~\ref{fig:weaktostrong} the symmetric  and antisymmetric
parts of the correlations $S_{q_y}^{+-}$ at $m^z=0.25$ for
different coupling ratios $\gamma =\infty,2,1,0.5,0$.

At $\gamma\rightarrow\infty$ (Fig.~\ref{fig:weaktostrong}.1), the chains forming the ladder correspond to two decoupled Heisenberg chains. In this case the symmetric and antisymmetric correlations are
identical $S_0^{+-}=S_\pi^{+-}$ and are equivalent to the correlation $2S^{+-}$ of the single chain~\cite{Muller_spinchain_dyncor} with magnetization per spin $m^z/2=0.125$. A complex low energy continuum exists with zero energy branches~\cite{giamarchi_book_1d,chitra_spinchains_field,Muller_spinchain_dyncor} at momenta $q=\pm\pi m^z,\pi$ similar to that discussed in Sec.~\ref{sec:lowenergyexcitations}.
In contrast, in the strong coupling limit ($\gamma \rightarrow0$) (Fig.~\ref{fig:weaktostrong}.5.b)
the symmetric correlations vanish and the antisymmetric
part corresponds to the single chain correlation $2S^{+-}$ with
anisotropy $\Delta=1/2$ and magnetization per spin $m^z-1/2$ (see the spin chain mapping in Sec.~\ref{sec:spinchainmap}). The antisymmetric part consists of a low energy continuum with branches at momenta $q=(1\pm2 m^z)\pi,\pi$ (Sec.~\ref{sec:lowenergyexcitations}). Note, that a bosonization description of the low energy
sectors of both extreme regimes can be formulated ~\cite{chitra_spinchains_field,giamarchi_ladder_coupled,furusaki_correlations_ladder} (Sec.~\ref{sec:luttingerliquidcorr}).

In the following we discuss the evolution between these two limits.
In the antisymmetric correlation (cf.~Fig.~\ref{fig:weaktostrong}.2-5.b) a low energy continuum exists  at all couplings with a zero energy excitation branch at $q=\pi$. These low energy excitations correspond mainly to the excitations with $\Delta S=\Delta M^z=-1$. This has been pointed out for the weak coupling limit~\cite{Muller_spinchain_dyncor,chitra_spinchains_field}. They become the transitions $|t^+\rangle\rightarrow|s\rangle$ with the same quantum numbers in the decoupled bond limit.
Additionally the upper part of the excitation spectrum at weak coupling, which mainly corresponds to excitations with ~\cite{Muller_spinchain_dyncor} $\Delta S=0,1$ and $\Delta
M^z=-1$ splits from the lower part of the spectrum and moves to higher energy while increasing the coupling. This upper part evolves to a high energy excitation branch which corresponds in the decoupled bond limit to the $|s\rangle\rightarrow|t^-\rangle$ transition, i.e.~single triplet excitations of type (ii) (Sec.~\ref{sec:singletripletLL}) approximately at\footnote{Note that in the strong coupling limit the energy scales set by $J_\perp$ and $h^z$ become very close, such that this position is in agreement with the ones previously discussed for the strong coupling limit. }
$2J_\perp$.

The properties of the zero energy excitation branch at $q=\pi$ show a smooth transition between the two limits~\cite{giamarchi_ladder_coupled,chitra_spinchains_field}.
For example the slope of the lower edge continuum which is determined by the LL velocity $u$ decreases smoothly from its value for the Heisenberg chain to the lower value for the anisotropic spin chain with $\Delta=1/2$ in the strong coupling limit.
In contrast to this smooth change, the presence of a finite value of $J_\perp$ leads to the formation of a gap in the
incommensurate low energy branches~\cite{chitra_spinchains_field} at $q=\pm\pi m^z$. With increasing coupling strength $J_\perp$ new low energy branches at momenta $q=\pi(1\pm 2m^z)$ become visible~\cite{giamarchi_ladder_coupled,furusaki_correlations_ladder}. The weight of these gapless branches is very small for small coupling and increases with stronger coupling ~\cite{giamarchi_ladder_coupled}.

In contrast to the antisymmetric part, the
symmetric part $S_0^{+-}$ becomes gapped when the interladder coupling $J_\perp$ is turned on. The lowest energy excitations remain close to the momenta $q=\pm\pi m^z$ in agreement with Ref.~\cite{furusaki_correlations_ladder}. They connect to the single triplet excitations of type (ii) (Sec.~\ref{sec:singletripletLL}) which are approximately at an energy $J_\perp$. While increasing $\gamma$ the higher part of the spectrum starts to separate from the main part and evolves to a branch of high energy two-triplet excitations of type (iii) (Sec.~\ref{sec:twotripletLL}). These are located at approximately $3J_\perp$.
Our computed spectra for $\gamma=2,1,0.5$ presented in
Fig.~\ref{fig:weaktostrong}.2-4.a clearly show this behavior. In Fig.~\ref{fig:weaktostrong}.4.a the highest two-triplet excitations cannot be seen anymore since their spectral weight is too low.

\section{Influence of the weak interladder coupling on the excitation spectrum}

Up to now we only discussed the excitations of a single spin ladder and neglected the weak interladder coupling $J'$ usually present in real compounds.

Deep inside of the spin liquid phase, the correlations for a single ladder are dominated by high energy single or multi triplet excitations as discussed in Sec.~\ref{sec:spinliquidexcitations}. The presence of a small interladder coupling $J'$ causes a dispersion in the interladder direction with an amplitude of order $J'$. This effect can be evaluated for independent triplet excitations using a single mode approximation~\cite{Auerbach_book_magnetism}. However, for the compound BPCB the interladder coupling is so small that present day experiments do not resolve this small broadening~\cite{Thielemann_INS_ladder,Savici_BPCB_INS}.

In contrast in the gapless phase the effect of the interladder coupling can change considerably the excitations.
In particular, below the transition temperature to the 3D-ordered phase, a
Bragg peak appears at $q=\pi$ in the transverse dynamical functions
$S^{\pm\mp}_\pi$ due to the transverse antiferromagnetic long range order presented in Secs.~\ref{sec:mean-field} and~\ref{sec:orderparameter}. As discussed in Ref.~\cite{schulz_coupled_spinchains},
this Bragg peak is surrounded by gapless Goldstone modes and it has been
measured in the compound BPCB~\cite{Thielemann_ND_3Dladder} by ND experiments described in Sec.~\ref{sec:neutron_diffraction} (see Fig.~\ref{fig:ND_measurements}). Additional high energy modes are predicted to occur in the transverse $S^{\pm\mp}_\pi$ and longitudinal $S^{zz}_0$~\cite{schulz_coupled_spinchains}.
It would be interesting to compute the excitations using random phase approximation analogously to Ref.~\cite{schulz_coupled_spinchains} in combination with the computed dynamical correlations for the single ladder in order to investigate the effect of a weak interladder coupling in more detail. However, this goes beyond the scope of the present work and
will be left for a future study.

\section{Inelastic neutron scattering (INS)}\label{sec:INS}

The inelastic neutron scattering (INS) technique is
a direct probe for dynamical spin-spin correlation functions. Measurements have been performed on the compound BPCB in the spin liquid
phase~\cite{Savici_BPCB_INS,Thielemann_INS_ladder} (low magnetic
field) and in the gapless regime~\cite{Thielemann_INS_ladder}.
Modeling the compound BPCB by two inequivalent uncoupled ladders  oriented along the two rung
vectors ${\bf d}_{1,2}$~\eqref{equ:rung_orientation} (see Fig.~\ref{fig:structure}) the magnetic INS cross section~\cite{lovesey_neutron_scattering} is given by the formula
\begin{multline}\label{equ:laddercross_section}
\frac{d^2\sigma}{d\Omega dE'}\propto \frac{q'}{q}|F(\mathbf{Q})|^2\left\{
4\left(1-\frac{{Q^z}^2}{\mathbf{Q}^2}\right)
\left[c(\mathbf Q)\cdot S^{zz}_0+s(\mathbf Q)\cdot S^{zz}_\pi\right]
\right. \\
\left.+
\left(1+\frac{{Q^z}^2}{\mathbf{Q}^2}\right)
\left[c(\mathbf Q)\cdot\left(S^{+-}_0+S^{-+}_0\right)+s(\mathbf Q)\cdot\left(S^{+-}_\pi+S^{-+}_\pi\right)
\right]\right\}
\end{multline}
with $c(\mathbf Q) =\sum_{i=1,2}\cos^2(\mathbf{Q}\cdot \mathbf{d}_i/2)$ and $s(\mathbf Q)=\sum_{i=1,2}\sin^2(\mathbf{Q}\cdot \mathbf{d}_i/2)$.
Here $\mathbf{Q}=(Q^x,Q^y,Q^z)=\mathbf{q}-\mathbf{q}'$ is the momentum transferred to the sample ($\mathbf{q}$, $\mathbf{q}'$ are the ingoing, outgoing neutron momenta, respectively) and $\omega=E-E'$ is the transferred energy  ($E$, $E'$ are the ingoing, outgoing neutron energies). The correlations $S^{\alpha\beta}_{q_y}$ are defined at zero temperature in Eq.~(\ref{equ:correlation1}) and evaluated at a momentum $q={\bf Q}\cdot {\bf a}$ along the ${\bf a}$ unit cell vector (momentum along the ladder direction) and energy $\omega$. The magnetic form factor $F({\bf{Q}})$ of the $\mathrm{Cu^{2+}}$ and the ratio $q'/q$ are corrected in the experimental data.

The INS cross section \eqref{equ:laddercross_section} is
directly related to a combination of different correlation functions
$S_{q_y}^{\alpha\beta}$ with weights depending on the transferred momentum ${\bf Q}$ and the
magnetic field orientation. In the model definition (see
Sec.~\ref{sec:model}), the magnetic field ${\bf h}$ is pointing
along the $z$ direction. Additionally, the two rung vectors~\eqref{equ:rung_orientation} of BPCB have identical components along the unit cell vectors ${\bf a}$ and ${\bf c}$. Hence, aligning the field to the ${\bf b}$ unit cell
vector and tuning ${\bf Q}$ in the ${\bf a^\star c^\star}$-plane
(${\bf a^\star}$, ${\bf b^\star}$ and ${\bf c^\star}$ are the
reciprocal vectors of ${\bf a}$, ${\bf b}$ and ${\bf c}$) allows one
to keep constant the prefactors in front of the correlations in Eq.~\eqref{equ:laddercross_section} scanning the ${\bf
a^\star}$-momentum with the condition ${\bf Q}\cdot{\bf
d_i}=0\text{ or }\pi$, for both $i=1,2$, to target the symmetric or
antisymmetric part, respectively.

\begin{figure}[h!]
\begin{center}
\includegraphics[width=0.7\linewidth]{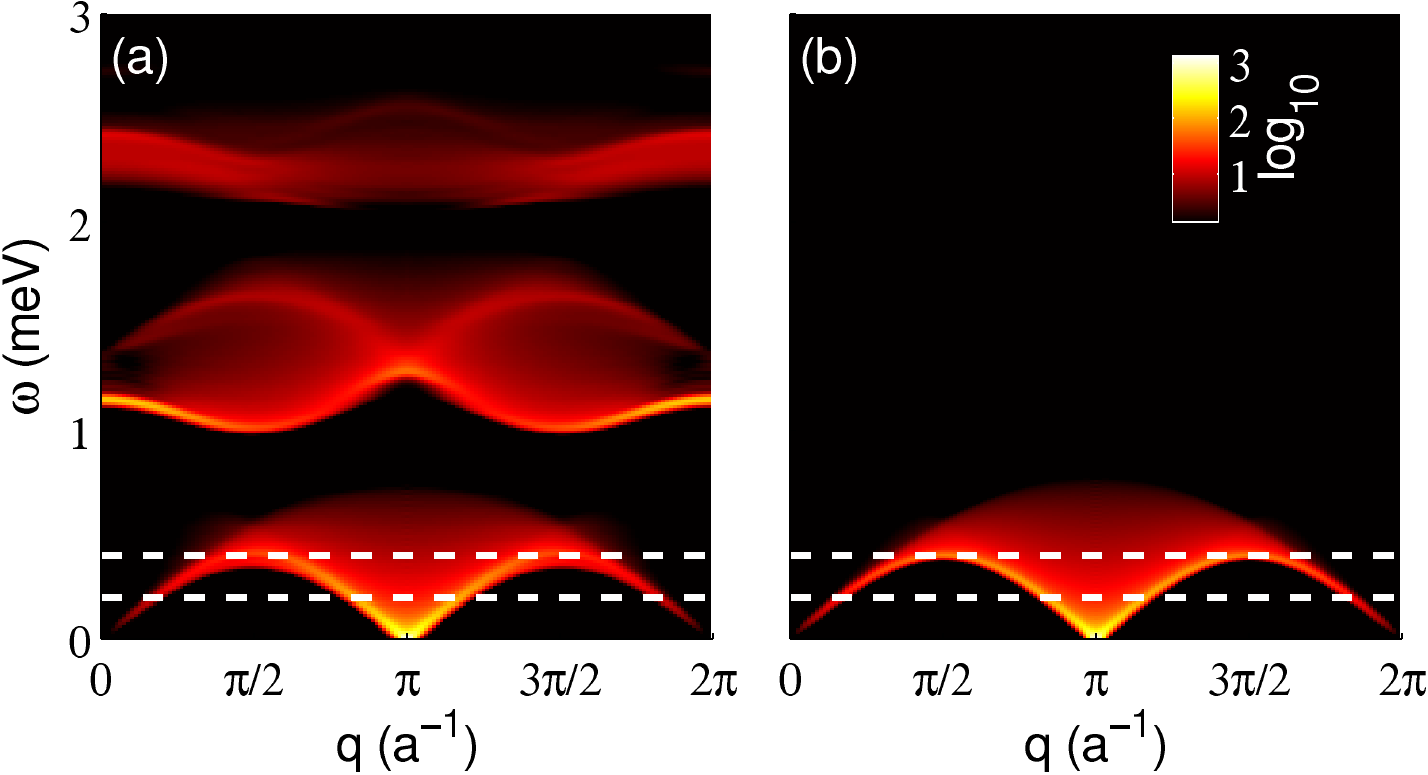}
\end{center}
\caption{Theoretical momentum-energy dependent INS cross section for BPCB with $\mathbf{Q}\cdot\mathbf{d}_i=\pi$ ($i=1,2$) and $q={\bf Q}\cdot {\bf a}$ at $m^z=0.5$ in (a) a ladder system and (b) the spin chain mapping. The horizontal dashed lines correspond to the constant energy scans at $\omega=0.2,0.4 ~{\rm meV}$ shown in Fig.~\ref{fig:experimentalINS1}. [Taken from Ref.~\cite{Bouillot_ladder_statics_dynamics}]\label{fig:theoreticalINS1}}
\end{figure}

\begin{figure}[h!]
\begin{center}
\includegraphics[width=\linewidth]{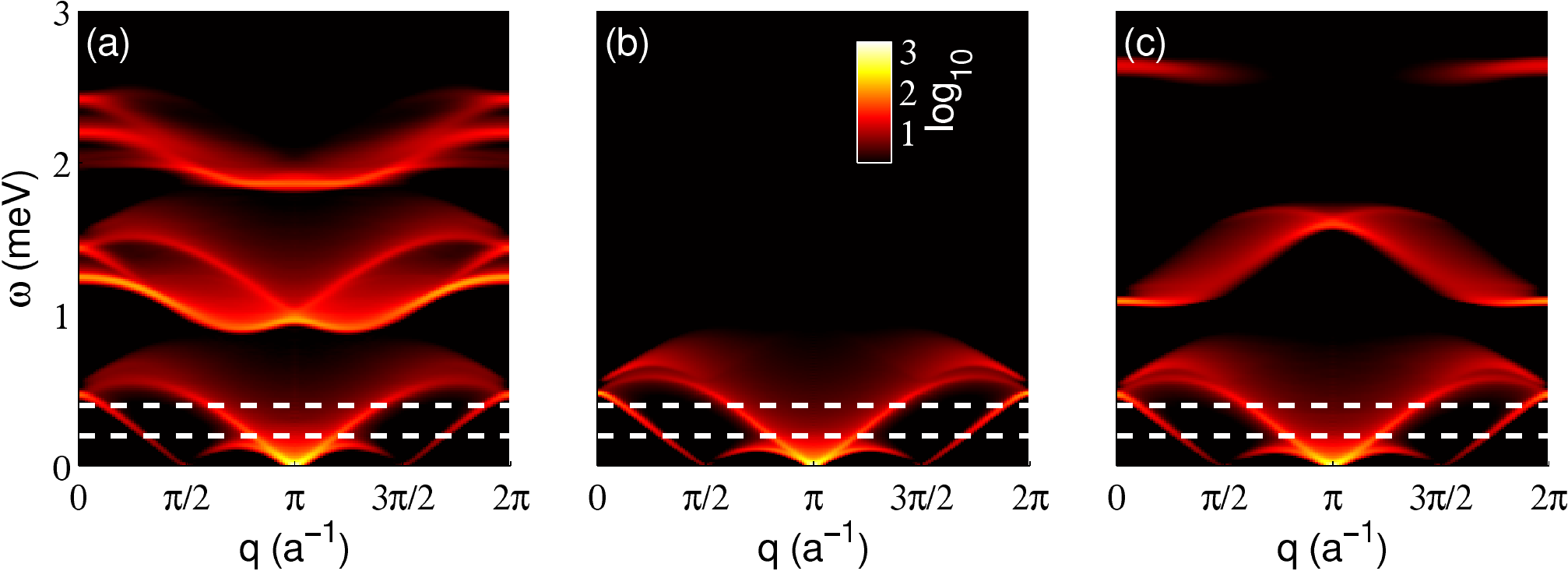}
\end{center}
\caption{Theoretical momentum-energy dependent INS cross section for BPCB with $\mathbf{Q}\cdot\mathbf{d}_i=\pi$ ($i=1,2$) and $q={\bf Q}\cdot {\bf a}$, (a) at $m^z=0.25$ (b) at $m^z=0.25, 0.75$ in the spin chain mapping, (c) at $m^z=0.75$. The horizontal dashed lines correspond to the constant energy scans at $\omega=0.2,0.4 ~{\rm meV}$ plotted in Fig.~\ref{fig:experimentalINS2}. [Taken from Ref.~\cite{Bouillot_ladder_statics_dynamics}]\label{fig:theoreticalINS2}}
\end{figure}

We focus here on the antisymmetric
part for which the low energy spectra have already been
studied experimentally and theoretically~\cite{Thielemann_INS_ladder}.
Theoretically the focus so far lay on the description of the low energy excitations by the spin chain mapping.
In contrast, we compute here the INS cross section \eqref{equ:laddercross_section} for the full ladder at $m^z=0.25,0.5,0.75$ using the correlations presented in Sec.~\ref{sec:gapless_dynamics}. The results are shown in Figs.~\ref{fig:theoreticalINS1} and~\ref{fig:theoreticalINS2} and are compared to the results from the spin chain mapping.

As expected from expression \eqref{equ:laddercross_section}, the INS cross section contains the different excitations
present in the spectra of $S_\pi^{zz}$ and $S_\pi^{\pm\mp}$ (cf.~Figs.~\ref{fig:zzcorrelationmz}.b, \ref{fig:pmcorrelationmz}.b and~\ref{fig:mpcorrelationmz}.b):
\begin{itemize}
 \item[(a)] The low energy continuum originates from the transversal correlations $S_\pi^{\pm\mp}$. It is qualitatively well described by the spin chain mapping that presents a symmetry with respect to half magnetization.
 \item[(b)] The continuum of excitations at energy $\sim J_\perp$ comes from the longitudinal correlation $S_\pi^{zz}$ (not present in the spin chain mapping).
 \item[(c)] The continuum of excitations at energy $\sim 2J_\perp$ stems from the transversal correlation $S_\pi^{+-}$ (not present in the spin chain mapping).
\end{itemize}

\begin{figure}[h!]
\begin{center}
\includegraphics[width=0.7\linewidth]{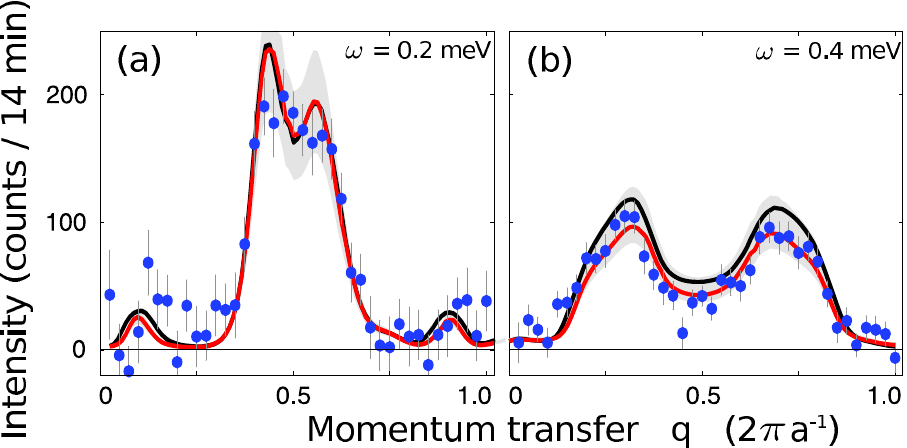}
\end{center}
\caption{Inelastic neutron scattering intensity measured along
  ${\bf a^\star}$ of BPCB~\cite{Thielemann_INS_ladder} with the momentum $\pi$ in the rung direction (${\bf Q}\cdot{\bf
d_i}=\pi$) at $h^z=10.1~{\rm T}$ ($m^z\approx0.5$) and $T=250~{\rm mK}$ after subtraction of the zero-field background. In each panel, fixed energy scans (shown by white dashed lines in Fig.~\ref{fig:theoreticalINS1}) are plotted: (a) $\omega=0.2~{\rm meV}$, (b) $\omega=0.4~{\rm meV}$. The circles correspond to the experimental data. The red (black) solid lines are the $m^z=0.5$ theoretical data for the ladder (the spin chain mapping) convolved with the instrumental resolution. The shaded bands indicate the error bar in the experimental
determination of a single proportionality constant valid for all fields, energies, and wave vectors. The width of these areas combines the statistics of all scans with uncertainties
in the exact magnetization values at the chosen
fields and in the convolution procedure. [Taken from Ref.~\cite{Bouillot_ladder_statics_dynamics}]
\label{fig:experimentalINS1}}
\end{figure}

\begin{figure}[h!]
\begin{center}
\includegraphics[width=0.7\linewidth]{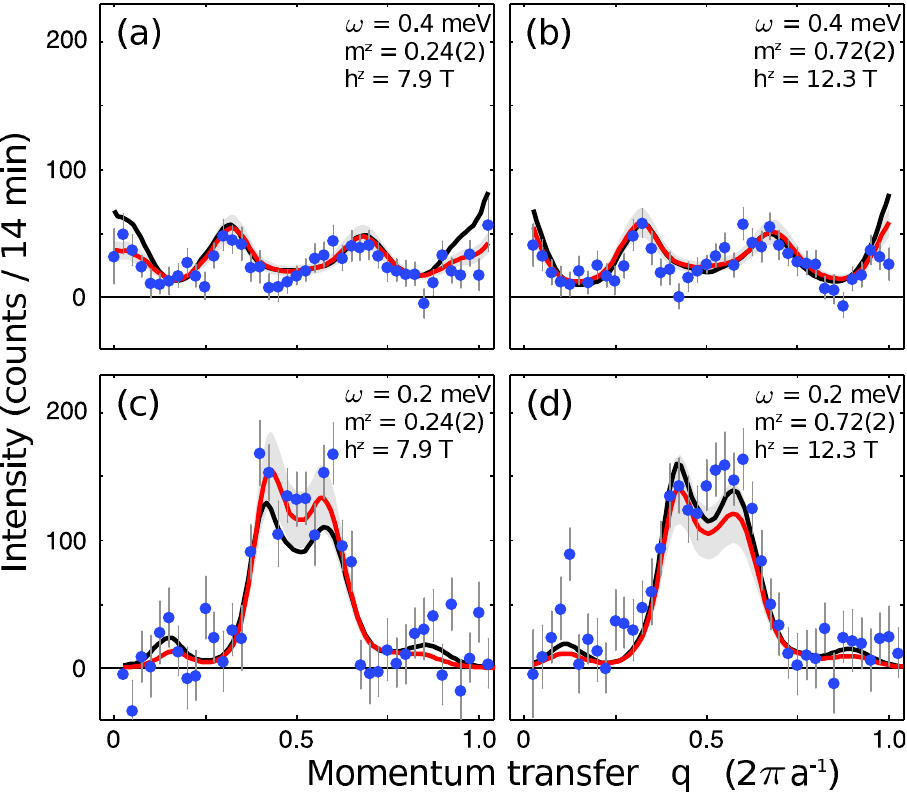}
\end{center}
\caption{Inelastic neutron scattering intensity measured along $\bf a^\star$ of BPCB~\cite{Thielemann_INS_ladder} with a $\pi$ momentum along the rung direction (${\bf Q}\cdot{\bf
d_i}=\pi$) at $T=250~{\rm mK}$ after subtraction of the zero-field background. In each panel, cuts at fixed energy (shown by white dashed lines in Fig. \ref{fig:theoreticalINS2}) are plotted: (a) $\omega=0.4~\mathrm{meV}$ and $m^z = 0.24$,
(b) $\omega=0.4~\mathrm{meV}$ and $m^z = 0.72$,
(c) $\omega=0.2~\mathrm{meV}$ and $m^z = 0.24$,
(d) $\omega=0.2~\mathrm{meV}$ and $m^z = 0.72$. The circles correspond to the experimental data. The solid red (black) curves are the theoretical data for the ladder (the spin chain mapping) convolved with the instrumental resolution. The shaded bands indicate the error bar in the experimental
determination of a single proportionality constant valid for all fields, energies, and wave vectors. The width of these areas combines the statistics of all scans with uncertainties
in the exact magnetization values at the chosen
fields and in the convolution procedure. [Taken from Ref.~\cite{Bouillot_ladder_statics_dynamics}]\label{fig:experimentalINS2}}
\end{figure}

The main features of the low energy continuum (a) are well covered by the spin chain mapping~\cite{Thielemann_INS_ladder}. However, slight differences between the low
energy excitations in the spin ladder and the spin chain are still visible in Fig.~\ref{fig:theoreticalINS2} (cf.~also Sec.~\ref{sec:lowenergyexcitations}). These differences can even be distinguished in the experimental data in Figs.~\ref{fig:experimentalINS1} and~\ref{fig:experimentalINS2} where some cuts at fixed energy $\omega=0.2,0.4\ {\rm
meV}$ are plotted. The
INS measured intensity is directly compared to the theoretical
cross section (\ref{equ:laddercross_section}) computed for the
ladder and the spin chain mapping at $m^z=0.24,0.5,
0.72$ convolved with the instrumental resolution. The amplitude is fixed by fitting one proportionality constant for all fields, energies, and wave vectors.

The scans at fixed energy present peaks when the lower edge of the continua (related to the correlations $S_\pi^{\pm\mp}$) is crossed (see dashed white lines in Figs.~\ref{fig:theoreticalINS1} and~\ref{fig:theoreticalINS2}). As one can see, the theoretical curves for the ladder
and the spin chain both reproduce well the main features in the experimental data and only small differences are present:
\begin{itemize}
 \item[-] The spectral weight intensity at $m^z=0.5$ and $\omega=0.4~{\rm meV}$ (in Fig.~\ref{fig:experimentalINS1}.b) is slightly overestimated by the spin chain mapping.
\item[-] The height of the two central peaks at $m^z=0.24$ and $\omega=0.2~{\rm meV}$ (in Fig.~\ref{fig:experimentalINS2}.c) is underestimated by the spin chain mapping.
\end{itemize}

Whereas the low energy excitations (a) only showed a slight asymmetry with respect to the magnetization, a very different behavior can be seen in the high energy part (b)-(c). Indeed, the
high energy part of the INS cross section (Fig.~\ref{fig:theoreticalINS2}) is very asymmetric with respect to half magnetization. As we discussed in
Sec.~\ref{sec:gapless_dynamics}, these excitations are due
to the high energy triplets that can be excited in $S_\pi^{zz}$
and $S_\pi^{+-}$ (see Fig.~\ref{fig:zzcorrelationmz}.b and
Fig.~\ref{fig:pmcorrelationmz}.b) and are totally neglected in
the spin chain mapping. Their corresponding spectral
weight is of the same order than the low energy spectra, and
thus should be accessible experimentally. It would be
very interesting to have an experimental determination of this
part of the spectrum, since as we have seen this part contains characteristic information on the system itself and is related to itinerant systems via the various mappings.

\subsection{Neutron diffraction (ND)}\label{sec:neutron_diffraction}

\begin{figure}[h!]
\begin{center}
\includegraphics[width=0.9\linewidth]{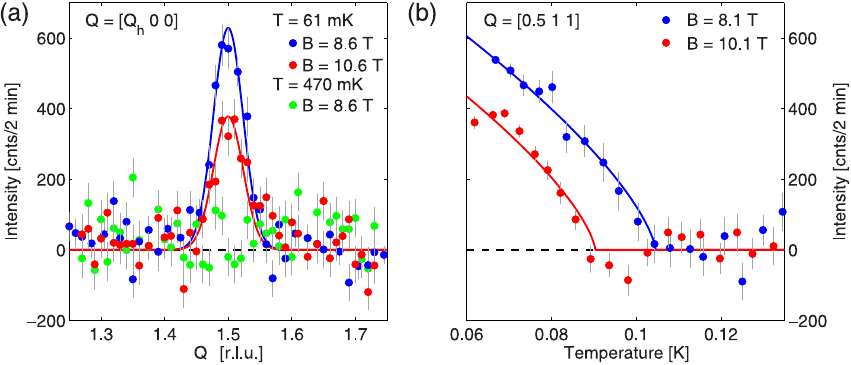}
\end{center}
\caption{Neutron diffraction measurements on BPCB at several temperatures $T$ and applied magnetic fields $B$ ($h^z$ in our notation) in the gapless regime. (a) Momentum scan $\mathbf{Q}$ along $\bf a^\star$ of BPCB across an antiferromagnetic Bragg peak after subtraction of a flat back-ground measured in the spin liquid phase at $h^z = 6\ \text{T}$ and $T = 63\ \text{mK}$. (b) Temperature dependence of the Bragg peak intensity, demonstrating
the onset of  3D long range order at $T_c(h^z)$. Solid lines are fits
using the 3D-XY exponent $2\beta\approx0.7$~\cite{zinn_book,nikuni00_tlcucl3_bec}. [Taken from Ref.~\cite{Thielemann_ND_3Dladder}]\label{fig:ND_measurements}}
\end{figure}

The so-called neutron diffraction (ND) is an elastic neutron scattering process without energy transfer ($\omega =0$). Assuming a non-degenerate ground state, which is the case for the (weakly coupled) spin-$1/2$ ladder \eqref{equ:spinladderhamiltonian} and \eqref{equ:coupledladdershamiltonian} (without frustration (see Ref.~\cite{Auerbach_book_magnetism})), the cross section \eqref{equ:laddercross_section} leads to
\begin{multline}\label{equ:ND_cross_section}\frac{d^2\sigma}{d\Omega dE'}\propto \frac{q'}{q}|F(\mathbf{Q})|^2\left\{2\left(1-\frac{{Q^z}^2}{\mathbf{Q}^2}\right)
\left[c(\mathbf Q)\cdot (m_0^z(q))^2+s(\mathbf Q)\cdot (m_\pi^z(q))^2\right]
\right. \\
\left.+
\left(1+\frac{{Q^z}^2}{\mathbf{Q}^2}\right)
\left[c(\mathbf Q)\cdot\left(m_0^{xy}(q)\right)^2+s(\mathbf Q)\cdot\left(m_\pi^{xy}(q)\right)^2\right]
\right\}.
\end{multline}
The magnetic orders $m_{q_y}^z(q)=\langle S^z_{q_y}(q)\rangle$ and $m_{q_y}^{xy}(q)=\sqrt{\langle S^x_{q_y}(q)\rangle^2 +\langle S^y_{q_y}(q)\rangle^2}$ computed for the ground state, with momentum $q$ ($q_y$) along the leg (rung), are parallel and perpendicular to the magnetic field orientation, respectively. Hence, following Eq.~\eqref{equ:ND_cross_section}, at low temperature, ND provides a quantitative measurement of the order parameters. It has been used to measure  on BPCB the longitudinal magnetization (Fig.~\ref{fig:mz2_ND}) and the transverse long range order in the 3D phase. For the latter, Fig.~\ref{fig:ND_measurements}.a shows the intensity of the ND experiments in the gapless regime versus the momentum $\mathbf{Q}$ along $\bf a^\star$. The appearance of a Bragg peak at the momentum $q=\mathbf{Q}\cdot\mathbf{a}=\pi$ for temperatures below $T_c(h^z)$ is the signature of the existence of a staggered long range order. Following the ND cross section~\eqref{equ:ND_cross_section}, the intensity of the peak is proportional to $\left(m_\pi^{xy}(\pi)\right)^2$ and is used to extract the 3D staggered transverse order $m_a^x(h^z)$ shown in Fig.~\ref{fig:orderparameter_exp}. In Fig.~\ref{fig:ND_measurements}.b, the intensity of the peak plotted versus the temperature at fixed magnetic field $h^z$ clearly shows the onset of the 3D order at $T_c(h^z)$. These ND measurements from Ref.~\cite{Thielemann_ND_3Dladder} are used to extract the magnetic field dependence of the critical temperature $T_c(h^z)$ shown in Figs.~\ref{fig:INS_3D} and~\ref{fig:orderparameter_exp}.

\chapter{Conclusions and perspectives}\label{sec:conclusions}

\section{Conclusions}
In this work we have investigated the thermodynamic and dynamic
properties of weakly coupled spin ladders in a magnetic field. Combining a LL analytical theory and DMRG numerical
techniques, both generalized using a mean field approximation while considering the interladder coupling, we were able to explore
characteristic physical properties of the system in all the regimes of the phase diagram shown in Fig.~\ref{fig:phasediagram}. In addition to the theoretical analysis we compared our findings to experimental results on
the compound BPCB ($\mathrm{(C}_5\mathrm{H}_{12}\mathrm{N)}_2\mathrm{CuBr}_4$) and obtained excellent agreement. This compound appears to be a very good realization of such ladder systems in which other possible effects such as frustration, anisotropic interaction or Dzyloshinskii-Moriya term are very small. The excellent low-dimensionality of this compound and the ideal range of its coupling parameters lead to a clear separation of the energy scales between the different phases which were all accessed experimentally.

As thermodynamic quantities we computed the magnetization and specific
heat of the system as a function of temperature and magnetic
field. The extension of the DMRG technique to finite
temperature allowed us to compute these quantities with an
excellent accuracy. In the gapless phase the low energy part
of the specific heat agreed well with the prediction of the
LL theory. At higher
temperatures, the numerical solution was needed to capture the
precise structure of the peaks in the specific heat, that
reflect the presence of the excited states in the ladder.
The comparison of the theoretical calculations with the measured
magnetization and specific heat proved to be remarkable. This good agreement confirms that the ladder model is indeed
a faithful description of the compound BPCB. It also gives
direct access, via the extrema in magnetization and peaks in the
specific heat to the approximate region of applicability of the Luttinger
liquid description. For BPCB we found
an extended region lying approximatively one order of magnitude above the 3D order transition temperature. The large domain of validity of the LL description for BPCB opened the way for a more detailed experimental investigation of this regime.

For the low energy dynamics we used a combination of the numerical techniques to determine the
Luttinger liquid parameters and then the analytical description
based on Luttinger liquids to compute the dynamical spin-spin
correlation functions. This allowed us to extract the NMR
relaxation rate and the one dimensional antiferromagnetic
transverse susceptibility. If the ladders are very weakly
coupled, which is the case in the considered material, the divergence
of the susceptibility leads to a three dimensional
antiferromagnetic order at low temperatures in the direction
transverse to the applied magnetic field. We computed this
transition temperature and the order parameter at zero
temperature. Comparison with the measured experimental
quantities both by NMR and neutron diffraction proved again to be
remarkable. This excellent agreement between theory and
experiment for these quantities as a function of the magnetic
field allows to \emph{quantitatively} test the Luttinger liquid
theory. It shows that several different correlations are
indeed fully described by the knowledge of the two Luttinger
liquid parameters (and the amplitudes relating the microscopic
operators to the field theoretical ones). This is something that had not
really been tested previously since either the microscopic
interactions were not known in detail leaving the Luttinger
parameters as adjustable parameters, or only one correlation
function could be measured in a given experiment, not allowing
to test for universality of the description. 

We also gave a detailed analysis of the dynamical spin-spin
correlations, for $T=0$ using the time dependent DMRG method,
for a wide range of energies and all momenta. The excitations reveal a lot of important information on the system and are well suited to characterize it. In particular we showed the interesting evolution of the excitations in the system with the magnetic field and the coupling strengths. Quite interestingly the intermediate energy
part can be related to the excitations of a t-J model and shows thereby features of itinerant systems. We also showed that the
dynamical correlations of the ladder posses characteristic high
energy features that are clearly distinct from the
corresponding spectrum for spin chains.

The numerical calculation is efficient for the high and intermediate energy
part of the spectrum for which the Luttinger liquid description
cannot be applied. We showed that the two methods, numerics and
LL have enough overlap, given the accuracy of our calculation
so that we can have a full description of the dynamical
properties at all energies. This allowed us to use each of the method in
the regime where it is efficient. In particular, in this thesis
we did not push the numerical calculations to try to obtain the exact behavior at low energies or the very fine structure effects,
but focused on the high and intermediate energy regime suitable for the existing and future INS experiments. We used the analytical description coupled
to the numerical determination of the Luttinger parameters to obtain an accurate low
energy description. We made the connection between our
results and several analytical predictions. In particular at
intermediate-low energy our calculation agreed with the
Luttinger liquid prediction of incommensurate points and
behavior (divergence, convergence) of the correlations.

We compared our numerical results with existing INS data on
the compound BPCB and found excellent agreement. It is
rewarding to note that the resolution of our dynamical
calculation is, in energy and momentum, at the moment better
than the one of the experiment. The comparison between theory
and experiment is thus essentially free of numerical errors and thus allows for precise tests of the models.
Given the current resolution of the INS
experiment it is difficult to distinguish in the low energy
part of the spectrum the difference between the dynamical
correlation of the true ladder and the one of an anisotropic
spin $1/2$ system, which corresponds to the strong rung
exchange limit. Additionnal experiments would be
desirable in this respect. An alternative route is to probe experimentally the
high energy part of the spectrum, since the predicted high energy excitations contain many characteristic features of the underlying model.

In connection with the compound BPCB and also on the conceptual side, there are several points which remain to be investigated:

\begin{itemize}
\item An improvement of the description of the quasi one
dimensional systems, by including in a mean field way the
effect of the other ladders in the numerical study, has been partly performed in this thesis. Nevertheless the extension to the dynamical
quantities remains to be done. This is specially important close
to the quantum critical points $h_{c1}$ and $h_{c2}$ where the
interladder coupling becomes crucial and the system undergoes a
dimensional crossover between a one dimensional and a higher
dimensional (three dimensional typically) behavior.
Understanding such a crossover is a particularly challenging
question since the system goes from a description for which a picture of essentially free fermions applies (in the one
dimensional regime) to one for which a description in terms of
essentially free bosons (the three dimensional regime) applies.

\item As discussed in Sec.~\ref{sec:thermo_BPCB} small discrepancies between the 3D order parameter measured on BPCB at finite temperature and the computed mean field value at zero temperature remain unexplained. In order to clarify these deviations an investigation of the temperature effects on the order parameter would be necessary. In addition a more precise description of the 3D structure going beyond the mean field approximation, already partially done in Ref.~\cite{Thielemann_ND_3Dladder} using a quantum Monte-Carlo technique, including eventually frustrating interladder couplings would probably help to understand the discrepancies.

\item As briefly mentioned in Secs.~\ref{sec:bpcb} and~\ref{sec:thermo_BPCB}, recent ESR measurements on BPCB \cite{cizmar_esr_bpcb} confirmed by a refined theoretical analysis~\cite{shunsuke_ladder_anisotropy}
 have pointed out small anisotropies in the exchange couplings of this compound. 
These deviations are very small and thus do not change in an essential way the results for the thermodynamic quantities and the dynamical ones presented in this thesis. However at lower energy scales and in some other observable they will lead to deviations compared to the ideal Heisenberg model which deserve further investigations. These investigations could also help to understand the small deviations of the experimental measurements on BPCB from the theoretical predictions discussed in Sec.~\ref{sec:thermo_BPCB}.

\item An extension of the dynamical results to finite temperature would be desirable. This could be used to study different
effects such as the interesting shifts and damping of the triplet modes
with temperature that have been observed in three dimensional gapped system~\cite{Ruegg_quantum_statistics_triplons}. Similar effects have been also predicted in the excitation continuum of spin-$1/2$ chains~\cite{Barthel_spectrum_1D}.

\item Recently, the presence of non LL edge singularities due to the band curvature of the dispersion relation have been pointed out in spin-$1/2$ chains~\cite{pustilnik_edge_exponent,Pereira_edge_singularities,Cheianov_treshold_singularities}. A very precise numerical analysis of the dynamical correlations close to the continuum threshold of the spin-$1/2$ ladder excitations would be required in order to investigate such fine deviations from the LL description.

\end{itemize}

\section{Outlook}

Our combined numerical-analytical methods approach is quite general and could be extended to explore many other systems. In effect, various ladder structures including frustration, dimerization or long range interaction remain to be explored. Focussing on the dynamical properties, which are difficult to access, our theoretical approach could be extended to study the excitations of these more complex systems.

Motivated by the impressive description of BPCB with a simple spin-$1/2$ ladder model  several other ladder compounds with different coupling ratios $\gamma$ have been synthesized. One of the most promising compound is $\mathrm{(C_7H_{10}N)_2CuBr_4}$ (DIMPY) \cite{shapiro_DIMPY}. This material has been investigated experimentally using INS and specific heat measurements~\cite{hong_INS_DIMPY,schmidiger_INS_DIMPY}. These first experiments agree on the ladder structure. Interestingly, DIMPY has a coupling ratio  
$\gamma\approx 2$ and thus belongs to the opposite weak coupling limit compared to BPCB. A more detailed theoretical analysis of this weak coupling regime would thus help to characterize this compound.

An other possible direction would consist in looking at the effects of disorder or doping on the ladder compounds. The former could lead to interesting physics such as a Bose glass phase. Indication of such a phase has been already observed experimentally in other magnetic compounds~\cite{hong_bose_glass} and investigated theoretically~\cite{nohadani_bose_glass,fisher_random_transverse,orignac_2spinchains,giamarchi_quantum_pinning}. In contrary, the latter would in principle lead to superconductivity~\cite{sigrist_ladder_superconductivity}. Further theoretical investigations in connection with these two experimental realizations are of course strongly required for a correct interpretation of the experiments.


\appendix


\chapter{Strong coupling expansion of a single spin ladder}\label{sec:tjmodelmapping}

In this appendix we show how the spin ladder Hamiltonian~\eqref{equ:spinladderhamiltonian} at strong coupling ($\gamma\ll1$) can be expressed in bosonic operators acting on single bonds introduced in Ref.~\cite{sachdev_bot}. This representation classifies the excitations with respect to their energy. We first derive perturbatively an effective system based on this Hilbert space organization by energy sectors. We introduce the Schrieffer-Wolff transformation that maps the physical system to the effective one, and approximate the effective system using a strong coupling expansion. We evaluate the rung densities of the ground state in the spin liquid, and derive an effective theory for the gapless regime. Furthermore we evaluate the corrections of the LL parameters from the spin chain mapping.

\section{Strong coupling expansion}

The four-dimensional Hilbert space on each rung $l$ is spanned by the states $\ket{s},$ $\ket{t^+},$ $\ket{t^0},$ and $\ket{t^-},$ (cf.~Eqs.~\eqref{equ:smult} and~\eqref{equ:tmult}), obtained by applying the boson creation operators $s^\dagger_l,$ $t^\dagger_{l,+},$ $t^\dagger_{l,0},$ and $t^\dagger_{l,-}$ to a vacuum state. A hardcore boson constraint applies on each rung $l$, i.e.
\begin{equation}\label{equ:hcfull}
\varrho_{l,s}+\varrho_{l,+}+\varrho_{l,0}+\varrho_{l,-}=1
\end{equation}
where $\varrho_{l,s}=s^\dagger_ls_l$ and $\varrho_{l,k}=t^\dagger_{l,k}t_{l,k}$ are the density operators and $k=\pm,0$.

While the Hamiltonian on the rung $H_\perp$~\eqref{equ:Hperp_parallel} is quadratic in the boson operators
\begin{equation}\label{equ:Hbosonperp}
H_\perp=\sum_{l=1}^L \left[(1-h^z/J_\perp)\varrho_{l,+} + \varrho_{l,0}+ (1+h^z/J_\perp)\varrho_{l,-}\right] -\frac34L,
\end{equation}
the chain Hamiltonian $H_\parallel$~\eqref{equ:Hperp_parallel} is quartic, and its structure is quite complex. The advantage of the boson representation reveals itself when considering the case of small $\gamma$. In that case we perform a Schrieffer-Wolff transformation of the spin ladder Hamiltonian~\eqref{equ:spinladderhamiltonian}
\begin{equation}\label{equ:heff}
H_\text{eff}=e^{i\gamma A}He^{-i\gamma A}.
\end{equation}
The Hermitian operator $A$ can be expanded in powers of $\gamma$
\begin{equation}\label{equ:Aexpansion}
A= A_1+\gamma A_2+\cdots.
\end{equation}
Thus $H_\text{eff}$ can be written in orders of $\gamma$ as
\begin{equation}\label{equ:heffexp}
J_\perp^{-1} H_\text{eff} = H_\perp+\gamma  H^{(1)} + \gamma^2  H^{(2)}+ \cdots,
\end{equation}
where
\begin{align}\label{equ:H1_def}
&H^{(1)} = H_\parallel +i[A_1,H_\perp], \\
&H^{(2)} = i[A_2,H_\perp] -\frac12 [A_1,[A_1,H_\perp]]+i [A_1,H_\parallel],\label{equ:H2_def}
\end{align}
etc. Through this expansion, the unitary transformation $e^{i\gamma A}$ can be perturbatively determined computing the $A_k$ recursively in order to eliminate the transitions between the energy sectors of excitations in $H_{\text{eff}}$. Since the first term $J_\perp H_\perp$ in Eq.~\eqref{equ:heffexp} leads to a separation of excitations on the order of the energy scale $J_\perp$ (cf.~Fig.~\ref{fig:phasediagram}.a) the decoupled bond limit provides an effective Hilbert space that contributes to each energy sector. The second term $J_\parallel H^{(1)}$ causes broadening of these bands on the order of $J_\parallel$ and can induce a complex structure within the energy bands. To obtain the desired expansion up to the first order in $\gamma$ we choose
\begin{equation}\label{equ:A1bos}
A_1=\frac{i}{4}\sum_{l}
s_l^\dagger s_{l+1}^\dagger \left(t_{l,0} t_{l+1,0} - t_{l,+}t_{l+1,-}  - t_{l,-} t_{l+1,+}\right)+\text{h.c.},
\end{equation}
where h.c.\ stands for the Hermitian conjugate.

\section{Rung state density in the spin liquid}

In the spin liquid phase, the decoupled bond limit provides the effective ground state
$\ket{0_{\text{eff}}}=\ket{s\cdots s}$ which is related to the physical ground state by
\begin{equation}
\ket{0}=e^{-i\gamma A}\ket{0_{\text{eff}}}.
\end{equation}
So the triplet density of the ground state $\langle \rho_k\rangle$ (with $k=\pm,0$) is given by
\begin{equation}\label{equ:tripletdensity}
\langle \rho_k\rangle=\langle\varrho_{l,k}\rangle=\bra{s\cdots s}e^{i\gamma A}\varrho_{l,k}e^{-i\gamma A}\ket{s\cdots s}.
\end{equation}
Using Eq.~\eqref{equ:A1bos}, and keeping only the non-vanishing corresponding terms in~\eqref{equ:tripletdensity} up to second order we get
\begin{equation}\label{equ:tripletdensity2}
\langle \rho_k\rangle\cong \gamma^2\bra{s\cdots s}A_1\varrho_{l,k} A_1\ket{s\cdots s}=\frac{\gamma^2}{8}.
\end{equation}
In the case of the compound BPCB (see Sec.~\ref{sec:bpcb}) this expansion gives $\langle \rho_k\rangle\cong0.01$, and due to the hardcore boson constraint (Eq.~\eqref{equ:hcfull}) $\langle \rho_s \rangle=\langle\varrho_{l,s}\rangle\cong0.97$. Even though we took into account only the first order term for $A$ in Eq.~\eqref{equ:Aexpansion}, this approximation of the triplet density differs from the direct numerical computations (in Fig.~\ref{fig:tripletdensity}) by only $\sim20\%$.

\section{Effective Hamiltonian in the gapless regime}

The first order term $H^{(1)}$ of the effective Hamiltonian~\eqref{equ:heffexp} is computed substituting~\eqref{equ:A1bos} into~\eqref{equ:H1_def}. This leads to $H^{(1)}$ in the form
\begin{equation}\label{equ:1storderH}
H^{(1)} =\sum_{k=0}^4 \underbrace{\frac12\sum_{l} \mathcal{H}_{k,l}^{(1)}}_{=H_k^{(1)}}
\end{equation}
where
\begin{equation} \label{equ:H0}
\mathcal{H}_{0,l}^{(1)}=s_{l+1}^\dagger t_{l,+}^\dagger   t_{l+1,+}s_{l}+ \frac12 \varrho_{l+1,+} \varrho_{l,+} + \text{h.c.},
\end{equation}
\begin{equation}\label{equ:HJperp}
\mathcal{H}_{1,l}^{(1)}=s_{l+1}^\dagger t_{l,0}^\dagger t_{l+1,0} s_{l}
+  t_{l+1,+}^\dagger t_{l,0}^\dagger t_{l+1,0} t_{l,+} + \text{h.c.},
\end{equation}
\begin{multline}\label{equ:H2Jperp}
\mathcal{H}_{2,l}^{(1)}=s_{l+1}^\dagger  t_{l,-}^\dagger t_{l+1,-} s_{l}  - \frac12\left( \varrho_{l+1,+}\varrho_{l,-}
+ \varrho_{l+1,-}\varrho_{l,+}\right)\\
+ t_{l+1,0}^\dagger t_{l,0}^\dagger t_{l+1,-}t_{l,+}
+ t_{l+1,0}^\dagger t_{l,0}^\dagger t_{l+1,+}t_{l,-} +\text{h.c.},
\end{multline}
\begin{equation}\label{equ:H3Jperp}
\mathcal{H}_{3,l}^{(1)}= t_{l+1,0}^\dagger t_{l,-}^\dagger t_{l+1,-}t_{l,0} +\text{h.c.},
\end{equation}
and
\begin{equation}\label{equ:H4Jperp}
\mathcal{H}_{4,l}^{(1)}=\varrho_{l+1,-}\varrho_{l,-}.
\end{equation}
Here we regrouped the terms such that each $J_\perp H_k^{(1)}$ acts on  the corresponding energy sector $kJ_\perp$, $k=0,1,\ldots,4$ in the gapless regime.  Note that in each sector $A_1=0$ such that to the given order in $\gamma$ the Hamiltonian~\eqref{equ:spinladderhamiltonian} corresponds to the effective Hamiltonian.

\subsection{Low energy sector}\label{sec:sec_sector_0}
When focusing on the low energy sector, the $\ket{s}$ and the $\ket{t_+}$ modes dominate the behavior and we can assume a vanishing density of $\ket{t_0}$ and $\ket{t_-}$ triplets. Thus the hardcore boson constraint~\eqref{equ:hcfull} simplifies to
\begin{equation}
\varrho_{l,s} +\varrho_{l,+}  =1
\end{equation}
and the rung Hamiltonian~\eqref{equ:Hbosonperp} to
\begin{equation}\label{equ:Hbpreduced}
H_\perp = (1-h^z/J_\perp) \sum_{l} \varrho_{l,+} -\frac34L.
\end{equation}
Furthermore the only contribution to the first order term in $\gamma$ comes from $H_0^{(1)}$. Taking this into account we obtain from Eq.~\eqref{equ:heffexp} for the Hamiltonian~\eqref{equ:spinladderhamiltonian} in the lowest energy sector
\begin{equation}\label{equ:H_sector_0_init}
 H= J_\perp H_\perp +J_\parallel H_0^{(1)},
\end{equation}
where $H_\perp$ is given by Eq.~\eqref{equ:Hbpreduced} and $H_0^{(1)}$ by Eq.~\eqref{equ:H0}. Following Ref.~\cite{mila_ladder_strongcoupling}, we map the two low energy modes onto the two states of a pseudo spin (Eq.~\eqref{equ:Hsreduced}) and replace the boson operators $s^\dagger$ and $t_+^\dagger$ with the spin-$1/2$ operators
\begin{equation}\label{equ:operatormapping}
\tilde S_l^+=t^\dagger_{l,+}s_{l}, \quad \tilde S_l^- = s^\dagger_{l} t_{l,+}, \quad \tilde S_l^z = \varrho_{l,+} -\frac12.
\end{equation}
The effective Hamiltonian is the XXZ spin-$1/2$ chain Hamiltonian, Eq.~\eqref{equ:strongcouplinghamiltonian}.

\subsection{Sector of energy $J_\perp$}\label{sec:sec_sector_Jperp}
The effective Hilbert space of the $J_\perp$ energy sector corresponds to a single $\ket{t^0}$ triplet excitation lying in a sea of singlets $\ket{s}$ and triplets $\ket{t^+}$. The effective Hamiltonian up to first order in $\gamma$ is given by
\begin{equation}\label{equ:H_sector_Jperp_init}
H= J_\perp H_\perp +J_\parallel \left(H_0^{(1)}+H_1^{(1)}\right).
\end{equation}
The excitation $\ket{t^0}$ can be interpreted as a single hole excitation in a spin chain formed by $|s\rangle$ and $|t^+\rangle$. Each rung state of this sector is identified with
\begin{equation}
|\tilde\downarrow\rangle =|s\rangle,  \quad |\tilde\uparrow\rangle = |t^+\rangle,\quad |0\rangle = |t^0\rangle. \label{equ:Hsreduced1}
\end{equation}

In this picture the Hamiltonian \eqref{equ:H_sector_Jperp_init} can be mapped onto the anisotropic t-J model
\begin{equation}
H_\text{t-J}= H_\text{XXZ} + H_\text{t} + H_\text{s-h}+\epsilon.
\end{equation}
where $\epsilon=(J_\perp+h^z)/2$ is an energy shift.

The hopping term
\begin{equation}
H_\text{t}=\frac{J_\parallel}{2}\sum_{l,\sigma}\left(c_{l,\sigma}^\dagger c_{l+1,\sigma}^\phd +c_{l+1,\sigma}^\dagger c_{l,\sigma}^\phd \right)
\end{equation}
stems from the term $J_\parallel H_1^{(1)}$ in Eq.~\eqref{equ:H_sector_Jperp_init}. Here $c^\dagger_{l,\sigma}$ ($c_{l,\sigma}$) is the creation (annihilation) operator of a fermion with pseudo spin $\sigma=\tilde\uparrow,\tilde\downarrow$ at the site $l$. Note that although we are dealing here with spin states, it is possible to 
faithfully represent the three states of each site's Hilbert space ($\ket{s}$, $\ket{t^+}$, $\ket{t^0}$) using a fermionic representation. 

Additionally, a nearest neighbor spin dependent density-density term arises
\begin{equation}
H_\text{s-h}=-\frac{J_\parallel}{4}\sum_l\left[n_{l,h}n_{l+1,\tilde\uparrow}+n_{l,\tilde\uparrow}n_{l+1,h}\right].
\end{equation}
Here $n_{l,h}$ is the density operator of the hole at site $l$.
This term stems from the nearest-neighbour interaction between the $\ket{t_+}$ triplets, i.e.~the second term in Eq. \eqref{equ:H0}.
Mapping this term onto a spin chain in the presence of a hole leads to an interaction term between the hole and the spin up state. Note that this is in contrast to the usual mapping onto a spin chain without a hole in which case the term would only cause a shift in energy.

\section{Second order perturbation and Luttinger liquid parameters}\label{sec:secondorder_LL}

The second order term $H^{(2)}$~\eqref{equ:H2_def} of the expansion~\eqref{equ:heffexp}, contains a huge amount of terms. Nevertheless considering the low energy sector, only the following terms
\begin{equation}\label{equ:H2}
H^{(2)}_0=-\frac{3}{8}\sum_{l}\varrho_{l,s}\varrho_{l+1,s}
-\frac{1}{8}\sum_{l}\left(t_{l-1,+}s_{l-1}^\dagger\varrho_{l,s}t^\dagger_{l+1,+}s_{l+1}+\text{h.c.}\right)
\end{equation}
are important. The first term in Eq.~\eqref{equ:H2} is a singlet density-density interaction which can be absorbed into the coupling of the XXZ chain, and the second term is a conditional hopping~\cite{picon_spindimer}. In order to study the effects of $H^{(2)}_0$ on the LL parameters $u$ and $K$ (see Fig.~\ref{fig:LLparameter}), we first replace the boson operators with the spin-$1/2$ operators (Eq.~\eqref{equ:operatormapping}). So the Hamiltonian~\eqref{equ:spinladderhamiltonian} in the low energy sector becomes
\begin{equation}\label{equ:Hxxz_2nd_order}
H=H_\text{XXZ}-\frac{1}{8}\sum_{l}\left[\tilde S_{l-1}^- \left(\frac{1}{2}-\tilde S^z_l\right)\tilde S_{l+1}^++{\rm h.c.}\right]+\text{const},
\end{equation}
where $H_\text{XXZ}$ is the XXZ spin-$1/2$ chain Hamiltonian (Eq.~\eqref{equ:strongcouplinghamiltonian}) with the corrected parameters
\begin{equation}\label{equ:Hxxz_newparam}
\Delta^{(2)} =  \frac{1}{2}-\frac{3}{8}\gamma\quad,\quad \tilde {h^z}^{(2)}=\tilde h^z-\frac{3}{8}J_\perp\gamma^2
\end{equation}
up to the second order in $\gamma$. For the BPCB parameters (Eq.~\ref{equ:couplingratio2}) $\Delta^{(2)}\cong 0.4$ instead of $\Delta=0.5$ for the spin chain mapping (first order approximation). The LL parameters $u$, $K$ and $A_x$ of $H_\text{XXZ}$ with the anisotropy $\Delta^{(2)}$ are computed, and we treat the conditional hopping term by approximating $1/2-\tilde S^z_l\cong 1/2-\tilde m^z$ (mean field approximation). The remaining term is then bosonized using the expression~\eqref{equ:luttingeroperator1} for $\tilde S^{\pm}(r=l)$. It leads to the corrected LL parameter $\tilde u$ and $\tilde K$ of the Hamiltonian~\eqref{equ:Hxxz_2nd_order} through the relations
\begin{equation}\label{equ:newLLparam2}
\left\{
\begin{array}{lll}
\tilde u\tilde K&=&u K + 2\pi\gamma^2J_\perp A_x\left(1/2-\tilde m^z\right)\\
\tilde u/\tilde K&=&u/K
\end{array}
\right..
\end{equation}
The corrected $\tilde u$ and $\tilde K$ are plotted in Fig.~\ref{fig:LLparameter} and clearly show the asymmetric signature of the full ladder parameters induced by the conditional hopping term in~\eqref{equ:Hxxz_2nd_order}. Note, that the lack of convergence $K\rightarrow 1$  when $m^z=\tilde m^z+1/2\rightarrow0$ is obviously an artifact of the mean field approximation $1/2-\tilde S^z_l\cong 1/2-\tilde m^z$.



\chapter{Static correlations in a finite size Luttinger liquid}\label{sec:staticcorrfinite}

In this appendix we recall the analytical expressions of the static correlation functions of the spin-$1/2$ chain~\eqref{equ:strongcouplinghamiltonian} and the spin-$1/2$ ladder~\eqref{equ:spinladderhamiltonian} models computed for finite size systems using the LL theory (Sec.~\ref{sec:luttinger_liquid}) in Ref.~\cite{hikihara_LL_ladder_magneticfield}. These correlations allow for a more precise determination of the LL parameters by comparison with the numerical DMRG computations (see Sec.~\ref{sec:LLparameters}) taking into account the boundary effects neglected in the infinite size expressions (Eqs.~\ref{equ:xxcorrelationsimplify} and~\ref{equ:zzcorrelationsimplify}) but which are obviously present in the numerical computations.


\section{Static correlation functions}

In the following, we give the relations used for the spin chains only, since from these the relations for the spin ladders can be easily inferred using the spin chain mapping (Sec.~\ref{sec:spinchainmap})
\begin{multline}\label{equ:spinchainmappingLL}
 m^z\rightarrow \tilde m^z +\frac{1}{2},\quad
\langle S^z_{l,0} \rangle\rightarrow\langle \tilde S^z_l \rangle+\frac{1}{2},\\
\langle S^x_{l,\pi} S^x_{l',\pi}\rangle\rightarrow2\langle \tilde S^x_l \tilde S^x_{l'}\rangle,
\quad \langle S^z_{l,0} S^z_{l',0}\rangle\rightarrow\langle \tilde S^z_l \tilde S^z_{l'}\rangle+\frac{1}{2}\left(\langle \tilde S^z_l\rangle+\langle \tilde S^z_{l'}\rangle\right)+\frac{1}{4}.
\end{multline}
The other correlations $\langle S^x_{l,0} S^x_{l',0}\rangle$ and $\langle S^z_{l,\pi} S^z_{l',\pi}\rangle$ are not present in the spin chain approximation. Using a weak coupling approach~\cite{furusaki_correlations_ladder} or a direct DMRG computation~\cite{hikihara_LL_ladder_magneticfield} these correlations are shown to vanish exponentially when $|l-l'|\rightarrow\infty$ leading to the gapped spectra of the corresponding energy-momentum correlation functions shown in Figs.~\ref{fig:pmcorrelationmz}.a, \ref{fig:mpcorrelationmz}.a and~\ref{fig:zzcorrelationmz}.b.

The LL analytical expressions of the correlation functions $\langle \tilde S^x_l\tilde S^x_{l'}\rangle$, $\langle \tilde S^z_l\tilde S^z_{l'}\rangle$ and the local magnetization $\langle \tilde S^z_l\rangle$ for a system of length $L$ are
\begin{align}
\langle \tilde S^x_l\tilde S^x_{l'}\rangle&= \frac{f_{1/4K}(2l)f_{1/4K}(2l')}{f_{1/2K}(l-l')f_{1/2K}(l+l')}\left[A_x(-1)^{l-l'}\right.\notag\\
&+\sqrt{2A_x B_x}\ \mathrm{sgn}(l-l')\left(\frac{(-1)^l\cos(ql')}{f_K(2l')}
-\frac{(-1)^{l'}\cos(ql)}{f_K(2l)}\right)\notag\\
&\left.
-\frac{B_x}{f_K(2l)f_K(2l')}\left(\cos[q(l+l')]\frac{f_{2K}(l-l')}{f_{2K}(l+l')}+\cos[q(l-l')]\frac{f_{2K}(l+l')}{f_{2K}(l-l')}\right)\right]\label{equ:xxcorrelationfurusaki}\\
\langle \tilde S^z_l\tilde S^z_{l'}\rangle&=
\frac{q}{2\pi}\left[\frac{q}{2\pi}+\sqrt{2A_z}\left(\frac{(-1)^l\sin(ql)}{f_K(2l)}+\frac{(-1)^{l'}\sin(ql')}{f_K(2l')}\right)\right]\notag\\
&-\frac{K}{2\pi^2}\left(\frac{1}{f_2(l-l')}+\frac{1}{f_2(l+l')}\right)\notag\\
&+\frac{A_z(-1)^{l+l'}}{f_K(2l)f_K(2l')}\left(\cos[q(l-l')]\frac{f_{2K}(l+l')}{f_{2K}(l-l')}-\cos[q(l+l')]\frac{f_{2K}(l-l')}{f_{2K}(l+l')}\right)\notag\\
&+\frac{K\sqrt{2A_z}}{\pi}\left(\frac{(-1)^l\cos(ql)}{f_K(2l)}[g(l+l')+g(l-l')]\right.\notag\\
&\left.+\frac{(-1)^{l'}\cos(ql')}{f_K(2l')}[g(l+l')-g(l-l')]\right)
\label{equ:zzcorrelationfurusaki}\\
\langle\tilde S^z_l\rangle&=\frac{q}{2\pi}+\sqrt{2A_z}\frac{(-1)^l\sin(ql)}{f_K(2l)}\label{equ:zmagnetizationfurusaki}
\end{align}
with
\begin{equation}
f_\alpha(x)=\left[\frac{2(L+1)}{\pi}\sin\left(\frac{\pi|x|}{2(L+1)}\right)\right]^\alpha,\quad
g(x)=\frac{\pi}{2(L+1)}\cot\left(\frac{\pi x}{2(L+1)}\right)
\end{equation}
and $q=2\pi L\tilde m^z/(L+1)$. We can verify that in the limit of
infinite system size ($L\rightarrow\infty$) and far from the boundaries ($|l-L/2|\ll L$ and $|l'-L/2|\ll L$) the LL correlations Eqs.~\eqref{equ:xxcorrelationfurusaki} and~\eqref{equ:zzcorrelationfurusaki} simplify to the well known power law decay Eqs.~\eqref{equ:xxcorrelationsimplify} and~\eqref{equ:zzcorrelationsimplify}, and the local magnetization~\eqref{equ:zmagnetizationfurusaki} becomes constant, $\langle \tilde S^z_l\rangle= \tilde m^z$.







\bibliographystyle{h-physrev}
\renewcommand{\bibname}{References}
\bibliography{totphys,thispaper}







\end{document}